\documentstyle[12pt,epsf]{article}

\makeatletter

\@addtoreset{equation}{section}

\makeatother

\newcommand{\tQ}{{\widetilde{Q}}}

\newcommand{\out}[1]{{\bf N \& A, I took this out:} {\it #1}}
\renewcommand{\out}[1]{}

\def\reff#1{(\ref{#1})}

\newcommand{\ba}[1]{\begin{array}{#1}}
\newcommand{\ea}{\end{array}}

\newcommand{\be}{\begin{equation}}
\newcommand{\ee}{\end{equation}}
\newcommand{\bea}{\begin{eqnarray}}
\newcommand{\eea}{\end{eqnarray}}
\newcommand{\beann}{\begin{eqnarray*}}
\newcommand{\eeann}{\end{eqnarray*}}
\newcommand{\spod}{\preceq}

 \newcommand{\tr}{{\rm Tr}\,}

\newcommand{\up}{+} \newcommand{\down}{-}

\newcommand{\ovb}{\overline{b}}

\def\phicl{\Phi^{\rm cl}}
\def\OO{{\rm O}}
\def\size{{\rm s}} 
\def\bydef{:=} 
\def\defby{=:} 
\def\re{{\rm Re\,}}
\def\const{{\rm const\,}}
\def\idty{{\leavevmode{\rm 1\ifmmode\mkern -5.4mu\else
\kern -.3em\fi I}}} \def\Ibb #1{ {\rm I\ifmmode\mkern -3.6mu\else\kern
-.2em\fi#1}} \def\Ird{{\hbox{\kern2pt\vbox{\hrule height0pt depth.4pt
width5.7pt \hbox{\kern-1pt\sevensy\char"36\kern2pt\char"36} \vskip-.2pt
\hrule height.4pt depth0pt width6pt}}}}
\def\Irs{{\hbox{\kern2pt\vbox{\hrule height0pt depth.34pt width5pt
\hbox{\kern-1pt\fivesy\char"36\kern1.6pt\char"36} \vskip -.1pt \hrule
height .34 pt depth 0pt width 5.1 pt}}}}
\def\reff#1{(\ref{#1})} 
 \def\bbbz{{\mathchoice
{\hbox{$\sf\textstyle Z\kern-0.4em Z$}} {\hbox{$\sf\textstyle
Z\kern-0.4em Z$}} {\hbox{$\sf\scriptstyle Z\kern-0.3em Z$}}
{\hbox{$\sf\scriptscriptstyle Z\kern-0.2em Z$}}}}
 \def\ibb
#1{\leavevmode\hbox{\kern.3em\vrule height 1.5ex depth -.1ex width
.2pt\kern-.3em\rm#1}}   \def\Rl
{{\Ibb R}}

\def\QED{{\hspace*{\fill}{\vrule height 1.8ex width 1.8ex }\quad} \vskip
0pt plus20pt}  
 \def\A1n{A_1\otimes\cdots\otimes A_n} 

 \def\Ran{{\rm Ran}\,}    

\def\tr{\mathop{\rm tr}\nolimits} 

\def\phi{\varphi}            
\def\epsilon{\varepsilon}    

\def\A{{\cal A}} \def\B{{\cal B}} \def\C{{\cal C}} 
 \def\HH{{\cal H}}

\def\bbbc{{\mathchoice {\setbox0=\hbox{$\displaystyle    C$}\hbox{\hbox
to0pt{\kern0.4\wd0\vrule height0.9\ht0\hss}\box0}}
{\setbox0=\hbox{$\textstyle    C$}\hbox{\hbox
to0pt{\kern0.4\wd0\vrule height0.9\ht0\hss}\box0}}
{\setbox0=\hbox{$\scriptstyle    C$}\hbox{\hbox
to0pt{\kern0.4\wd0\vrule height0.9\ht0\hss}\box0}}
{\setbox0=\hbox{$\scriptscriptstyle    C$}\hbox{\hbox
to0pt{\kern0.4\wd0\vrule height0.9\ht0\hss}\box0}}}}

\let\zed=\bbbz 
\def\L{\Lambda}
\def\lat{\zed^\nu}
\def\up{+}
 \def\down{-}
\def\tua{t_\up}
\def\tda{t_\down}
\def\tu#1{t_{\up}^{#1}}
\def\td#1{t_{\down}^{#1}}
\def\tas#1#2{t_{#1}^{#2}}
\def\nn#1{\left\langle #1\right\rangle}
\def\PP{\ ^{(1)}\!P}
\def\Tt{\underline{t}}

\begin{document}
\bibliographystyle{plain}

\title{Effective Hamiltonians and Phase Diagrams for Tight-Binding Models}

\author{
  \\
  {\normalsize Nilanjana Datta}\\[-1.5mm]
{\normalsize\it Institut de Physique Th\'eorique, EPFL}\\[-1.5mm]
{\normalsize\it CH-1015 Ecublens, Lausanne, Switzerland}\\[-1.5mm]
{\normalsize \tt datta@dpmail.epfl.ch}\\[-1mm]
  {\protect\makebox[5in]{\quad}}  
  \\[-1mm] \and {\normalsize Roberto Fern\'andez\thanks{Researcher of
      the National Research Council (CONICET), Argentina}\ 
    \thanks{Work partially supported by CNPq (grant 301625/95-6),
      FAPESP 95/0790-1 (Projeto Tem\'atico ``Fen\^omenos cr\'\i ticos
      e processos evolutivos e sistemas em equil\'\i brio'') and FINEP
      (N\'ucleo de Excel\^encia ``Fen\'omenos cr\'\i ticos em
      probabilidade e processos estoc\'asticos'' PRONEX-177/96).}}\\[-1.5mm]
{\normalsize\it Instituto de Estudos Avan\c{c}ados, 
Universidade de S\~ao Paulo}\\[-1.5mm]
{\normalsize\it Av.\ Prof.\ Luciano Gualberto, 
Travessa J, 374 T\'erreo} \\[-1.5mm]
{\normalsize\it 05508-900 - S\~{a}o Paulo, Brazil}\\[-1mm]
{\normalsize\tt rf@ime.usp.br}\\[-1mm]
  {\protect\makebox[5in]{\quad}}  
  \\[-1mm] \and
  {\normalsize J\"urg Fr\"ohlich}\\[-1.5mm]
  {\normalsize\it Institut f\"ur Theoretische Physik,
  ETH-H\"onggerberg}\\[-1.5mm] 
  {\normalsize\it CH-8093 Z\"urich, Switzerland}\\[-1.5mm]
  {\normalsize \tt juerg@itp.phys.ethz.ch} \\[-1mm]
  {\protect\makebox[5in]{\quad}}  
  \\[-2mm]
}

\maketitle

\begin{abstract}

We present rigorous results for several variants of the Hubbard model
in the strong-coupling regime.  We establish a mathematically
controlled perturbation expansion which shows how previously proposed
effective interactions are, in fact, leading-order terms of well
defined (volume-independent) unitarily equivalent interactions.  In
addition, in the very asymmetric (Falicov-Kimball) regime, we are able
to apply recently developed phase-diagram technology (quantum
Pirogov-Sinai theory) to conclude that the zero-temperature phase
diagrams obtained for the leading classical part remain valid, except
for thin excluded regions and small deformations, for the full-fledged
quantum interaction at zero or small temperature.  Moreover, the phase
diagram is stable upon the addition of arbitrary, but sufficiently
small, further quantum terms that do not break the ground-states
symmetries.  This generalizes and unifies a number of previous results
on the subject; in particular published results on the
zero-temperature phase diagram of the Falikov-Kimball model (with and
without magnetic flux) are extended to small temperatures and/or small
ionic hopping.  We give explicit expressions for the first few orders,
in the hopping amplitude, of these equivalent interactions, and we
describe the resulting phase diagram.  Our approach, however, yields
algorithms to compute arbitrary high orders.

\end{abstract}

\tableofcontents

\section{Introduction}

\subsection{Scope of the paper}

In this paper we present rigorous results for several variants of the
Hubbard model in the strong-coupling regime.  For a general class of
Hubbard-type models, we construct effective Hamiltonians with explicit
exchange interaction terms by using unitary conjugations obtained from
convergent perturbation expansions.  Furthermore, we determine ground
states and low-temperature phases in the asymmetric (Falicov-Kimball)
regime.  Our results are applications of tools recently developed in
\cite{datferfro96,datetal96} for the rigorous study of quantum
statistical mechanical lattice systems.  These tools can be applied to
systems with finite-volume Hamiltonians of the form $H_\L =H_{0\L}
+V_\L$, where: (1) the dominant part, $H_{0\L}$, is ``classical''
---in the sense of being diagonal in a tensor-product basis (see
Section \ref{prelim} below)--- and has a known energy spectrum with
spectral gaps uniformly positive in the volume $\L\subset\lat$, and
(2) the perturbation, $V_\L$, must be small in a sense to be made
precise below, but can be rather general. In particular, it may
involve interactions of infinite range as long as some
exponential-decay condition [\reff{expo} below] is satisfied.  For the
(one-band) Hubbard-type models to be studied in this paper,
$H_{0\Lambda}$ involves only on-site terms, and the perturbation is
just nearest-neighbor hopping. These models rank among the simplest
applications of our conjugation method.  To illustrate a slightly more
complicated application, we also analyze the 3-band Hubbard model.

The methodology used in this paper has two parts: (I) a
\emph{rigorously controlled perturbation expansion}, the ``conjugation
method'', adapted to the statistical mechanical setting, and (II) a
theory of stability of phases and phase diagrams known as
\emph{quantum Pirogov-Sinai theory}.  Part (I) yields a perturbative
expansion with mathematically controlled error terms, which agrees, to
leading orders, with well known expansions
\cite{bul66,harlan67,klesei73,tak77,chaspaole77,chaspaole78,grojoyric87,macgiryos88}.
Part (II) allows us to extend ---and rederive in a more systematic
fashion--- rigorous results on low-temperature phase diagrams for
models such as the Falicov-Kimball model
\cite{kenlie86,gruetal90,grujedlem92,lebmac94,ken94,mesmir96,gruetal97}.

In the statistical mechanics of classical lattice systems, there exist
efficient methods and well tested intuitions to determine and describe
ground states and equilibrium states. The mathematical study of
quantum lattice systems is less advanced. The loss of the
probabilistic framework ---noncommuting observables, complex-valued
expectations--- necessarily makes an analysis more abstract, and often
there is a divorce between the simple and concrete intuitions
practitioners resort to and a mathematical approach suitable for rigorous
proofs.

The study of quantum perturbations of classical systems offers an
excellent opportunity to establish a bridge between the classical and
quantum formalisms and to extend techniques and intuitions
from the classical realm to the quantum realm in a mathematically 
controlled manner.  This is, in essence, what our 
methods accomplish.  

Of course, the scope of such methods is limited. They are inadequate
to rigorously study strong {\em{quantum-mechanical}} correlations,
such as those appearing in Fermi liquids or superconductors, or to
analyze the spontaneous breaking of continuous symmetries. To be more
specific, our perturbation expansions and quantum Pirogov Sinai theory
converge only if the quantum perturbation $V_\Lambda$ is small, in a
sense explained later in this paper, as compared to the classical
Hamiltonian $H_{0\Lambda}$. But the perturbation expansion enables us
to decompose the Hamiltonian $H_\Lambda$ into an {\em{effective}}
classical Hamiltonian and a quantum perturbation in a way that takes
into account how the original perturbation $V_\Lambda$ lifts
degeneracies in the energy spectrum of the original classical
Hamiltonian $H_{0\Lambda}$. In particular, it enables us to show how
---as a consequence of the Pauli principle--- {\em{effective exchange
    interactions}} are generated, starting from a Hamiltonian with
{\em{spin-independent}} interactions, such as the Hubbard Hamiltonian.
The key idea underlying our perturbation technique is to perturbatively
construct a {\em{unitary conjugation}} of the original Hamiltonian
which block-diagonalizes it up to some finite order in the quantum
perturbation. It is analogous to Nekhoroshev's method in classical
mechanics. The goal of the method is to unitarily conjugate the
original Hamiltonian to an effective Hamiltonian of a form that
enables us to apply quantum Pirogov-Sinai theory. As a result, we are
able to study genuine quantum effects.

The main purpose of this paper is to illustrate the perturbation
technique (I) and 
quantum Pirogov Sinai theory (II) by analyzing concrete models, such as
Hubbard-type models, of interest in condensed matter physics. Further
applications can be found in \cite{datetal96,frorey96}.  

We emphasize that the two methods, (I) and (II) above, used in this
paper are independent of each other. The perturbation method can be
applied whenever the leading classical Hamiltonian has an appropriate,
essentially volume-independent discrete low-energy spectrum and its
ground states simultaneously minimize all local contributions to the
classical Hamiltonian (technically, the classical Hamiltonian is assumed to
be defined by a so-called $m$-potentials \cite{holsla78} ---see
Section \ref{pert}), and the quantum perturbation is ``small'' (see
Section \ref{pert}). It may or may not happen that the transformed
Hamiltonian, the ``effective'' Hamiltonian, is in a form suitable for
the application of quantum Pirogov Sinai theory.  If it is, then we
are able to rigorously investigate the low-temperature phase diagram
in certain regions of the space of thermodynamic parameters.  But even
if it is not in a form enabling us to apply method (II), it may
nevertheless yield new insight into the structure of ground states or
low-temperature equilibrium states (by enabling us to e.g. appeal to
``educated guesses''). As mentioned, the effective
Hamiltonian may, for example, include exchange interactions between
spins -- absent in the original Hamiltonian -- that enable one to
determine, at least heuristically, the type of magnetic ordering of
the ground states. We are confident that our perturbation method will
be useful in providing suitable effective Hamiltonians as starting
points for alternative approaches to the study of phase diagrams, such
as renormalization group methods.

A number of delicate concepts are involved when designing
mathematically controlled approaches.  Some of them are discussed
below (Sections \ref{forexp} and \ref{forpir}).  We would like to
start here with a general remark related to the infinite-volume limit:
As we are after statistical mechanical properties
(phases, phase transitions), we have to pass, eventually,
to the thermodynamic limit. This implies that we need to
cope with Hamiltonians for arbitrarily
large volumes $\L$, e.g. arbitrarily large squares.  To achieve this, 
we shall adopt the following two ``principles'':

\begin{itemize}

\item[(P1)]  We shall attempt to establish estimates or bounds that 
hold \emph{uniformly}
for sufficiently large volumes. 

\item[(P2)] Instead of working at the level of Hamiltonians, we shall
  work with so-called \emph{interactions}, which are families of
  \emph{local} operators $\left\{\Phi_X : X\subset\lat\hbox{
      finite}\right\}$ (so-called $|X|$-body terms), in terms of which
  the Hamiltonians are given by $ H_\Lambda = \sum_{X \cap \Lambda \ne
    \emptyset} \Phi_X$.  Interactions are defined {\em{without}}
  reference to the total volume of the system. Hence operations
  performed on interactions automatically apply to all finite-volume
  Hamiltonians.
\end{itemize}

Next, we proceed to introducing some of the models studied in this
paper and to summarizing our main results.

\subsection{Models and results}

\subsubsection{One-band Hubbard model}

We shall  study the one-band Hubbard model and
some of its variants.  These are models of two types of
fermions ---which, for concreteness, will be called spin-up and
spin-down electrons, but which can also be considered to be different
species of particles, like ions and spin-polarized electrons ---whose
quantum dynamics, on a finite square lattice $\Lambda$, is governed
by the Hamiltonian  
\bea
H_\Lambda &=& U \sum_{x \in \Lambda} n_{x +} \, n_{x -} 
-  \mu_{ +} \sum_{x \in \Lambda} n_{x +} -
\mu_{ -} \sum_{x \in \Lambda} n_{x -} \nonumber\\
&& \quad {} - \sum_{\nn{xy}}\sum_{\sigma =+, -} [\tas{\sigma}{[xy]}\,
c^{\dagger}_{x\sigma} c_{y\sigma} + \tas{\sigma}{[yx]}\,
c^{\dagger}_{y\sigma} c_{x\sigma}] \;.
\label{gen}
\eea Here $c^{\dagger}_{x\sigma}$ and $c_{x\sigma}$ are electron
creation- and annihilation operators satisfying the usual
anticommutation relations [see \reff{anticom1}-\reff{anticom2}].  The
number operator of an electron of spin $\sigma =+, -$ at the site $x$
is $n_{x \sigma} = c^{\dagger}_{x\sigma} c_{x\sigma}$.  The chemical
potentials of the electrons with up-spin and down-spin are denoted by
$\mu_{ +}$ and $\mu_{ -}$, respectively.  The symbol $\nn{xy}$ (resp.\ 
$[xy]$) denotes an unordered (resp.\ ordered) pair of nearest neighbor
sites of the lattice.

The Hubbard model is the simplest model believed to embody the
essential physics governing strongly correlated electron liquids. It
exhibits the interplay between Coulomb repulsion and the kinetic
energy term of electrons.  The model is widely studied in the context
of phenomena such as magnetism and high-$T_c$ superconductivity, which
are believed to result from such an interplay. It is a lattice model
of interacting fermions in the tight--binding approximation.  The
model retains only that part of the (screened) Coulomb repulsion which
manifests itself when two electrons of opposite spin occupy the same
lattice site, the strength of this interaction being given by the
coupling constant $U$.  The kinetic energy term describes the hopping
of electrons between nearest-neighbor sites.  We allow a
direction-dependent hopping: the hopping amplitude of electrons of
spin $\sigma$ from a site $x$ to a site $y$ is denoted by the symbol
$\tas{\sigma}{[yx]}$.  For the Hamiltonian to be self-adjoint,
$\tas{\sigma}{[yx]}$ must be the complex-conjugate of
$\tas{\sigma}{[xy]}$.  Complex hopping amplitudes are encountered in
the presence of an external magnetic field.

If $\tas{+}{[yx]}=\tas{-}{[yx]}$ the Hamiltonian \reff{gen} does not
change under a rotation of the spin quantization axis. Hence the model
has an SU($2$) symmetry.  This is the case for the standard one--band
Hubbard model for which $\tas{+}{[yx]}=\tas{-}{[xy]}=t$, for all
nearest neighbor sites $x,y$.  The highly asymmetric regime
$\left|\tas{+}{[yx]}\right|\ll\left|\tas{-}{[yx]}\right|$ will be
called the \emph{Falicov-Kimball regime}.  The limiting case
$\tas{+}{[yx]}=0$ constitutes the Falicov-Kimball model (with magnetic
flux \cite{gruetal97}).  In this setting, up-spin electrons are
interpreted as {\em{static ions}}, while down-spin electrons are
viewed as scalar quantum-mechanical particles.  If the magnetic flux
is zero, $\tas{-}{[yx]}$ is real and independent of $x,y$.

We study the model \reff{gen} in the strong coupling limit $U\gg
\left|\tas{\sigma}{[yx]}\right|$.  This corresponds to the situation
in which the Coulomb repulsion is much larger than the bandwidth of
(uncorrelated) electrons. The large on-site repulsion forbids double
occupancy of a site at zero temperature. In this limit, the terms in
the first line on the RHS of \reff{gen} act as a leading classical
part, $H_{0\Lambda}$, while the hopping term can be treated as a
perturbation, $V_\Lambda$.  As we shall see, this gives rise to a
relatively simple application of our perturbation scheme, due to the
on-site character of the leading interaction.

\begin{figure}
\vspace{1cm}
\epsfxsize=\hsize
\epsfbox{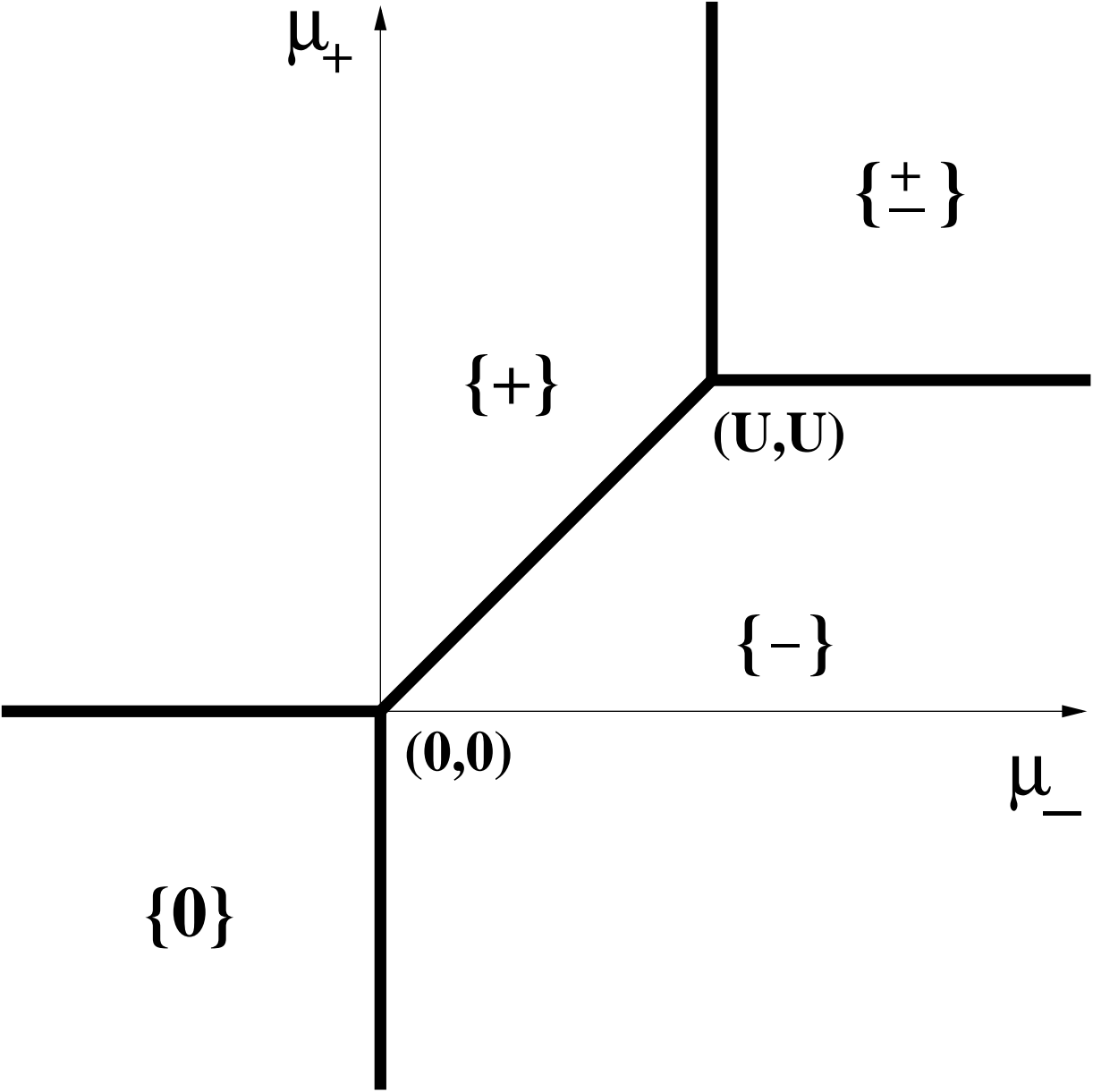}
\vskip 0.2in
\caption{Zero-temperature phase diagram for $H_{0\Lambda} = U \sum_{x
    \in \Lambda} n_{x +} \, n_{x -}  -  \mu_{ +} \sum_{x \in \Lambda}
  n_{x +} - \mu_{ -} \sum_{x \in \Lambda} n_{x -}$.  Ground states are
  defined by spin configurations, which at each site, are as denoted
  in curly brackets}

\label{fig1}
\end{figure}

To discuss our results, we first present the zero--temperature phase
diagram of the dominant part, $H_{0\Lambda}$, of the Hamiltonian, in
the plane of the chemical potentials.  This is a purely classical
interaction, and ground states are tensor products of single-site spin
states minimizing each on-site term.  It is given in Figure
\ref{fig1}. The symbols $\{+\}$, $\{-\}$, $\{\pm\}$ and $\{0\}$ are
used to denote ground states of $H_{0\Lambda}$ in which each lattice
site is occupied by an up-spin electron, a down-spin electron, two
electrons of opposite spins and no electron, respectively. All the
boundary lines in this phase diagram are lines of infinite degeneracy;
e.g.  for $0<\mu_{+} = \mu_{-} < U$ the ground state is infinitely
degenerate, because all singly occupied configurations are equally
likely.  The point $\mu_+ = \mu_- = 0$ (orig in) and the point $\mu_{+}
= \mu_{-}= U$ also correspond to infinitely many ground states. At the
origin, each site is either empty or singly occupied, whereas, at the
point $(U, U)$, each site is either singly- or doubly occupied.
Next, we describe our results for the Hubbard model.

\paragraph{(I) Controlled perturbation expansion}\mbox{}

\begin{figure}
\vspace{1cm}
\epsfxsize=\hsize
\epsfbox{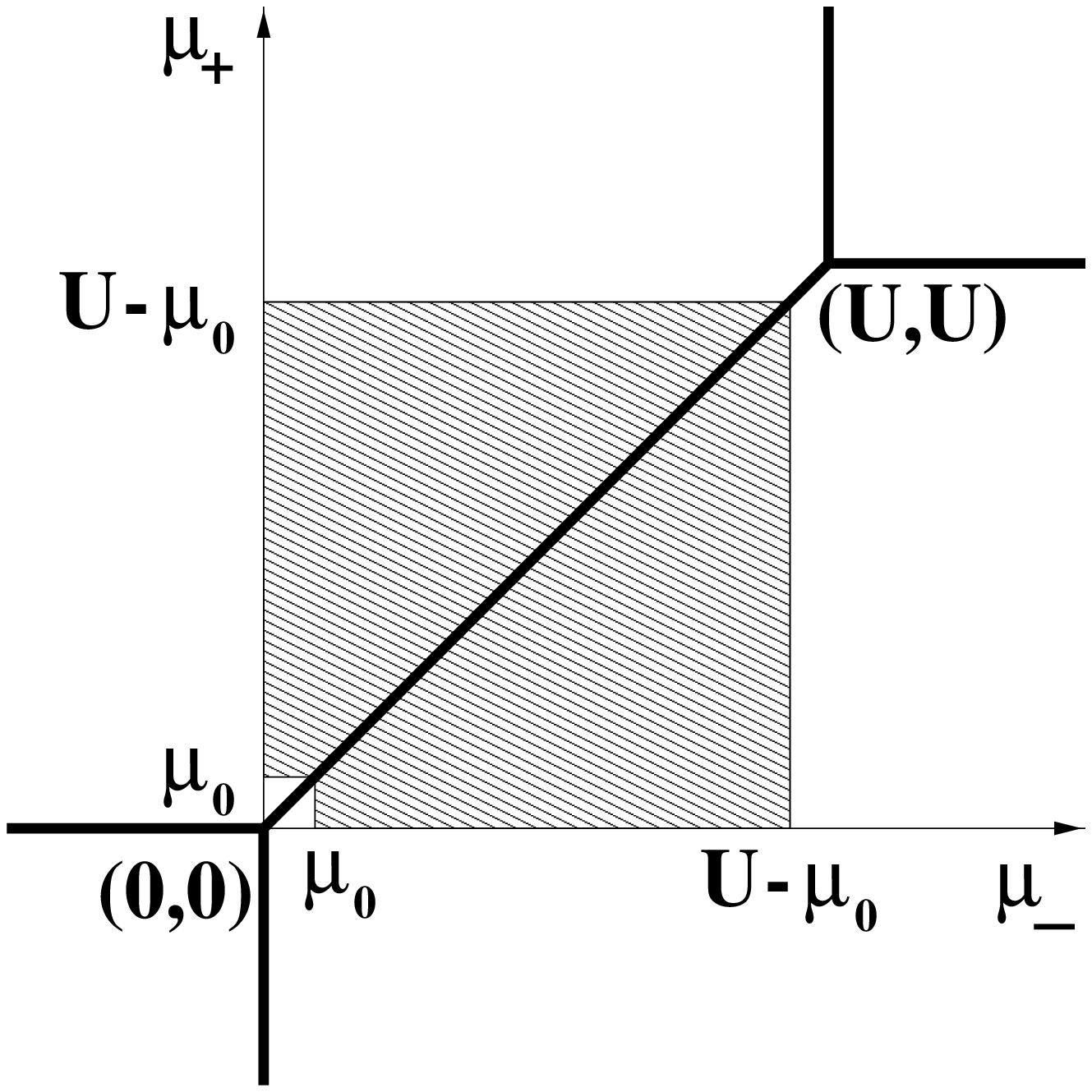}
\vskip 0.2in
\caption{Shaded region to which we restrict most of the analysis of
  the generalized Hubbard model \reff{gen}.}
\label{shaded}
\end{figure}

We restrict our attention to those values of the chemical potentials
for which the spectrum of the Hamiltonian $H_{0\Lambda}$ can be
decomposed into two spectral bands separated by an energy gap which is
large compared to the hopping amplitudes.  We discuss mainly the
half-filled regime (Section \ref{perthub}), for which the lower band
corresponds to the subspace of states with singly occupied sites.  In
this regime, to have a large-enough spectral gap, we must consider a
region in the positive quadrant in the plane of chemical potentials in
which $\mu_0 < \mu_{+}, \mu_{-} < U-\mu_0$, where
$\mu_0 \gg \left|\tas{\sigma}{[yx]}\right|$.
This is the shaded region in 
the plane of chemical potentials shown in Figure \ref{shaded}. 

For values of the chemical potentials in this shaded region, we
determine, for each finite order $n$, 
an interaction $\left\{\Phi^{(n)}_X\right\}$ that is
\begin{itemize}
\item[(R1)] \emph{equivalent} to the one corresponding to \reff{gen},
  i.e.\ it gives rise to Hamiltonians related to \reff{gen} by a
  unitary transformation, and
\item[(R2)] \emph{block-diagonal} to order $t^n/U^{n-1}$, where
  $t=\sup_{x,y} \left\{\left|\tas{-}{[yx]}\right|,
  \left|\tas{+}{[yx]}\right|\right\}$, in the sense that the matrix elements
  of the operators $\Phi^{(n)}_X$ between the lower and higher
  bands (of $H_{0\Lambda}$) are of the prescribed order.  
\end{itemize}

Transformations of this sort have been devised before, see for
instance
\cite{bul66,harlan67,klesei73,tak77,chaspaole77,chaspaole78,grojoyric87,macgiryos88},
though not in a manner suited for rigorous analysis, as we comment
upon below, in Section \ref{forexp}.  Our expressions, obtained with
the technique developed in \cite{datetal96}, yield, when restricted to
the lower band, the well known leading order terms: Heisenberg
interaction + terms of order $t^4/U^3$ favoring valence-bond states
(and breaking the degeneracy among those) + terms of order $t^6/U^5$ +
$\ldots$ In particular, they coincide exactly with those of
\cite{macgiryos88}, in the half-filled regime.  (Our approach can be
viewed as a mathematical justification of the series proposed there.)
Nevertheless, each transformed interaction
$\left\{\Phi^{(n)}_X\right\}$ has terms, $\Phi^{(n)}_X \ne 0$, with
arbitrarily large $|X|$ but correspondingly small order in $t/U$,
which have been largely ignored in the literature.  They are, however,
crucial if one wants to determine whether 
$\left\{\Phi^{(n)}_X\right\}$ is indeed an honest interaction or only
some auxiliary object corresponding to the first few terms in an
asymptotic series.  Our approach is designed to settle such issues.
In fact, it yields complete mathematical control over the transformed
interactions (and the unitary transformations giving rise to them):
\begin{itemize}
\item[(a)]  We obtain explicit bounds for the error terms, and show
  that each transformed interaction $\left\{\Phi^{(n)}_X\right\}$ is
  well defined in the strong-coupling regime. (In fact, it is
  exponentially summable.)  We provide explicit estimations of the
  radius of convergence, $t_0(n)$, and of the exponential decay of the terms
  $\Phi^{(n)}_X$ in the diameter of $X$. 
\item[(b)]  We provide explicit algorithms to construct the full
  interaction $\left\{\Phi^{(n)}_X\right\}$, not only its first few
  orders (restricted to the lowest band).  
\end{itemize}

As we work in terms of interactions, all the expressions, bounds and
estimations are, of course, independent of the volume.
For the Falicov-Kimball regime, our analysis implies the various
estimations obtained, by more cumbersome means, in previous
studies of the model 
\cite{kenlie86,gruetal90,grujedlem92,lebmac94,ken94,mesmir96,gruetal97}
(see also the review \cite{grumac96}).  
\medskip

Similar calculations, that we leave to the reader, can be done in the
region of the plane of chemical potentials close to the origin. The
low-lying band is now formed by states with no doubly 
occupied sites.  This yields a family of transformed interactions 
$\left\{\widetilde\Phi^{(n)}_X\right\}$ whose restrictions to the lower
band yield the well known perturbation series often associated with the keyword
``Gutzwiller projections''.  In fact, these series, which start with the
very popular $t$-$J$ interaction plus (usually neglected) three-site
terms, are the ones usually reported in the physics literature (see
e.g.\ \cite{harlan67,chaspaole77,chaspaole78,macgiryos88}).  The
half-filled case discussed above is obtained after a further
projection onto the half-filled band.  Our formalism has, in relation
to previous publications, the advantages (a) and (b) described above.

\paragraph{(II) Phase diagram at low temperatures}\mbox{}

\begin{figure}
\vspace{1cm}
\epsfxsize=\hsize
\epsfbox{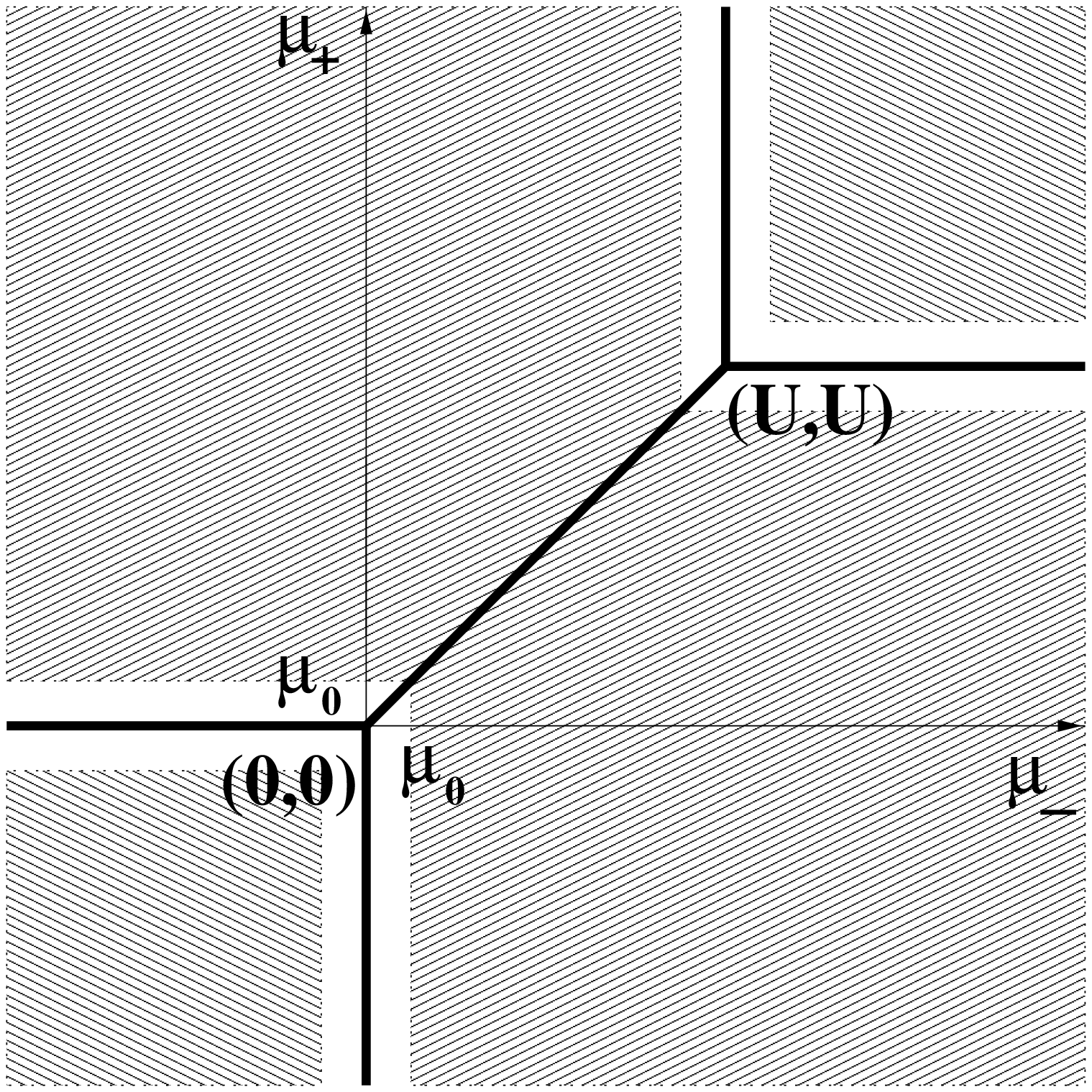}
\vskip 0.2in
\caption{Region (shown shaded) where low-temperature properties of the
  model \reff{gen} are determined in the strong-coupling regime}
\label{fig3}
\end{figure}

We are able to infer low-temperature properties of the model
\reff{gen}, in the strong coupling regime, in the shaded region of
Figure \ref{fig3}, namely away from the manifolds of coexistence of
ground states, and at and near the segment along the line
$\mu_+=\mu_-$, as long as we stay clear of its infinitely degenerate
endpoints.  As we will show later by using contour arguments, we
restrict our attention to dimensions $d\ge2$ and hopping amplitudes
$\tas{\sigma}{[yx]}$ that are \emph{periodic} with respect to
translations of the sites $x,y$. Of course, rigorous control of the
ground states and low-temperature equilibrium states {\em{on the
    ``unperturbed'' coexistence line}} $\mu_+ = \mu_-$ with the help
of contour expansions will be achieved only for the models
{\em{without}} continuous $SU(2)$- symmetry.
We shall prove the following results:
\begin{itemize}
\item[(R3)] Inside the (open) regions where $H_{0\Lambda}$ has a
  unique ground state, we apply the cluster expansion developed in
  \cite[Section 6]{datferfro96} to conclude that the ground state is
  stable for small hopping amplitudes and at low temperatures.  That
  is, in these ranges, one finds an equilibrium state for which the
  observables have expectations close to their values for the
  corresponding ground state.  Such a state can be constructed as the
  limit of a sequence of states associated with finite boxes with a
  ground state configuration as boundary condition, and the cluster
  expansion allows for a visualization (in a very precise sense) of
  the state as ``ground state + quantum and thermal fluctuations''.
\item[(R4)] For the Falicov-Kimball regime,
  $\left|\tas{+}{[yx]}\right|\ll\left|\tas{-}{[yx]}\right|$, we can
  apply the Pirogov Sinai approach developed in
  \cite{datferfro96,datetal96} in conjunction with the transformed
  interactions $\left\{\Phi^{(n)}_X\right\}$ described above, to
  analyze the shaded region of Figure \ref{fig3} at or close to the
  line $\mu_+=\mu_-$, for strong Coulomb coupling $U$.  In this way we
  recover and extend a number of previously published results, prove
  some expected features of the low-temperature phase diagram and open
  the way for further systematic analysis of the latter.
\end{itemize}
In more detail, the results alluded to in (R4) are as follows.

\paragraph{Falicov-Kimball model without flux} ($\tas{-}{[yx]}=t$ for
all $x,y$)
\begin{itemize}
\item[(R4.1)] As soon as $t\neq 0$, the quantum hopping breaks the
  infinite-degeneracy present on the line $0 < \mu_+ = \mu_- < U$.
  Instead, two N\'eel-ordered ground states appear on this line,
  and in its neighborhood, and remain stable for low temperatures.
  For the Falicov-Kimball model itself (i.e.\ when $\tas{+}{[yx]}$ is
  exactly zero), this fact has been proven in \cite{kenlie86} for
  chemical potentials on the line itself, and in \cite{lebmac94} for
  potentials in its vicinity.  Our methods show that the same is true
  even if one adds some small ionic hopping (a fact that also follows
  from the treatment in \cite{mesmir96}) or, indeed, any other
  translation-invariant quantum perturbation, which could be
  long-range but must decay exponentially; see \reff{expo}.
\item[(R4.2)] By considering the transformed interaction,
  $\left\{\Phi^{(2)}_X\right\}$, we can actually show that, for low
  temperatures and $t/U$ small, a phase diagram as in Figure
  \ref{fig4a} appears for the Falicov-Kimball model:
\begin{figure}
\vspace{1cm}
\epsfxsize=\hsize
\epsfbox{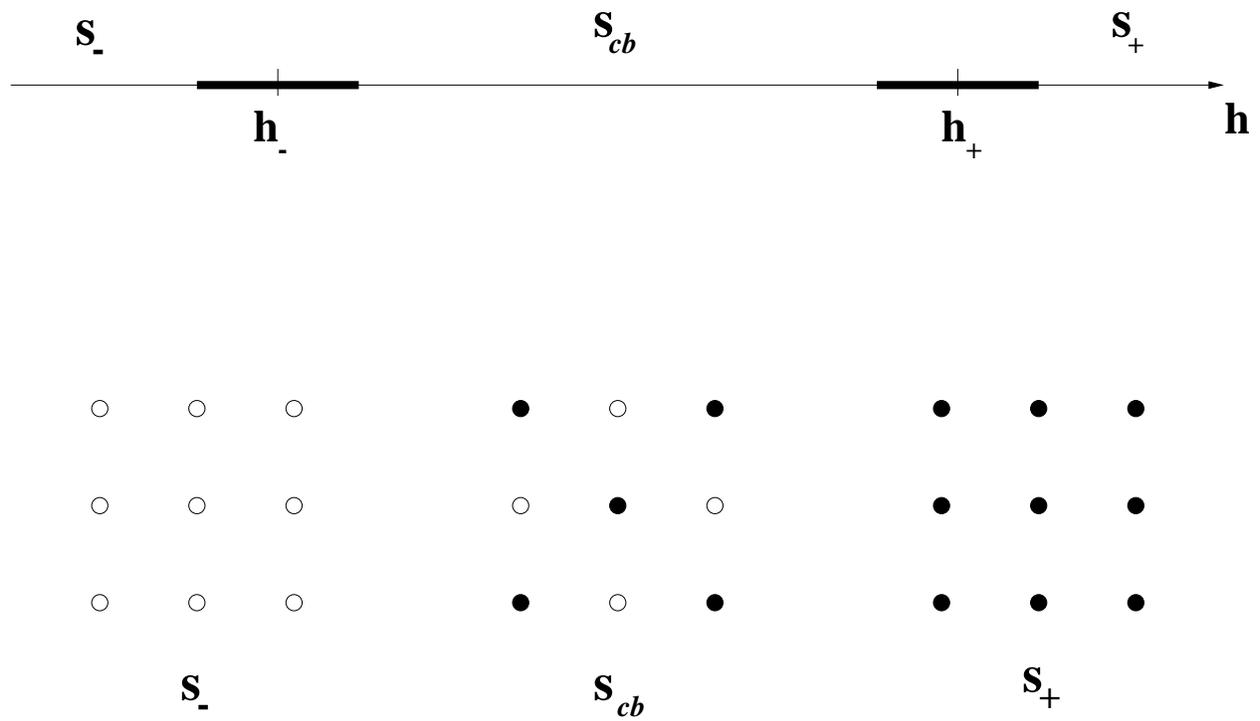}
\vskip 0.2in
\caption{Phase diagram at zero and low temperatures of the Falicov-Kimball
  model to order $t^2/U$, as a function of $h=\mu_+ - \mu_-$.  The
  labels $S_+$, $S_-$ and $S_{cb}$ refer to states that are
  quantum and thermal fluctuations of the indicated configurations or
  its translations (open circles correspond to ``$+$'' particles and
  closed circles to ``$-$'' particles).  Thick lines represent excluded
  regions of width $\OO(t^4/U^3)$ located around $h_{\pm}=\pm 4t^2/U + %
  \OO(\varepsilon_{\beta t})$.  The small correction
  $\varepsilon_{\beta t}$ is discussed in Sections \protect
  \ref{stabdiag} and \protect \ref{phase} below}
\label{fig4a}
\end{figure}
\begin{figure}
\vspace{1cm}
\epsfxsize=\hsize
\epsfbox{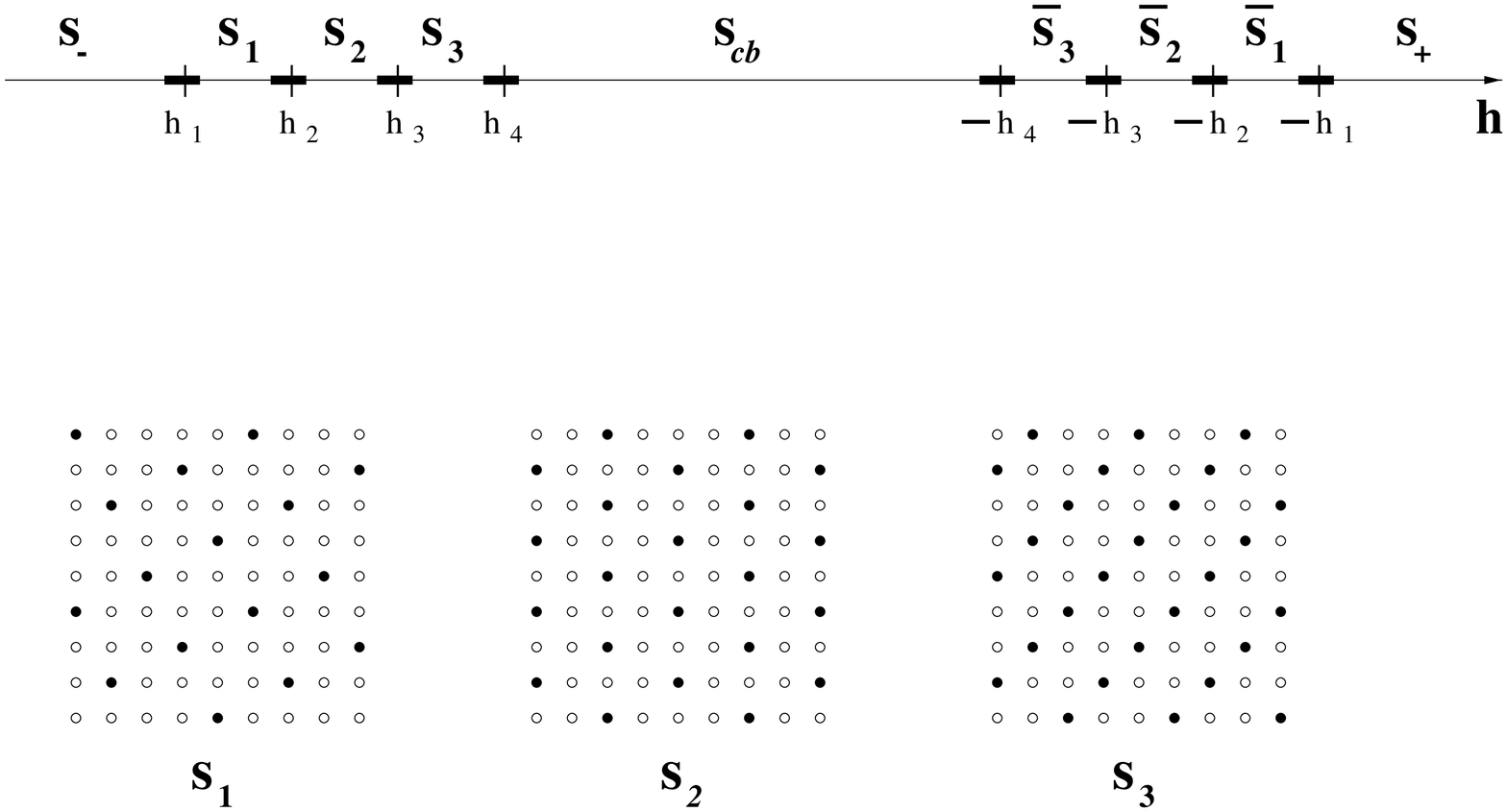}
\vskip 0.2in
\caption{Phase diagram at zero and low temperatures of the Falicov-Kimball
  model to order $t^4/U^3$, as a function of $h=\mu_+ - \mu_-$.
  Besides the states of Figure \ref{fig4a}, there appear those
  corresponding to quantum and thermal fluctuations of rotations and
  tranlations of the configurations $S_i$ (depicted) and $\overline
  S_i$ (obtained from the $S_i$ by a
  $\bullet\longleftrightarrow\circ$ interchange).  The 
  excluded regions (thick lines) have width $\OO(t^6/U^5)$ and their
  location are determined by the values $h_1 = - 4t^2/U - 4 t^4/U^3 +
  \OO(\varepsilon_{\beta t})$,   $h_2 = - 4t^2/U + 16 t^4/U^3+
  \OO(\varepsilon_{\beta t})$, $h_3= - 4t^2/U  + 48 t^4/U^3 +
  \OO(\varepsilon_{\beta t})$ and $h_4 = -4t^2/U + 84 t^4/U^3+
  \OO(\varepsilon_{\beta t})$}
\label{fig4b}
\end{figure}
There are two ``strips'' of width of order $t^4/U^3$, located around
the lines $\mu_+-\mu_- = \pm 4t^2/U$ for zero temperature, and some
smooth deformation [Section 3.2.1] of them for small nonzero
temperatures, where nothing can be said.  Between them, the states are
N\'eel ordered, while towards the upper left the state is (a thermal
perturbation of the ground state) all-``+'', and to the bottom right
it is all-``$-$''.  Such a phase diagram has already been established
at zero temperature \cite{gruetal90,ken94}, but the proof of its
stability for small temperatures escaped the methods available up to
now.  Recently, however, this stability has been established by
contour methods, related to ours, but replacing our partial
diagonalization by a study of ``restricted ensembles''
\cite{kotuel98}.
\item[(R4.3)] Resorting to the next order in the transformation of
  interactions, i.e., to $\left\{\Phi^{(4)}_X\right\}$, we can see in
  more detail what goes on inside the two excluded regions discussed
  above.  For $d=2$, each region splits into four ``strips'', now of
  width of order $t^6/U^5$, that require further investigation (Figure
  \ref{fig4b}), and, between them, a finite number of states appear,
  which are quantum and thermal perturbations of certain
  configurations with different proportions of ``+'' and ``$-$''
  particles.  At zero temperature, these features have been formally
  calculated in \cite{grujedlem92} and rigorously established in
  \cite{ken94} and, by a cluster-expansion method, in \cite{mesmir96}.
  Our formalism shows that they are also present at small nonzero
  temperature, except that the excluded manifolds suffer a smooth (and
  small) deformation.
\item[(R4.4)] At the next order, $t^6/U^{5}$, each of the eight
  precedent strips is expected to open up into eight finer strips
  outside of which there is only a finite ground state degeneracy.
  Our formalism would then yield stability of this structure when the
  temperature is increased or further small quantum perturbations are
  added.  In fact, a ``devil's staircase'' is expected to appear in
  this way, to successively larger orders $t^n/U^{n-1}$.  While we do
  not work out any details we think that this paper provides enough
  tools for a motivated reader to pursue these issues: It provides
  algorithms for the higher-order transformed interactions
  $\left\{\Phi^{(n)}_X\right\}$, the certainty that the remainders are
  rigorously controlled to the specified order, and conditions to
  predict stability.  Of course, in order to get precise results, one
  has to come up with an accurate determination of the ground states
  of the classical leading part of the transformed interactions.  This
  tends to be the hardest part of the work.  The beautiful method
  introduced in \cite{ken94} (see also the application in
  \cite[Appendix A]{gruetal97}) could be very helpful at this point.
\end{itemize}

An advantage of our methods is their robustness with respect to
perturbations: All the results sketched above remain valid if we add
an arbitrary, but small enough, quantum perturbation that respects the
symmetries connecting the different ground states (typically invariant
under translations and $90^\circ$-rotations).  In particular, they
extend to the Falicov-Kimball \emph{regime} (i.e.
$\left|\tas{+}{[yx]}\right|$ small, but not necessarily zero), and to
models including additional perturbations, which can possibly be
long-range but must decay exponentially; see \reff{expo}.

Our methods are not suited for the study of the phase diagram of the
$SU(2)$-symmetric Hubbard model, because continuous symmetry breaking
is accompanied by the appearance of gapless modes.  In particular, we
cannot apply our phase-diagram technology to the effective
Hamiltonians that would be obtained via Gutzwiller projections
($t$-$J$ interaction + higher orders).  But we emphasize that the
perturbation technique (I) is applicable in such situations.

\paragraph{Falicov-Kimball model with flux} [$\tas{-}{[yx]}=t
\exp(i\theta_{yx})$, with $\theta_{yx} \in (0,2\pi)$]

The previous results can be extended to the model with flux, starting
from the study of ground states presented in \cite{gruetal97}.  For
this case, the results (R4.1) and (R4.2) remain valid, because the
properties involved are independent of the flux.  Regarding (R4.3),
the positions of the excluded regions do depend on the flux, but the
ground states around them do not (see Section 3.2.1 in
\cite{gruetal97}).  Our methods show that these ground states remain
stable at small temperature and/or under the addition of small
translation- and rotation-invariant quantum perturbations (e.g.\ ionic
motion).

\subsubsection{3-band Hubbard model}
\label{spia.3}

\begin{figure}
\vspace{1cm}
\epsfxsize=\hsize
\epsfbox{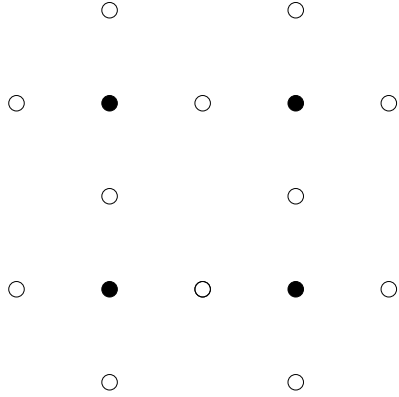}
\vskip 0.2in
\caption{Lattice for the 3-band Hubbard model.  The open circles
  denote oxygen sites while the closed circles denote copper sites}
\label{fig5}
\end{figure}

As an illustration of the use of our perturbation technique for models
whose leading interaction is not strictly on-site, we consider the
$3$--band Hubbard model, which was first introduced by Emery to
describe the behaviour of the charge carriers in the $\mathrm{CuO}_2$
planes of cuprates exhibiting high--$T_c$ superconductivity. The model
is defined on a two--dimensional lattice, a unit cell of which
contains one copper and two oxygen atoms (Figure \ref{fig5}).

The copper atoms form a square lattice, and there is an oxygen atom 
between each nearest neighbor copper--copper pair. 

The charge carriers are spin--$1/2$ holes which can hop between the
$d$--orbital of a copper atom and the $p$--orbital of an adjacent
oxygen atom. We shall refer to this as the hopping of holes between a
copper site and an oxygen site.  The Hamiltonian governing the model
defined on a finite lattice $\Lambda := A \cup B$, where $A$ denotes
the sublattice of copper atoms and $B$ denotes the sublattice of
oxygen atoms, is given as follows:
\bea H_\Lambda &=& \biggl\{U_d \sum_{x \in \Lambda\atop{x \in A}}
n^d_{x +} \, n^d_{x -} + U_p \sum_{y \in \Lambda\atop{y \in B} }
n^p_{y +} \, n^p_{y -} + U_{pd} \sum_{\nn{xy} \in \Lambda\atop{x\in
    A\,,y \in B}}
n^d_{x} \, n^p_{y}\nonumber\\
&&{}+ \epsilon_d \sum_{x \in \Lambda\atop{x \in A}} n_{x}^d + \epsilon_p
\sum_{y \in \Lambda\atop{y \in B}} n_{y}^p \biggr\} + \biggl\{
t_{pd}\,\sum_{\nn{xy}\atop{x\in A\,,y \in B}}\sum_{\sigma =+,-} 
[p^{\dagger}_{y\sigma}
d_{x\sigma} + d^{\dagger}_{x\sigma} p_{x\sigma}]\biggr\} \nonumber\\
&=:& H_{0\Lambda} + V_\Lambda({t_{pd}})\;.
\label{3hub}
\eea 
The number operator for a hole at the site $x \in A$ is $n_{x}^d =
(n_{x+}^d + n_{x-}^d)$, where $n_{x \sigma}^d = d^{\dagger}_{x\sigma}
d_{x\sigma}$, for $\sigma \in \{+, -\}$, with $d_{x\sigma}^{\dagger}$
and $d_{x\sigma}$ denoting fermion creation- and annihilation
operators for a hole of spin $\sigma$ at the site $x$. The
corresponding number operator for a hole at the site $y \in B$ is
$n_{y}^p =(n_{y+}^p + n_{y-}^p)$ where $n_{y \sigma}^p =
p^{\dagger}_{y\sigma} p_{y\sigma}$, with $p^{\dagger}_{y\sigma}$ and
$p_{y\sigma}$ the fermion creation- and annihilation operators
for a hole of spin $\sigma$ in the $p$--orbital of an oxygen atom.
The coefficients $U_d$ and $U_p$ denote the strengths of the on--site
repulsions between two holes of opposite spin when they occupy a
copper site and an oxygen site, respectively. In addition to this
on--site interaction, two holes experience a Coulomb repulsion of
strength $U_{pd}$ when they occupy adjacent copper and oxygen sites.
The on--site energies for the copper and the oxygen are denoted by the
symbols $ \epsilon_d$ and $ \epsilon_p$, respectively, and the charge
transfer gap is given by 
\be \Delta := \epsilon_p - \epsilon_d.
\label{charge}
\ee 
The hopping amplitude of a hole between a copper site and an oxygen
site is denoted by $t_{pd}$ and is proportional to an overlap integral
of atomic orbitals. Often a small additional term representing the
direct hopping of a hole between the $p$--orbitals of two oxygen atoms
is included in the Hamiltonian, but we neglect it here.

Band-structure calculations of cuprates exhibiting
high--$T_c$ superconductivity show that it is not unreasonable to study
this model assuming 
\be
\Delta > 0\quad;\quad U_d \gg  U_p >U_{pd} > 0\quad;\quad U_d \gg  \Delta
\quad;\quad t_{pd} >0,
\ee
and considering the {\em{strong coupling regime}} $U_d \gg t_{pd}$.
(For instance, in \cite{hybschchr89,hybetal90}, the estimated values
are $\Delta =3.6 {\rm eV}$, $U_d=10.5 {\rm eV}$, $U_p=4 {\rm eV}$,
$U_{pd}=1.2 {\rm eV}$ and $t_{pd} = 1.3 {\rm eV}$.)  We observe that:
(i) we can treat $V_\Lambda({t_{pd}})$ as a perturbation of
$H_{0\Lambda}$; (ii) since $\Delta > 0$, holes prefer to reside on
copper sites rather than on oxygen sites, and (iii) since $U_d$ is
positive and large, double occupancy of holes at a copper site is
energetically unfavorable.

We restrict our attention to the situation in which the total number
of holes, $N_h$, in the lattice $\Lambda$ is equal to the total number
of copper sites $|A|$: 
\be N_h = |A|.
\label{qf}
\ee 
For this choice, a ground-state configuration of $H_{0\Lambda}$ has a
single hole at each copper site, there being no holes at the oxygen
sites.  [Equivalently, we could introduce suitable ``chemical
potentials'' and work in the region of parameter space where the
ground states have this property.]  The hole at a copper site can,
however, be either spin-up or spin-down. Hence, the ground state of
$H_{0\Lambda}$ has a $2^{|A|}$--fold spin degeneracy.

By resorting to our perturbation expansion, we obtain an equivalent
interaction that only involves copper atoms, which includes an
antiferromagnetic Heisenberg term.  This does not settle the question
of whether the hopping term $V_\Lambda({t_{pd}})$ lifts the
spin--degeneracy, because, as mentioned above, our methods to
determine stable phases do not work in the situation where a
continuous symmetry [here $SU(2)$] can be broken spontaneously.
Nevertheless, our methods show that the transformed interaction is
mathematically well defined and provide error estimates which should
be valuable in future studies.  The form of the transformed
interaction is well known. (Indeed, the fact that the leading terms
coincide with those of the transformed interaction for the one-band
model is an argument favoring the study of the latter rather than of
the more cumbersome 3-band model).

It should be noted that the perturbative analysis of models with 
classical interactions that are not on-site is more difficult, technically, 
than the one of models with single-site classical interactions, such as
the single-band Hubbard model.  Our presentations in Sections 
\ref{pert} and \ref{3band} will serve to illustrate this point.

\section{The Perturbation Technique}
\label{pert}

\subsection{Foreword}
\label{forexp}

In quantum mechanics, Rayleigh-Schr\"odinger perturbation theory can
be formulated as the perturbative construction of unitary conjugations
transforming a Hamiltonian of the form $H = H_0 + t V$ into a new
Hamiltonian, $H' = H_0'(t) + V'(t)$, with the property that $H'_0(t)$
is block-diagonal (in the eigenbasis of $H_0$), and $V'(t)$ is a
perturbation of order $t^{n+1}$. The operators $H'_0(t)$ and $V'(t)$
are given in terms of power series, and it is important to analyze the
convergence properties of these series.  Kato's analytic perturbation
theory \cite{kat66,reesim78} offers criteria that guarantee
convergence: Suppose we attempt to block-diagonalize the Hamiltonian
$H$ with respect to a spectral projection of $H_0$ corresponding to a
bounded subset of the spectrum of $H_0$ separated from its complement
by an energy gap $\delta > 0$. Let us assume that the perturbation $V$
is relatively bounded with respect to $H_0$, e.g. in the sense that
there exist finite (non-negative) constants $a$ and $b$ such that
\begin{equation}
\|V \psi\| \;\le\; a\, \|H_{0}\psi\| + b\,\|\psi\|,
\label{rel0}
\end{equation}
for all vectors $\psi$ in the domain of $H_{0}$.  Then perturbation theory for
the block-diagonalization of $H$ with respect to the given spectral
projection converges, provided 
\be
|t| a < 1, \quad \frac{|t|}{\delta} (a+b) < C,
\ee
where $C$ is a constant that depends on the choice of the spectral 
projection.

Unfortunately, this result cannot be applied directly to the study of
systems of quantum statistical mechanics. The reasons are as follows.

\begin{itemize}
\item[(SM1)] First, we do not just want to block-diagonalize a single
  Hamiltonian $H$ but a {\em{family}} of Hamiltonians $\{H_\Lambda\}$
  indexed by an increasing sequence of finite regions $\Lambda$ of an
  infinite lattice.  In this setting, Kato theory is {\em{not}}
  applicable, because the norms $\|H_{0\L}\|$ and $\|V_\L\|$ are
  typically proportional to the volume $|\L|$, and this usually
  implies that the constant $b$ in \reff{rel0} is proportional to the
  volume $|\L|$, too. As a consequence, the radius of convergence of
  plain Rayleigh-Schr\"odinger perturbation series tends to zero, as
  $|\L|$ tends to $\infty$. This makes it impossible to establish
  facts valid in the thermodynamic limit by straightforward use of
  Rayleigh-Schr\"odinger-Kato analytic perturbation theory.
  ({\em{Even}} if $H_{0\L}$ has a {\em{finite}} ground state
  degeneracy, {\em{uniformly}} in $\Lambda$, a naive application of
  Kato theory does not prove that the perturbation series for the
  perturbed ground state energy density, for example, converges
  uniformly in $\Lambda$.)

\item[(SM2)] Second, to study the phase diagram of the system, the
  transformed Hamiltonians $H'(t)$ must be expressed again in terms of
  {\em local}\/ interactions.  That is, the unitary conjugations must
  lead to a block-diagonalization at the level of local
  {\em{interactions}}, rather than just at the level of Hamiltonians.
  To satisfy this requirement, we find that these conjugations must
  themselves be defined by (exponentials of) sums of local operators.
  Furthermore, the original interaction and the local operators
  defining the conjugation must decrease exponentially in the size of
  the lattice region in which they are localized. 
\end{itemize}

Keeping in mind these requirements, it is natural to search for
transformations whose generators $S$ are sums of local terms, $S=\sum
S_Y$, where each $S_Y$ is localized in the region $Y$, in the sense
that it commutes with local operators $\Phi_X$, provided $X$ and $Y$
are disjoint. To achieve this, we incorporate two crucial ingredients
in our expansion:
\begin{itemize}
\item[(I1)] Defining \emph{local} ground states (of the leading part
  $H_{0\L}$), we decompose each term $\Phi_X$ according to how it acts
  on states that are local ground states only in the vicinity of $X$.
  This is achieved by introducing suitable \emph{local} projectors
  ($P^0_Y$, $P^1_Y$ and $P^2_Y$ below).
\item[(I2)] We introduce \emph{protection zones} to decouple
  sufficiently far separated excitations.  If we normalize the
  $H_0$-energy of local ground states to be 0 then the $H_0$-energy of
  a configuration of local excitations (local deviations from ground
  states) separated by protection zones is additive in the
  $H_0$-energies of the excitations.
\end{itemize}

Ingredient (I2) is irrelevant for the discussion of models, such as
the one-band Hubbard model, where the leading interaction is on-site.
In such models, the $H_0$-energy of excitations localized at different
sites is additive. But this is untypical, and in the study of more
general models (I2) plays an important role. Regarding ingredient
(I1), one finds that it has been used by some authors
\cite{chaspaole77,grojoyric87,macgiryos88,esketal94}, but it has been
ignored by others \cite{bul66,klesei73,tak77,chaspaole78,tas96}.  The
latter use ``global'' projections involving all sites in a given
region $\Lambda$, which lead to volume-dependent expansions with a
shrinking convergence radius as mentioned in (SM1).  Nevertheless,
they appear to obtain correct perturbative results for certain
infinite-volume expectation values and for (intensive) thermodynamic
quantities, like ground-state energy densities. The problem with their
methods is that they do not offer a handle to {\em{proving}}
(rigorously) that, indeed, their predictions are correct or to
determine whether the corresponding series are {\em{convergent}}
rather than just asymptotic.  Expansion schemes such as ours, yielding
volume-independent effective interactions, are suitable to answer such
mathematical questions.  To accomplish this, they must, however, be
supplemented with an appropriate construction of ground- and Gibbs
states for the transformed Hamiltonian. This is not an easy matter, in
general, and it is here that quantum Pirogov-Sinai theory will enter
the scene.

We are able to provide a complete picture of the models analyzed in
Section \ref{phase}, below, where Gibbs states are constructed through
quantum Pirogov-Sinai theory.  In regions of stable phases, the
cluster expansion methods we use imply the analyticity of expectations
and of thermodynamic potentials as functions of the couplings.  For
those models, therefore, we are able to prove that the series obtained
for expectations of local observables and for intensive quantities are
well defined and convergent.

\subsection{Notation, definitions and preliminary results}
\label{prelim}
In this section, we review the different elements of our general
perturbation technique.  (For further details see \cite{datetal96}.)

\subsubsection{Bases and operators}

\paragraph{The Hilbert space.}

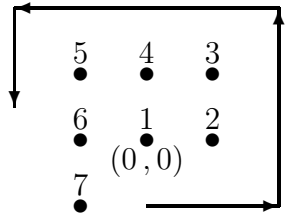
\begin{figure}
\begin{center}
\begin{picture}(100,75)(-50,-25)
\put(0,0){\circle*{5}}
\put(0,8){\makebox(0,0){$1$}}
\put(0,-8){\makebox(0,0){$(0\,,0)$}}
\put(25,0){\circle*{5}}
\put(25,8){\makebox(0,0){$2$}}
\put(25,25){\circle*{5}}
\put(25,33){\makebox(0,0){$3$}}
\put(0,25){\circle*{5}}
\put(0,33){\makebox(0,0){$4$}}
\put(-25,25){\circle*{5}}
\put(-25,33){\makebox(0,0){$5$}}
\put(-25,0){\circle*{5}}
\put(-25,8){\makebox(0,0){$6$}}
\put(-25,-25){\circle*{5}}
\put(-25,-17){\makebox(0,0){$7$}}
\thicklines
\put(0,-25){\vector(1,0){50}}
\put(50,-25){\vector(0,1){75}}
\put(50,50){\vector(-1,0){100}}
\put(-50,50){\vector(0,-1){37.5}}
\end{picture}
\end{center}

\caption{Spiral order in $\zed^2$}
\label{spiral}
\end{figure}

We consider a quantum--mechanical system associated with a finite
subset $\Lambda$ of a $\nu$--dimensional lattice $\zed^\nu$.  To each
site $x$ in the lattice is associated a Hilbert space $\HH_x$. We
require that there be a Hilbert space isomorphism $\phi_x : \HH_x
\rightarrow \HH$, for all $x \in \zed^\nu$, $\HH$ being a fixed,
finite-dimensional Hilbert space.  The Hilbert space of the region
$\Lambda$ is given by the ordered tensor product space \be \HH_\Lambda
= \otimes_{x \in \Lambda} \HH_x.
\label{tenp}
\ee To avoid ambiguities in the definition of the tensor product
\reff{tenp}, we choose a total ordering (denoted by the symbol
$\spod$) of the sites in $\zed^\nu$. For convenience we choose an
order with the property that, for any finite set $X$, the set
$\overline{X}:= \{ z\in \zed^\nu, z \spod X\}$ of lattice sites which
are smaller than $X$, or belong to $X$, is finite.  For example, for
$\nu=2$ we can use the {\em{spiral order}} depicted in Figure
\reff{spiral}.
\smallskip

\paragraph{Tensor product basis of $\HH_\Lambda$.}

Let $I$ be an index set and let $\{ e_j\}_{j \in I}$ be an orthonormal
basis of $\HH$. Then $\{ e_{j}^x\}_{j \in I}$, where $ e_{j}^x =
\phi_x^{-1} e_j$, is an orthonormal basis of $\HH_x$. 

A configuration $\omega$ on $\Lambda$ is an assignment
$\{j_x(\omega )\}_{x \in \Lambda}$ of an element $j_x \in I$ to each 
$x \in \Lambda$. For $X \subset \Lambda$, let $\omega_X$ denote the
 restriction of the configuration $\omega$ to the subset $X$. 
The set of all configurations in $\Lambda$ is denoted by
\be
\C_\Lambda := \{ \omega\, |\, j_x(\omega) \in I, x\in \Lambda \}\;.
\ee
There is a one-to-one
correspondence between configurations $\omega$ in $\Lambda$ and
basis vectors 
\be e(\omega)\; := \; \otimes_{x \in \Lambda} e_{j_x(\omega)}^x
\ee
of $\HH_\Lambda$.  A tensor product basis of $\HH_\Lambda$ is given by 
$\{e(\omega)\}_{\omega \in \C_\Lambda}$.
\medskip

\paragraph{Local algebras of gauge--invariant operators.}

Let $\B(\HH_\Lambda)$ be the algebra of all bounded operators on
$\HH_\Lambda$. Let $\{U(\theta) \,|\, \theta \in {\bf{R}} \}$ be a 
one--parameter unitary group on $\HH_\Lambda$ (gauge group of the
first kind) with the property that the vectors $e(\omega)$ defined in
(3.5) are eigenvectors of $U(\theta)$, for all $\theta$. 
For any $X \subseteq \Lambda$, we define a local algebra 
$\A_X \subset \B(\HH_\Lambda)$ of gauge-invariant operators 
with the following properties:
\begin{itemize}
\item {If $\{ e(\omega) \}_{\omega \in \C_\Lambda}$ is an
    {\em{arbitrary}} tensor product basis of $\HH_\Lambda$ then 
\be
    \langle e(\omega')\,,\, a\, e(\omega) \rangle = 0,
\label{scalar}
\ee
unless $\omega'|_{\Lambda \setminus X} = \omega|_{\Lambda \setminus
  X}$, for {\bf all} operators $a \in \A_X$.
(Here $\langle\psi\,,\, \phi\rangle$ denotes the
scalar product of two vectors $\psi$ and $\phi$ in $\HH_\Lambda$.) }
\item{ $U(\theta)\, a\, U(\theta)^* = a$, for all $a \in \A_X$, for arbitrary
$X \subseteq \Lambda$; (gauge-invariance).}
\item{If $X \subset Y$ then $\A_X \subset \A_Y$.}
\item{If $X \cap Y = \emptyset$ then 
\be
\label{commut}
[a,b] = 0,
\ee
for any $a \in \A_X$ and any $b \in \A_Y$.}
\end{itemize}
\smallskip

\noindent
{\em{Example:}
Gauge-invariant polynomials in fermion creation- and annihilation
operators}\/.  Let 
\be
\{c_{x\sigma}, c_{x\sigma'}^\dagger\,|\, x \in \Lambda, \sigma = +,\,-\}
\ee
denote the usual fermion creation- and annihilation
operators satisfying the canonical anticommutation relations
\be
\{c_{x\sigma}, c_{y\sigma'}\} = \{c_{x\sigma}^\dagger,
c_{y\sigma'}^\dagger\}= 0, 
\label{anticom1}
\ee
and 
\be
\{c_{x\sigma}, c_{y\sigma'}^\dagger\}= \delta_{xy} \,
\delta_{\sigma\sigma'}.
\label{anticom2}
\ee
Of course, $\sigma$ is the spin index. We define 
\be
U(\theta) = \exp (i \theta N)\quad , \quad{\rm where}\quad N = \sum_{x
  \in \Lambda \atop{ 
\sigma \in \{ +,- \}}}  c_{x\sigma}^\dagger  c_{x\sigma}.
\ee
The local algebras $\A_X$, $X \subseteq \Lambda$, are defined by
\bea
\A_X &=& \{ a\,|\,\, a \,\, {\hbox{is an arbitrary polynomial
in}}\,\, c_{x\sigma},  
c_{x\sigma}^\dagger, \, x \in X,\, \sigma \in \{ + ,- \}, \nonumber\\
& & \quad  {\hbox{with
the property that}} \,\, U(\theta)\, a\, U(\theta)^* =a \}.
\eea

\paragraph{The algebra of local, gauge-invariant operators.}

Since $\A_X \subset \A_Y$, for $X \subset Y$, we may consider the
(inductive) limit $\bigvee_{X \nearrow \zed^\nu} \A_X$. The algebra
$\A$ is defined to be the closure of $\bigvee_{X \nearrow \zed^\nu}
\A_X$ in the operator norm and is called the algebra of all local,
gauge-invariant operators.

\subsubsection{Interactions and ground states}

\paragraph{Classical operators and -interactions.}

An operator $a$ is ``classical'' w.r.t. a {\em{given}} tensor product
basis $\{e(\omega)\}$ if and only if $e(\omega)$ is an eigenvector of
$a$, for all $\omega \in \C_\Lambda$. An interaction $\{\Phi_{0X}\}$
is said to be a ``classical interaction'' (w.r.t.
$\{e(\omega)\}_{\omega \in \C_\Lambda}$) if and only if the following
conditions hold:
\begin{enumerate}
\item{$\Phi_{0X} \in \A_X$ is a classical operator w.r.t $\{
e(\omega)\}$, for all $X \subseteq \Lambda$.} 
\item{More precisely, 
\be 
\Phi_{0X} e(\omega) = \Phi_{0X}(\omega) e(\omega)\;, 
\quad \Phi_{0X}(\omega) \in \Rl,
\ee
where $\Phi_{0X}(\omega)$ only depends on $\omega|_X$. 

[Note that 2. follows from 1. and from \reff{commut}].}
\end{enumerate}

\paragraph{$m$-potentials.}

Let $\Phi_0=\{\Phi_{0X}\}$ be a classical interaction.  We shall
always assume that there is at least one configuration, $\omega_0$,
minimizing all $\Phi_{0X}$, i.e.,
\be
 \Phi_{0X}(\omega_0) = \min_{\omega}  \Phi_{0X}(\omega),
\ee
for all $X$; (this assumption holds if $\Phi$ is given in terms of a 
so-called $m$-potential \cite{holsla78}).  We normalize
the operators $\Phi_{0X}$ such that $ \Phi_{0X}(\omega_0) =0$, for all $X$.
Thus
\be
\Phi_{0X}(\omega) \ge 0 \quad {\hbox{ and}} \quad 
\min_\omega \Phi_{0X}(\omega) =0,
\ee
for all $X$. 

\paragraph{Local ground states.}

A configuration $\omega$ is said to be a {\em{local ground-state
configuration}} for a classical interaction $\Phi_0$ and a subset $X$
of the lattice if 
\be
\Phi_{0Y}(\omega) = 0 \quad {\hbox{for all}} \quad Y \subseteq X.
\ee
The local Hamiltonians $H_{0X} = \sum_{Y \subseteq X} \Phi_{0Y}$ thus
have the property that 
\be
H_{0X}e(\omega) =0,
\label{gndnorm}
\ee
whenever $\omega$ is a local ground state configuration for $X$. A state
$e$ is said to be a {\em local ground state}\/ for a region $X$ if
$e$ is an arbitrary linear combination of the vectors $e(\omega)$,
where $\omega$ ranges over the local ground state configurations for
$X$. 

We shall always assume that the spectrum of $H_{0X}$ has a gap above 0
which is bounded from below by a positive $X$-independent constant,
for all bounded sets $X$.  This assumption can be derived from
suitable assumptions on the classical interaction $\Phi_0$ ($\Phi_0$
should be an $m$-potential with a finite number of periodic ground
states, see \cite{holsla78}).

\paragraph{The full interaction.}

We consider quantum lattice systems whose dynamics is encoded in an 
interaction $\Phi$ which is assumed to be of the form
\be
\Phi = \Phi_0 + Q(\lambda),
\label{inter}
\ee 
where $\Phi_0 = \{\Phi_{0X}\}$ is a finite-range classical
interaction given in terms of an $m$-potential, and $Q(\lambda) =
\{Q_X(\lambda)\}$ is a perturbation interaction with $\lambda=
\{\lambda_1, \ldots, \lambda_k\}$ a family of real or complex
perturbation parameters. For brevity, we shall say that a set $X$ is
a \emph{classical bond} if $\Phi_{0X}\neq 0$, and a \emph{quantum
  bond} if $Q_X(\lambda)\neq 0$.  The operators $\Phi_{0X}$ and
$Q_X(\lambda)$ belong to the local algebra $\A_X$. A polynomial
$p(\lambda)$ is said to be of degree $n$ if \be p(\lambda) = \sum
c_{n_1\ldots n_k} \lambda_1^{n_1} \ldots \lambda_k^{n_k}, \ee with
$c_{n_1\ldots n_k} = 0$ whenever $\sum_{j=1}^k n_j >n$. We assume that
the leading order in $\lambda$ of the interaction $Q(\lambda)$ in
\reff{inter} is of degree $1$ (i.e., linear in $\lambda$).

We also assume that the interaction  $Q(\lambda) = \{Q_X(\lambda)\}$ is
either of finite range or decays exponentially in the size of its
support, i.e., that there exist an $r >0$ such that \be \sum_{X \ni 0}
\sqrt{{\rm{tr}}[Q_X^*(\lambda) Q_X(\lambda)]} e^{rs(X)} < \infty,
\label{expo}
\ee 
where $s(X)$ denotes the cardinality of the smallest connected subset
of the lattice which contains $X$. The interaction  $Q(\lambda)$ will 
be treated as a perturbation. 
\medskip

{\em{Remark}}.  Roughly speaking, we shall always split the
Hamiltonian $H_\Lambda$ into a sum of an unperturbed operator
$H_{0\Lambda}$ and a perturbation $V_\Lambda(\lambda)$ in such a way
that all low-lying eigenstates of $H_{\Lambda}$, corresponding to
eigenvalues separated from the rest of the spectrum of $H_\Lambda$ by
a positive ($\Lambda$-independent) gap, correspond to {\em{degenerate
    ground states}} of $H_{0\Lambda}$. More precisely, we choose the
classical interaction $\Phi_0$ in such a way that all local low-energy
eigenstates of the local Hamiltonians $H_{0X}$, for $X \subset
\Lambda$ are {\em{exactly degenerate}} in energy; small
{\em{degeneracy-\-lifting}} terms are systematically put into the
interaction $Q(\lambda)$ defining the perturbation
$V_\Lambda(\lambda)$. In a general exposition of our methods, this is
a very convenient way of ensuring that $H_0$-energies of sufficiently
far separated local excitations are {\em{additive}}. (Of course, in a
concrete model, additivity of $H_0$-energies of local excitations may
hold for independent reasons; see e.g. Section \ref{perthub}.)

\subsubsection{Projection operators.  Protection zones}

\paragraph{``Diagonal'' and ``off-diagonal'' operators.}

Given a partition of unity 
\be {\bf{1}} = \sum_{j=1}^N P_j \quad ,
\quad P_i P_j = \delta_{ij} P_j, \quad P_i^* =P_i, \ee 
and an operator
$Q$, we set 
\be Q^{ij} = P_i Q P_j.  
\ee 
We will call the operators $Q^{ii}$ ``diagonal'', and the operators
$Q^{ij}$, $i \ne j$, ``off--diagonal'' (with respect to the given
partition of unity).

\paragraph{Protection zones}
To deal properly with the local nature of different operators we
define, for any $x \in \zed^\nu$,a so-called {\em{R-plaquette centered
    at x}} : \be W_x := \{ y\in \zed^\nu : |y_i - x_i | \le R, \,
{\rm{for}} \, 1 \le i \le \nu\},
\label{wx}
\ee
where $R$ is the range of the interaction $\Phi_0$. 
For any finite set $X \subset \zed^\nu$, its covering by $R$-plaquettes is
denoted by
\be
B_X := {\bigcup}_{x \in X} W_x.
\label{bx}
\ee
The set $B_X\setminus X$ can be interpreted as a {\em protection
zone}\/ around $X$.

\paragraph{Local projections.}

We introduce some special projection operators on the Hilbert space
$\HH_{\Lambda}$:
\begin{enumerate}
\item {If $Y \subset \Lambda$ then $P^0_{Y}$ is the orthogonal projection 
onto local ground states for $Y$. If $Z \subseteq Y$ then
$P^0_{Z} \supseteq P^0_Y$.}
\item {The orthogonal projection onto the space of states which are 
ground states on $B_X \setminus X$, but fail
to be ground states on $X$:
\be
P^1_{B_X} := P^0_{B_X \setminus X} - P^0_{B_X}.
\label{p1}
\ee
Hence $P^1_{B_X}$ projects onto the space of states which have an excitation 
{\em{localized}} in $X$.
The set ${B_X \setminus X}$ acts as a ``protection zone'' introduced to ensure
the additivity of energies of disconnected excitations [see \reff{add}
below].}
\item {The projection onto states with excitations in the ``protection zone''
${B_X \setminus X}$:
\be
P^2_{B_X} := {\bf{1}} - P^0_{B_X \setminus X},
\label{p2}
\ee
where ${\bf{1}}$ is the identity operator.}
\end{enumerate}

\paragraph{Additivity of excitation energies.}

For each finite $X \subset \zed^\nu$, we can decompose the Hamiltonian
$H_{0\Lambda}$ as follows:
\be
H_{0\Lambda} = {\overline{H}}_{0X} + {\overline{H}}_{0B_X^c} 
+ H_{0\, B_X \setminus X},
\label{decom}
\ee
where $B_X^c = \Lambda \setminus B_X$, 
\be
H_{0A} = \sum_{Y \subset A} \Phi_{0Y}.
\label{ll.1}
\ee
and
\be
{\overline{H}}_{0A} = \sum_{Y \cap A \ne \emptyset} \Phi_{0Y}
\label{hover}
\ee
If $\psi \in \Ran \, P^0_{B_X \setminus X}$ then
\bea
H_{0\Lambda}\psi &=& ({\overline{H}}_{0X} + H_{0\,B_X\setminus X} 
+ {\overline{H}}_{0B_X^c})\psi \nonumber\\
&=& ({\overline{H}}_{0X} + {\overline{H}}_{0B_X^c})\psi.
\label{add}
\eea
This follows, because $H_{0\,B_X \setminus X}\psi = 
H_{0\,B_X \setminus X}\,P^0_{B_X \setminus X} \psi=0$. Hence, on
 states $\psi$ corresponding to ground-state configurations in the 
region $B_X \setminus X$, energies are additive.

\subsubsection{Operator identities}

\paragraph{The Lie-Schwinger series}

With this expression we shall refer to the following expansion:
\begin{eqnarray} 
e^A B e^{-A} &=& B + [A,B] + \frac{1}{2!}[A,[A,B]] + \cdots \nonumber\\ 
&= & \sum_{n=0}^{\infty} {1\over n!} \,{\rm ad}^n A(B)\;. 
\label{schwing} 
\end{eqnarray} 
Here we use the notation
\begin{equation}
 {\rm ad}A(B) = [A,B]\,;\, {\rm ad}^2 A(B) = [A,[A,B]]\,;\,
{\rm ad}^n A(B) = [A, \,{\rm ad}^{n-1}\,(B)]
\end{equation}
and the convention
\begin{equation}
{\rm ad}^0 A(B) = B\;.
\end{equation} 

\paragraph{The ``${\rm ad}^{-1}$'' operation}

Given a self-adjoint operator $H$ whose spectrum consists only of
eigenvalues $E_i$, $i=1,2,\ldots$, and an operator $Q$ which is purely
``off-diagonal'' with respect to the partition of unity
given by the projections $P^i$ onto the eigenspaces corresponding
to each $E_i$, we define
\be
{\rm ad}^{-1}H(Q) = \sum_{ij} P^i \frac {Q}{E_i - E_j}  P^j.
\label{jf.0}
\ee
The right hand side is well defined, because $Q$ is an off-diagonal
operator. Of course, ${\rm ad}^{-1}$ is the operation inverse to ${\rm
  ad}$, i.e., 
\begin{equation} 
[H \,,\, {\rm ad}^{-1}H(Q)] = Q .
\label{pia1}
\end{equation}

\subsection{First-order perturbation theory}
\label{firstpert}
The first step in our perturbation technique consists in eliminating
``off--diagonal'' operators of lowest order in $\lambda$. Here
``off-diagonal'' refers to the partitions of unity \be {\bf{1}} =
P^0_{B_X} + P^1_{B_X} + P^2_{B_X}.
\label{p2u}
\ee
In the following,  we usually suppress the explicit dependence of the
operators $Q_X(\lambda)$ on $\lambda$.
We rewrite $Q_X \equiv Q_X(\lambda)$ as 
\be
Q_X = Q^{00}_{B_X} +  Q^{01}_{B_X} + Q^{R}_{B_X},
\label{qdecom}
\ee 
where $ Q^{00}_{B_X}$ is a  ``diagonal'' operator defined by 
\be
 Q^{00}_{B_X} := P^0_{B_X} Q_X P^0_{B_X},
\label{q00}
\ee
$ Q^{01}_{B_X}$ is an ``off--diagonal'' operator given by
\be
 Q^{01}_{B_X} := P^0_{B_X} Q_X P^1_{B_X} +  P^1_{B_X} Q_X P^0_{B_X},
\label{q01}
\ee
and $Q^{R}_{B_X}$ is the remainder
\be
 Q^{R}_{B_X} := P^2_{B_X} Q_X P^2_{B_X} + P^1_{B_X} Q_X P^1_{B_X}.
\label{qr}
\ee
Note that, since $Q_X \in \A_X$, the operators $ P^i_{B_X} Q_X P^2_{B_X}$
and $ P^2_{B_X} Q_X P^i_{B_X}$ vanish, for $i=0,1$. 

Using the decomposition \reff{qdecom} we can write the Hamiltonian
$H_{\Lambda}$ as follows:
\be
H_{\Lambda} =  H_{0\Lambda} + V^{00}_{\Lambda} +  V^{01}_{\Lambda} +
V^{R}_{\Lambda},
\label{hdecom}
\ee
where $V^{00}_{\Lambda}= \sum_{X \subset \Lambda}  Q^{00}_{B_X}$, 
$V^{01}_{\Lambda}= \sum_{X \subset \Lambda}  Q^{01}_{B_X}$,
and $V^{R}_{\Lambda}= \sum_{X \subset \Lambda}  Q^{R}_{B_X}$.

There is a slight subtlety, at this point, related to {\em{boundary
    conditions}}: The definition of the operators $P^i_{B_X}$,
$i=0,1,2,$ and of $Q^{00}_{B_X}$, $Q^{01}_{B_X}$, $Q^R_{B_X}$ for
regions $X$ with the property that $B_X \cap \Lambda^c$ is
{\em{non-empty}} must, in general, be modified in such a way that the
boundary conditions imposed on the configurations on $\Lambda^c$ are
properly taken into account. We shall not enter into a
detailed discussion of this (primarily technical and usually
straightforward) issue. In fact, we shall usually think of periodic
boundary conditions for which the issue does not arise.

For notational simplicity we henceforth suppress the subscript
$\Lambda$.  Following the guidelines explained in the Introduction, we
search to eliminate the first--order off--diagonal terms
$Q^{01}_{B_X}$, through a unitary transformation $U^{(1)}(\lambda)=
\exp(S_1(\lambda))$, where $S_1(\lambda)$ is of degree $1$ and is
given by a sum of local operators: \be S_1(\lambda):= \sum_X
S_{1B_X}(\lambda).
\label{sone}
\ee 
In the sequel we shall also suppress the explicit
$\lambda$-dependence of the operators $S_1(\lambda)$ and
$S_{1B_X}(\lambda)$.  In order to gain mathematical control of the
resulting expressions, it is essential \cite{datetal96} that each
$S_{1B_X}$ be a {\em local}\/ operator.

By the Lie-Schwinger series \reff{schwing}, the unitary operator
$U^{(1)}$ yields the transformed Hamiltonian 
\begin{eqnarray}  
H^{(1)} &:= & e^{S_1}\,H\, e^{-S_1} \nonumber\\ 
&=& H_0 +V^{00} + V^{01} + V^R\nonumber\\
&& \quad {} + \sum_{n \ge 1} \frac{1}{n!} {\rm
  ad}^{n}{S_1}(H_0 + V^{00} + V^{01} + V^R).
\label{errh.1}
\end{eqnarray}
To leading order, the operators $S_1$, $V^{00}$, $V^{01}$ and $V^R$
depend linearly  on the perturbation parameter $\lambda$. 
We wish to eliminate $V^{01}$. This leads to the condition 
\begin{equation}
{\rm ad}H_0(S_1)\,=\,V^{01}.
\label{errh.10.0}
\end{equation}
If this equation is satisfied, the Hamiltonian \reff{errh.1} becomes
---singling out the terms up to second order in $\lambda$---
\begin{eqnarray}  
H^{(1)} &= & H_0 +V^{00} + V^R + V_2\nonumber\\
&& \quad {} + \sum_{n \ge 2} \frac{1}{n!} {\rm
  ad}^{n}{S_1}(V^{00} + V^R + \frac{n}{n+1}V^{01}),
\label{errh.1.0}
\end{eqnarray}
with
\begin{equation}
  \label{pia.20}
  V_2 \bydef {\rm ad}S_1(V^{00} + V^R + \frac{ V^{01}}{2}).
\end{equation}

Condition \reff{errh.10.0} reads
\begin{equation}
\sum_X {\rm ad}H_0(S_{1B_X})\,=\,\sum_X Q^{01}_{B_X},
\label{errh.10}
\end{equation}
which, given \reff{pia1}, suggests the choice $S_{1B_X} = {\rm
  ad}^{-1}H_0 \left(Q^{01}_{B_X}\right)$.  The locality of the
operator $S_{1B_X}$ follows from \reff{add}, which implies that
\begin{equation}
{\rm ad}^{-1}H_0 \left(Q^{01}_{B_X}\right) \;=\;
{\rm ad}^{-1}{\overline{H}}_{0X}\left(Q^{01}_{B_X}\right).
\label{locglob}
\end{equation}
This is because the operator $Q^{01}_{B_X}$ vanishes on all states which are 
{\em{not}} local ground states for $H_{0{B_X}\setminus X}$ (see \reff{scalar}
and \reff{gndnorm}), whereas ${\overline{H}}_{0B_X^c}$
measures the energy of the configuration outside $X$; see definition
\reff{hover}.  Hence, \reff{errh.10} is satisfied if we choose
\begin{equation} 
S_{1B_X} =
{\rm ad}^{-1}{\overline{H}}_{0X} \left(Q^{01}_{B_X}\right).
\label{s1def}
\end{equation} 
Selfadjointness of ${\overline{H}}_{0X}$ and $Q^{01}_{B_X}$ implies
anti-selfadjointness of $S_{1B_X}$ and hence of $S_1$.  The identities
\reff{sone} and \reff{s1def} show that $S_1$ is given by a sum of
local operators.  The condition that $S_{1B_X}$ should be a local
operator (i.e., $S_{1B_X} \in \A_{B_X}$) motivates our introduction of
the ``protection zone'' $B_X \setminus X$.  In \cite{datetal96} it is
shown that the family $\{S_{1B_X}\}$ satisfies the required
summability condition if the initial interaction $\{Q_X\}$ does.

More precisely, if $P^i$ denotes the projection onto the 
{\em{eigenspace of ${\overline{H}}_{0X}$}}
corresponding to an eigenvalue $E_i$, $E_0 = 0 < E_1 < E_2 < \ldots$,
then from \reff{jf.0}
\be
S_{1B_X} = \sum_{ij} P^i \frac {Q^{01}_{B_X}}{E_i - E_j}  P^j.
\label{jf}
\ee
For $Q^{01}_{B_X}$ as in \reff{q01}, $i \ne j$ (and either $i=0$, 
or $j=0$\,); thus $E_i - E_j \ne 0$. Hence $S_{1B_X}$ is 
a well defined operator.  Let us point out that, with these
definitions, 
\begin{eqnarray}
V_2 &=& \sum_{X_1, X_2 \atop{B_{X_1} \cap X_2 \ne \emptyset}}  {\rm
  ad}S_{1{B_{X_1}}} (  Q^{00}_{{B_X}_2} + Q^R_{{B_X}_2} +  
\frac{ Q^{01}_{{B_X}_2}}{2}) \nonumber\\
&=:&  \sum_{X_1, X_2 \atop{B_{X_1} \cap {X_2} \ne \emptyset}} 
V_{2B_{X_1 \cup X_2}}.
\label{v2}
\end{eqnarray} 

The transformed Hamiltonian \reff{errh.1} can be written in the form
\begin{equation}
H^{(1)}= K^{(1)} + R^{(1)},  \label{h2}
\end{equation}
where the operator $K^{(1)}$ contains the leading-order block-diagonal
contributions:
\be
K^{(1)} = H_0 + V^{00} + V^{R} + V_2^{00} + V_2^{R},
\ee
where the last two terms are sums of operators 
\begin{eqnarray}
V_{2B_{X_1 \cup X_2}}^{00} &\bydef &
P^0_{B_{X_1 \cup X_2}} V_{2B_{X_1 \cup X_2}} P^0_{B_{X_1 \cup X_2}}\\
V_{2B_{X_1 \cup X_2}}^{R} &\bydef &
P^1_{B_{X_1 \cup X_2}} V_{2B_{X_1 \cup X_2}} P^1_{B_{X_1 \cup X_2}}
+ P^2_{B_{X_1 \cup X_2}} V_{2B_{X_1 \cup X_2}} P^2_{B_{X_1 \cup X_2}}.
\label{pia.10}
\end{eqnarray}
This leading part $K^{(1)}$ basically corresponds to the first-order
expressions usually presented in the literature on perturbation
expansions.  The ``remainder'' $R^{(1)}$, on the other hand, starts
with the leading ``non-diagonal'' terms and includes all higher-order
contributions arising from the Lie-Schwinger series \reff{errh.1}
\begin{eqnarray}
\lefteqn{ R^{(1)} \;=\; V_2^{01} 
}\nonumber\\
&&+ \sum_{n \ge 2} \sum_{X_0, X_1, \ldots X_n:\atop {X_0 \cup \ldots \cup X_n 
\equiv \, {\rm{c.s.}} }}
\frac{1}{n!}\,  {\rm ad}S_{1 B_{X_n}}\Bigl( {\rm ad}S_{1 B_{X_{n-1}}}
\bigl(\ldots (Q^{00}_{{B_{X_0}}} + Q^R_{{B_{X_0}}} + 
\frac{n}{n+1}\,Q^{01}_{{B_{X_0}}})\ldots\bigr)\Bigr)\;,\nonumber\\
\label{r2}
\end{eqnarray}
where ${X_0 \cup \ldots \cup X_n 
\equiv \, {\rm{c.s.}} }$ denotes the condition 
\begin{equation}
X_i \cap B_{Y_{i-1}} \ne \emptyset \quad {\rm{for}} \quad 1\le i \le
n,
\label{eq.cs}
\end{equation}
with $Y_{i-1} = X_0 \cup \ldots \cup X_{i-1}$, and
\begin{equation}
V_{2B_{X_1 \cup X_2}}^{01} \; \bydef \;
P^0_{B_{X_1 \cup X_2}} V_{2B_{X_1 \cup X_2}} P^1_{B_{X_1 \cup X_2}}
+ P^1_{B_{X_1 \cup X_2}} V_{2B_{X_1 \cup X_2}} P^0_{B_{X_1 \cup X_2}}.
\label{pia.15}
\end{equation}
As shown in \cite{datetal96}, the series 
\reff{r2} converges, in fact absolutely and exponentially fast, if the
original interaction $Q$ satisfies \reff{expo}.

In fact, the
transformed Hamiltonian $ H^{(1)}$ comes from an interaction  
\be
\Phi^{(1)} = \Psi_0^{(1)} + {\tQ}^{(1)},
\label{phi2}
\ee where $\Psi_0^{(1)}$ and ${\tQ}^{(1)}$ are the interactions whose
terms yield $K^{(1)}$ and $R^{(1)}$ respectively.  Explicitly,
$\Psi_0^{(1)} = \{\Psi_{0{B_Y}}^{(1)}\}$ where the $\Psi_{0B_Y}$ are
nonzero only for $Y=X$ or $Y=X_1\cup X_2$ with $B_{X_1} \cap {X_2} \ne
\emptyset$ and $X$, $X_1$, $X_2$ (original) bonds and
take the values
\begin{eqnarray}
\Psi_{0{B_X}}^{(1)} &=& P^0_{B_X}\Phi_{0X}P^0_{B_X} + 
P^1_{B_X}\Phi_{0X}P^1_{B_X} + Q^{00}_{B_X} + Q^R_{B_X} +
V^{00}_{2B_{X}} + V^{R}_{2B_{X}}\\
\Psi_{0B_{X_1\cup X_2}}^{(1)} &=& V^{00}_{2B_{X_1 \cup X_2}}
+ V^{R}_{2B_{X_1 \cup X_2}}.
\end{eqnarray}
All the operators involved are ``diagonal'' with respect to the
partition of unity \reff{p2u} or the analogous partition with $X$
replaced by $X_1\cup X_2$. 
The interaction $\tQ^{(1)}$ has only terms of the
form $\{\tQ^{(1)}_{B_Y}\}$, where $Y$ is a ``c.s.''-connected set of
quantum bonds:
\begin{equation}
Y= \cup_{j=0}^n X_j \equiv \, {\rm{c.s.}} \;,\; n\ge 1\;,
\end{equation}
and
\be
\tQ^{(1)}_{B_Y} \;=\; \left\{\begin{array}{ll}
V^{01}_{2B_{X_0 \cup X_1}} & n=1\\[8pt]
\displaystyle
\frac{1}{n!}\,  {\rm ad}S_{1{B_X}_n}\Bigl( {\rm ad}S_{1{B_X}_{n-1}}
\bigl(\ldots (Q^{00}_{{B_X}_0} + Q^R_{{B_X}_0} + 
\frac{n}{n+1}\,Q^{01}_{{B_X}_0})\ldots\bigr)\Bigr) & n\ge 2\;.
\end{array}\right.
\label{q3}
\ee

The analysis of \cite{datetal96} shows that the operators
$\tQ^{(1)}_Z$, which make up the remainder $R^{(1)}$, decay
exponentially in the size of their supports.  More precisely, there
exists a positive constant $r_1>0$ such that \be \sum_{Z \ni 0}
\sqrt{{\rm{tr}}[\tQ_Z^{(1)*}(\lambda) \tQ_Z^{(1)}(\lambda)]}
e^{r_1s(Z)} \;=\; \OO(\lambda^2),
\label{expon}
\ee where $s(Z)$ denotes the cardinality of the smallest connected
subset of the lattice containing $Z$.  Furthermore, the part of
\reff{expon} involving ``00''-components is of order 3 or larger in
$\lambda$.  It is, therefore, natural to expect that this interaction
$\{\tQ^{(1)}_{B_Y}\}$ does not contribute to the ground-state energy of
the transformed Hamiltonian $H^{(1)}$, {\em{to second order in the
    perturbation parameter}} $\lambda$.  This fact is easy to see in
finite volume, but, as mentioned before, its verification in the
thermodynamic limit ---at the level of energy densities--- requires a
suitable construction of (infinite volume) ground states.  For this
reason we shall be able to confirm this expectation only in those
cases were the transformed interaction satisfies the hypotheses of the
quantum Pirogov-Sinai theory of Section \ref{sumhyp}.  (This excludes
models with a continuous symmetry.) 

The methods discussed in Section \ref{lowte} for the study of the
phase diagram rely on the choice of a suitable tensor product basis
$\{e(\omega)\}$ such that the operators $\Psi^{(1)}_{0B_Y}$ can be
decomposed into
\be
\Psi^{(1)}_{0B_Y} = \Phi^{(1)}_{0B_Y} + \Omega_{B_Y}^{(1)},
\label{om}
\ee 
where $\Phi^{(1)}_{0B_Y}$ is {\em{exactly}} diagonal in the tensor
product basis $\{e(\omega)\}$ ---i.e.,  $\Phi^{(1)}_{0B_Y}$
is {\it ``classical''} with respect to $\{e(\omega)\}$--- and
$\Omega_{B_Y}^{(1)}$ is a perturbation which is usually non-diagonal. 
Such a decomposition is always possible.
The optimal choice of $\{ e (\omega)\}$ is 
the one that renders the perturbation $\Omega_{B_Y}^{(1)}$ as small
as possible.  It is worth noting that the new unperturbed (classical)
interaction 
$\Phi_0^{(1)}$ is of {\em{finite range}}.

We finally set
\be
\Phi^{(1)} = \Phi_0^{(1)} + Q^{(1)},
\label{final1}
\ee
where 
\be
\Phi_0^{(1)} = \{\Phi^{(1)}_{0B_Y}\},
\ee
and
\be
Q^{(1)} = \{\tQ^{(1)}_{B_Y} + \Omega_{B_Y}^{(1)}\} .
\label{v1def}
\ee In the simplest examples, the only contribution to the ground
state energy of the transformed interaction $\Phi^{(1)}$, {\em{to
    second order in the perturbation parameter}} $\lambda$, arises
from the classical interaction $\Phi_0^{(1)}$.  In such cases one may
gain heuristic insight into the structure of ground- and
low-temperature states of $\Phi$ from the ground states of
$\Phi_0^{(1)}$.  For example, large degeneracies in the spectrum of
the Hamiltonian determined by $\Phi_0$ may turn out to be lifted in
the spectrum of the one determined by $\Phi_0^{(1)}$.  Hence this last
interaction, though also classical, ought to lead to more accurate
insight into low-temperature properties of the system than the
original $\Phi_0$.  The examples below illustrate instances where these
expectations can be rigorously confirmed.

\subsection{Second-order perturbation theory}
\label{secondpert}

Next, we search for a unitary transformation which removes the
off-diagonal part of the original perturbation, $\{Q^{01}_{B_Y}\}_Y$,
to order $|\lambda|^2$.  A possible approach (certainly not the only
one, see the discussion in Section \ref{ssyno} below), is to transform
the interaction $\Phi^{(1)}$ so as to eliminate the lowest order
off-diagonal part of the perturbation $Q^{(1)}$.  This leads one to
consider a unitary transformation of the form 
\be U^{(2)}(\lambda) =
e^{S_2(\lambda)}\, e^{S_1(\lambda)},
\label{pia.u}
\ee 
where $S_1 \equiv S_1(\lambda) = \sum_{X \subset \Lambda}
S_{1B_X}$ is defined by \reff{s1def}, while $S_2(\lambda)$ is
determined from the above requirement. The generator $S_2(\lambda)$
must have leading terms of degree $2$ and is required to be a sum of
local operators: 
\be S_2 \equiv S_2(\lambda) = \sum_{Y \subset
  \Lambda} S_{2B_Y}(\lambda).  
\ee

We notice that the choice \reff{pia.u} does \emph{not} amount to an
iteration of the first-order procedure, because we keep using the
partitions of unity determined by the spectral projections of the
\emph{original} classical part $\Phi_0$, rather than those
corresponding to the transformed $\Phi_0^{(1)}$.  We comment in
Section \ref{ssyno}, below, how an iterated method (Method 3) would
proceed.

To alleviate the forthcoming formulas let us suppress the explicit
$\lambda$-dependence of the 
operators $S_1(\lambda)$ and $S_2(\lambda)$. 
The unitary transformation yields the Hamiltonian
\begin{eqnarray}
H^{(2)} &:=& e^{S_2}\, e^{S_1}\,H \,e^{-S_1} \, e^{-S_2}\nonumber\\
&=& e^{S_2}\,H^{(1)} \,e^{-S_2},
\label{trsf22}
\end{eqnarray}
where $H^{(1)}$ is the Hamiltonian given by \reff{errh.1.0}.
Applying the Lie-Schwinger series we see that the term 
$V_2^{01}$ can be removed by defining
\begin{eqnarray}
S_2 &:=& {\rm ad}^{-1} H_0(V_2^{01}) \nonumber\\
&=& \sum_{X_1, X_2 \atop{X_1 \cap {B_{X_2}} \ne \emptyset}}
{\rm ad}^{-1} {\overline{H}}_{0{X_1 \cup X_2}}(V_{2B_{X_1 \cup
    X_2}}^{01}) \nonumber\\ 
&:=&  \sum_{X_1, X_2 \atop{X_1 \cap {B_{X_2}} \ne \emptyset}}
S_{2{B_{X_1 \cup X_2}}}. 
\label{s2}
\end{eqnarray}
With this choice, we obtain ---singling out the terms of up to fourth
order in $\lambda$--- 
\begin{equation}
H^{(2)} \;=\; H_0  + V^{00} + V^R + V_2^{00} + V_2^{R} + V_3 + V_4 + T
\label{hubfourth}
\end{equation} 
where
\begin{equation}
V_3 \;=\; {\rm ad}^2 S_1 (\frac{V^{00}}{2} + \frac{V^R}{2} + \frac{2
  V^{01}}{3!})
+ {\rm ad} S_2( V_2^{00} + V_2^R + \frac{ V_2^{01}}{2}) ,
\label{hubthird}
\end{equation} 
\begin{equation}
  \label{hubfourth.1}
  V_4 \;=\; {\rm ad}^3 S_1
( \frac{V^{00}}{3!} + \frac{V^R}{3!} + \frac{3 V^{01}}{4!}) ,
\end{equation}
and $T$ consists of all terms of order $n$
with $n \ge 5$.  

The transformed Hamiltonian $H^{(2)}$ can be written as
\be
H^{(2)} = K^{(2)} + R^{(2)},
\label{eff4}
\ee
where 
\begin{equation}
  \label{fla.20}
  K^{(2)} = K^{(1)} + V_3^{00} + V_3^{R} + V_4^{00} + V_4^{R}
\end{equation}
is given entirely in terms of an interaction
$\Psi_0^{(2)} \equiv \{\Psi_{0{B_Y}}^{(2)}\}$, 
with the property that  the operators $\Psi_{0{B_Y}}^{(2)}$ are
``diagonal'' with respect to the partitions of unity  
\be
 {\bf{1}} = P^0_{B_Y} + P^1_{B_Y} + P^2_{B_Y}. 
\label{part4}
\ee
and are of order $k$ in $\lambda$, with $0 \le k \le 4$.
All terms which are ``off--diagonal'' ---starting with $V_3^{01} +
V_4^{01}$---, and all ``diagonal'' terms   
of higher orders,  are included in the
remainder $R^{(2)}$, which is given in terms 
of an interaction $\tQ^{(2)}= \{\tQ^{(2)}_{B_Y}\}$.
The analysis of \cite{datetal96} shows that these operators satisfy
\be
\sum_{Z \ni 0} 
\sqrt{{\rm{tr}}[\tQ_Z^{(2)*}(\lambda) \tQ_Z^{(2)}(\lambda)]}
e^{r_2(Z)} \;=\; \OO(\lambda^3), 
\label{expon2}
\ee 
for some $r_2>0$, and that a similar sum involving
``00''-components is of order 5 or larger in $\lambda$.  

Further, as in \reff{om}, we can write 
\be
\Psi_0^{(2)} = \Phi_0^{(2)} + \Omega_0^{(2)}
\label{intd}
\ee 
where $\Phi_0^{(2)} \equiv \{\Phi_{0{B_Y}}^{(2)}\}$ is a finite
range classical interaction, i.e., diagonal in a tensor product basis
$\{e(\omega)\}$, and $\Omega_0^{(2)}$ is a generally non-diagonal
perturbation.  One expects that only the classical interaction
$\Phi_0^{(2)}$ contributes to the ground-state energy {\em{to order
    $4$ in $\lambda$}}\/.

\subsection{Synopsis of perturbation theory}\label{ssyno}

The above procedure can be iterated to any finite order in $\lambda$.  To
$n$-th order, we aim at constructing a unitary operator,
$U^{(n)}_\Lambda(\lambda)$,
such that
\be
 H^{(n)}_\Lambda (\lambda)=  U^{(n)}_\Lambda (\lambda) \,
H_\Lambda (\lambda) \, {U^{(n)}_\Lambda (\lambda)}^*
\label{z.80}
\ee
has the property that 
\be
 H^{(n)}_\Lambda (\lambda)= \sum_{Z \subset \Lambda}
 \Phi_Z^{(n)}(\lambda),
\ee
with 
\be 
 P^0_{B_Z} \Phi_Z^{(n)}(\lambda) P^1_{B_Z} +
 P^1_{B_Z} \Phi_Z^{(n)}(\lambda) P^0_{B_Z} = O(|\lambda|^{n+1})
\label{ordn}
\ee
for all $Z$. We need to ensure that the conjugation of an interaction with
exponential decay by the operator ${U^{(n)}_\Lambda (\lambda)}$ is
again an interaction with exponential decay.  In \cite{datetal96} we
present two possible structures for an operator $U^{(n)}_\Lambda (\lambda)$
compatible with this requirement (we drop the subscript $\Lambda$):
\medskip

\noindent
{\em Method 1:}
\be
\quad U^{(n)}_\Lambda (\lambda) = \prod_{j=1}^n
e^{S_j^{(1)}(\lambda)},
\label{z.85}
\ee

\noindent
{\em Method 2:}
\be
\quad U^{(n)}_\Lambda (\lambda) = \exp\bigl( \sum_{j=1}^n 
S_j^{(2)}(\lambda)\bigr),
\label{z.90}
\ee
\medskip

In both cases, 
\be
S_j^{(\alpha)}(\lambda) = \sum_Z S_{jZ}^{(\alpha)}(\lambda), \quad
\alpha= 1 \,\, {\rm{or}} \,\, 2.
\ee
\smallskip
\noindent
where the local operators $S_{jZ}^{(\alpha)}(\lambda)$ are of degree
$j$ in $\lambda$. The idea is to determine these operators 
{\em{recursively}} in such a way that the
condition \reff{ordn} holds, for $n=1,2,3, \ldots$.  Given 
$S_{1Y}^{(\alpha)}, \ldots, S_{kY}^{(\alpha)}$, for arbitrary $Y$,
\reff{ordn} uniquely fixes the ``off-diagonal'' contribution to 
$ S_{k+1\,Z}^{(\alpha)}$, for any $Z$ and for any $k < n$. Explicit 
formulas may be found in \cite{datetal96}. The method described in
Section \ref{secondpert} follows method 1.
\medskip

There is a third way of organizing the recursive perturbative 
construction of unitary operators $U^{(n)}(\lambda)$ approximately
block-diagonalizing the Hamiltonian $H$, inspired by Newton's method
and similar, in spirit, to the renormalization group strategy. Recall
that first order perturbation theory [see \reff{final1} in Section
\ref{firstpert}] yields a unitary operator $ U^{(1)}(\lambda)$ such
that 
\be
 U^{(1)}(\lambda) \, H \, {U^{(1)}(\lambda)}^* =: H^{(1)} =
H_0^{(1)} +  V^{(1)},
\ee
where $H_0^{(1)} = \sum_{Y} \Phi_{0B_Y}^{(1)}$, \ $\Phi_{0}^{(1)} =
\{\Phi_{0B_X}^{(1)}\}$ is a 
classical interaction, and the interaction giving rise to  $V^{(1)}:=
\sum_Y Q^{(1)}_{B_Y}$ has ``off-diagonal'' terms of order $2$ in
$\lambda$.  The third method to construct the conjugations is based
on keeping track of  the low-energy spectrum of the Hamiltonians
determined by $\{\Phi_{0B_Y}^{(1)}\}$
(rather than by $\{\Phi_{0B_Y}\}$),  and, in particular, of their
ground states.  This gives rise to a partition of unity
\be
{\bf{1}} = \PP^0_{B_X} +  \PP^1_{B_X} + \PP^2_{B_X},
\label{partnew}
\ee
which serves to decompose the operators $Q^{(1)}_X$ into
\be
Q^{(1)}_X = [Q^{(1)}_X]^{00} + [Q^{(1)}_X]^{01} +  [Q^{(1)}_X]^{R},
\ee
as in \reff{qdecom}, and hence 
\be
H^{(1)} = H^{(1)}_0 + [V^{(1)}]^{00} +  [V^{(1)}]^{01} +
[V^{(1)}]^{R},
\ee
in analogy to \reff{hdecom}.  We now proceed as in \reff{sone}
through \reff{jf} of Section \ref{firstpert}. This yields an operator 
\be
S^{(3)}_2(\lambda) = \sum_X S^{(3)}_{2B_X}(\lambda)
\ee
such that 
\be
H^{(2)} = e^{S^{(3)}_2} H^{(1)} e^{-S^{(3)}_2}
\ee
has the form
\be
H^{(2)} =H_0^{(2)} + V^{(2)},
\ee
where the perturbation $ V^{(2)}$ does not have any ``off-diagonal''
terms
of order $\le 2$, with respect to the partition of unity
\reff{partnew}. We may now proceed recursively, in the same manner,
eliminate off-diagonal terms to ever higher order in $\lambda$.

The analysis of
\cite{datetal96} remains valid for this case too, and provides
proofs of convergence and summability of the different interactions.

\section{Perturbation expansion for the one-band Hubbard model near
half-filling} 
\label{perthub}

\subsection{Preliminary remarks}

In this section we derive the effective Hamiltonians for the 
one-band Hubbard model, defined by the Hamiltonian \reff{gen}, near
half-filling by applying the 
perturbative unitary conjugation, Method 1, described in Section
\ref{ssyno}.  This model has a number of simplifying features that
makes it ideal for a non-trivial exemplary 
application of our methods.  The simplifications stem from the fact
that its leading classical interaction [formula \reff{onsite} below]
is a sum of on-site terms. First, this implies that every eigenvalue of 
$H_{0X}, X \subset \Lambda$ [defined as in \reff{ll.1}], is a sum of
on-site energies $\epsilon_x(s)$ [where $s$ 
denotes the on-site configuration and $x\in \Lambda$] which for this
model are 
\be
\epsilon_x(\uparrow) = - \mu_+ 
\;,\; \epsilon_x(\downarrow) = - \mu_-
\;,\; \epsilon_x(\emptyset)= 0
\;,\; \epsilon_x(\uparrow,\downarrow) = U -  \mu_+  - \mu_-\; .
\ee
Hence its zero-temperature phase diagram is easy to determine (Figure
\ref{fig1}). 

Second, the classical interaction has range
$R=0$, and there is no need for protection zones:
\be
B_Y=Y,
\label{noprot}
\ee
for arbitrary $Y \subset \Lambda$.

Below we shall consider, for the whole shaded region of Figure
\ref{shaded}, projections onto the entire band of single-occupancy
states.  These states are all simultaneously (local) ground states of
the classical interaction along the line $\mu_+=\mu_-$ (in the
complement of this line this degeneracy tends to be lifted).

We present the block-diagonalization of the Hubbard Hamiltonian only
to second (Section \ref{hubfirst}) and third order (Section
\ref{hubsec}) in the hopping amplitudes.  But we believe that the
following discussion, together with Section 4 of \cite{datetal96},
provide enough details for a motivated reader to pursue the procedure
to an arbitrary order.

\subsection{The original interaction}

As mentioned in the Introduction, it is necessary to implement our
perturbation scheme at the level of interactions instead of Hamiltonians.
The dominant term, $H_{0\Lambda}$,  of the Hamiltonian \reff{gen} 
can be expressed in terms of an
on--site interaction $\Phi_0 = \{\Phi_{0x}\}$ where
\begin{equation}
\Phi_{0x} = U n_{x +}  n_{x -} - \frac{h}{2} 
(n_{x +} - n_{x -}) - \frac{k}{2} 
(n_{x +} + n_{x -}),
\label{onsite} 
\end{equation}
with $x \in \Lambda$ and
\begin{equation}
h := \mu_{+} - \mu_{-},  \quad \quad k:=  \mu_{+} + \mu_{-}.  
\label{defhk}
\nonumber
\end{equation}
As in \reff{ll.1}, this interaction defines local Hamiltonians
\be
H_{0Y}=\sum_{x \in Y} \Phi_{0x}\;.
\ee
The quantum perturbation is the kinetic-energy term, and the
perturbation parameters are the hopping amplitudes $\Tt=\{\tas{+}{b},
\tas{-}{b},\tas{+}{\ovb},\tas{-}{\ovb}\}$, where $b=[xy]$ denotes an
ordered pair of nearest neighbor sites and $\ovb$ is the same
nearest neighbor pair, but with {\em{reverse}} ordering.  The
perturbation is assumed to be translation invariant, that is
$t_{\pm b}$ only depends on the direction, $y-x$, of $b$.
We shall write 
\begin{equation}
t \;=\; \max \{\tas{+}{b}, \tas{-}{b},\tas{+}{\ovb},\tas{-}{\ovb}\}.
\end{equation}
We write the quantum interaction as
\be
Q_X \;\equiv\; Q_X(\Tt) \;=\; Q_{[xy]}(\Tt) + Q_{[yx]}(\Tt),
\label{fl.1}
\ee
where $X$ denotes the {\em{unordered}} pair $\nn{xy}$ and
\be
Q_b \equiv Q_b(\Tt) = \sum_{\sigma = \pm} Q_{\sigma b}(\Tt)
\label{fl.2}
\ee
with
\be
Q_{\sigma b} \equiv Q_{\sigma b}(\Tt) := \tas{\sigma}{b} c_{x
  \sigma}^\dagger c_{y\sigma}.
\label{qsb}
\ee
Moreover, if the Hamiltonian is assumed to be self-adjoint then
\be
\tas{\pm}{b} = (\tas{\pm}{\ovb})^*.
\ee

\subsection{The local projections}
\label{locproj}

We start by introducing partitions of unity 
giving rise to appropriate decompositions of the hopping terms in the
Hubbard Hamiltonian into ``diagonal'' and ``off-diagonal''
operators.
The most interesting part of the shaded region of Figure
\ref{shaded} is the vicinity of the line of infinite degeneracy.
On this line, the ground states of $\Phi_0$ are
(linear combinations of) configurations with precisely one 
(up-spin or down-spin) electron occupying each site.
For $Y \subset \Lambda$, the orthogonal projections onto the subspace
spanned by the ground states of $H_{0Y}$ are of the form
\begin{equation}
P^0_Y := \prod_{x \in Y} P^0_x 
\label{projX}
\end{equation}
with
\begin{equation}
P^0_x := (n_{x+} - n_{x-})^2.
\label{pee0}
\end{equation}

As remarked above, as the classical interaction \reff{onsite} is
on-site, i.e., of zero range, there is no need for protection zones.
This implies, first, that we can define the projection $P^1_{0Y}$ by
setting 
\be 
{\bf 1} = P^0_Y + P^1_Y, \label{parteg} 
\ee 
and, second,
that we can use these projections $P^0_Y$ and $P^1_Y$ throughout the
whole shaded region of Figure \ref{shaded}. The operators $P^0_Y$ are
projections onto states with single occupancy.  This family of states
includes, therefore, ground states and low-lying excitations The
operators $P^1_Y$, on the other hand, project onto the complementary
subspace corresponding to of high-energy excitations, formed by states
with at least one site in $Y$ occupied by zero or two electrons.

\subsection{First-order perturbation theory}
\label{hubfirst}

Let us perform the first step of the perturbative analysis of Section
\ref{pert}, namely the block-diagonalization of the interaction to
second order in the perturbation parameters.  
In this section, $X$ is always used to denote {\em a pair of nearest
neighbor sites}\/.

We consider the perturbation defined in \reff{fl.1}--\reff{qsb} and
decompose it into the form \reff{qdecom}, $Q_X =  Q^{00}_{X} +
Q^{01}_{X} + Q^{R}_{X}$, using the projections $P^0_X$ and $P^1_X$
defined in \reff{projX}--\reff{pee0} ($P^2_Y=0$ for all
$Y\subset\Lambda$).  It is simple to check that
\be
Q_X^{00} = 0.
\ee

Our goal is to construct a unitary transformation, 
\be
U^{(1)}(t)=e^{S_1(t)},
\ee
with the property that 
$H_\Lambda^{(1)}(t):= U^{(1)}(t)H_\Lambda {U^{(1)}(t)}^*$ only
contains off-diagonal operators of order $2$ and higher in
$t$. Following the general ideas discussed in Section
\ref{firstpert}, we attempt to construct an operator
$S_1(t)$ of the form 
\begin{equation}
S_1 (t) \;=\; \sum_X S_{1X}(t) 
\;= \; \sum_{\sigma=+, -} \sum_b \tas{\sigma}{b} S_{1 \sigma b}.
\label{sone2}
\end{equation}
The formulas for the operators $ S_{1 \sigma b}$ can be inferred from
Section \ref{pert} [formula \reff{s1def}]: 
\be
S_{1 \sigma \nn{xy}} = {\rm{ad}}^{-1} H_{0\{xy\}}(c_{x\sigma}^\dagger
\, c_{y\sigma}). 
\ee
Conjugating the Hamiltonian $H$ of the asymmetric single-band Hubbard
model by the operator $e^{S_1(t)}$, yields an effective
Hamiltonian $H^{(1)}$ [see \reff{errh.1.0}] given in terms 
of a new interaction, $\Phi^{(1)}$, determined by \reff{h2}-\reff{phi2}
(with $B_X = X$).

Further, as explained in Section \ref{firstpert} [see \reff{om} ff.], 
to be able to apply the Pirogov-Sinai theory of quantum phase
diagrams, we need to decompose the operators $\Phi_X^{(1)}$ as in
\reff{final1}, namely in the form
\be
\Phi_X^{(1)} = \Phi_{0X}^{(1)} + Q_X^{(1)},
\ee
where $\{\Phi_{0X}^{(1)}\}$ is a finite range classical interaction
and $\{Q_X^{(1)}\}$ is a perturbation with
$[Q_X^{(1)}]^{01}=\OO(t^2)$.  For the Hubbard model, 
\be
\Phi_{0X}^{(1)} \;=\; \left\{\begin{array}{ll}
P^0_x\, \Phi_{0x}  P^0_x & \hbox{if } X=\{x\} \\[5pt]
\frac{1}{2} \,P^0_X \, {\rm{ad}}S_{1X}(Q^{01}_X) \,  P^0_X & 
\hbox{if } X \hbox{is a pair of n.n.}
\end{array}\right.
\label{this}
\ee 
It is not hard to see that the two-bond terms $V_{X_1 \cup
  X_2}^{00} = \frac{1}{2} [{\rm ad}S_{1{B_{X_1}}} (Q^{01}_{{B_X}_2}) +
S_{1{B_{X_2}}} (Q^{01}_{{B_X}_1})]^{00}$ are zero if $X_1\neq X_2$.
The second expression on the R.S. of \reff{this} involves operators
corresponding to an electron of a particular spin hopping from a singly
occupied site to a nearest neighbor site occupied by an electron of
opposite spin, thus momentarily creating a doubly occupied site and a
hole. They then restore the condition of single occupancy of the sites
by causing one of the electrons on the doubly occupied site to hop to
the neighboring hole.

In fact, the interaction \reff{this} coincides with the first-order
calculations often presented in the literature since the
initial observation of Anderson \cite{and59}.  It is conveniently
expressed in terms of the spin operators 
\begin{equation}
{\bf S}_x := \sum_{s, s'=+,-} c^\dagger_{x s} 
{\bf S}_{s s'} c_{x s'}
\label{spinop}
\end{equation}
where $ {\bf S} \equiv \frac{1}{2} (\sigma^1,\sigma^2, \sigma^3)$ and 
$\sigma^1,\sigma^2, \sigma^3$ are 
the standard Pauli matrices.  Note that 
\be
S_x^3 = \frac{1}{2} (n_{x+} - n_{x-}),
\label{sx3}
\ee
and
\begin{equation}
S_x^{\pm} := S_x^1 \pm iS_x^2 = c_{x \pm}^\dagger c_{x \mp}\;.
\label{spm}
\end{equation}
The operators $S_x^+$, $S_x^- = (S_x^+)^*$ and $S_x^3$ form a basis
of a representation of the Lie algebra of $SU(2)$ on the
four-dimensional Hilbert space $\HH_x = {\rm{span}} \{ \emptyset, 
\uparrow, \downarrow, \uparrow \downarrow\}$.
These operators annihilate the subspace spanned by
$\{\emptyset,\uparrow \downarrow\}$ and
act irreducibly (in the spin-$1/2$ representation) on the
subspace spanned by $\{ \uparrow, \downarrow\}$.  In particular,
\be
P_x^0\, S^3_x \,P_x^0 = S^3_x \quad,\quad
P_x^0\, S^\pm_x \,P_x^0 = S^\pm_x \;.
\label{nonec}
\ee

The operators ${\rm ad}S_{1 \sigma \ovb}(Q^{01}_{\sigma b})$, with 
$\sigma= +, -$, cause the same electron to hop in both 
steps of the process, whereas the operators 
${\rm ad}S_{1 \sigma \ovb}(Q^{01}_
{\sigma' b})$, with $\sigma \ne \sigma'$, cause electrons of opposite spins to
hop in the two steps. These processes can be expressed in terms of a 
sequence of creation and annihilation operators of the up-spin and
down-spin electrons 
\begin{eqnarray}
 P^0_X \,{\rm ad}S_{1 \sigma \ovb}(Q^{01}_{\sigma' b})\, P^0_X &=& 
-\,\frac{1}{U} \,
{{P^0_X}} \left[Q^{01}_{\sigma \ovb} Q^{01}_{\sigma' b} + 
Q^{01}_{\sigma' b}Q^{01}_{\sigma \ovb}\right]  {{P^0_X}}\nonumber\\
&=& -\frac{\tas{\sigma}{\ovb}\,\tas{\sigma'}{b}}
{U} {{P^0_X}} \Bigl[c^{\dagger}_{y\sigma}c_{x\sigma}c^{\dagger}_{x\sigma'}
c_{y\sigma'} + c^{\dagger}_{x\sigma}c_{y\sigma}c^{\dagger}_{y\sigma'}
c_{x\sigma'} \nonumber\\
&& {}+ c^{\dagger}_{y\sigma'}c_{x\sigma'}c^{\dagger}_{x\sigma}
c_{y\sigma} + c^{\dagger}_{x\sigma'}c_{y\sigma'}c^{\dagger}_{y\sigma}
c_{x\sigma} \Bigr]{{P^0_X}}. 
\label{long}
\end{eqnarray}
For $\sigma=\sigma'$, \reff{long} yields
\begin{eqnarray}
\sum_{\sigma=+, -}  P^0_X {\rm ad}S_{1 \sigma X}(Q^{01}_{\sigma X}) P^0_X
&=& - \,\frac{2\,(|\tas{+}{b}|^2 + |\tas{-}{b}|^2)}{U}{{P^0_X}}
\bigl[n_{y +} (1 - n_{x +})  + (1 - n_{y +})n_{x +}\bigr]{{P^0_X}} 
\nonumber\\
&=&  \frac{4 \,(|\tas{+}{b}|^2 + |\tas{-}{b}|^2)}{U} \Bigl( S^3_x
S^3_y  - \frac{P^0_X}{4}\Bigr).
\label{step1}
\end{eqnarray}
To arrive at \reff{step1}, we have made use of the identities
\begin{equation}
{{P^0_X}} n_{y \sigma}(1 - n_{x \sigma}) {{P^0_X}}
= {{P^0_X}} (1 - n_{y (-\sigma)}) n_{x (-\sigma)}{{P^0_X}}, 
\label{hoy.1}
\end{equation}
and 
\be
P^0_X \,\bigl[n_{x+} n_{y -} + n_{x -} n_{y+}\bigr]\,
P^0_X = \Bigl[- 2 S_x^3 S_y^3 + \frac{P^0_X}{2}\Bigr],
\label{hoy.2}
\ee
for $X = \{xy\}$.  In turn, this last identity follows from
\reff{sx3}, \reff{nonec} and the relations
\be
P^0_x\, n_{x \sigma}P^0_x =  P^0_x\, ( 1 - n_{x (-\sigma)})\,P^0_x
\label{fla.5}
\ee
and 
\be
P^0_x\, (n_{x+} + n_{x-})\, P^0_x = P^0_x\;.
\label{fla.6}
\ee

Similarly, for $\sigma \ne \sigma'$, \reff{long} yields 
\begin{eqnarray}
\sum_{\scriptstyle \sigma, \sigma'=+, - \atop \scriptstyle \sigma \ne
\sigma'}   
{\rm ad}S_{1 \sigma \ovb}(Q^{01}_{\sigma b})
&=& \frac{2\,( \tas{+}{\ovb}\,\tas{-}{b} +  \tas{-}{\ovb}\,
\tas{+}{b})}{U} 
\left[ S_y^+ S_x^- +  S_y^- S_x^+\right]\nonumber\\
&=&  \frac{4\,( \tas{+}{\ovb}\,\tas{-}{b} +  \tas{-}{\ovb}\,
\tas{+}{b})}{U} P^0_X\, \left[ S^1_x S^1_y +  S^2_x S^2_y \right]
P^0_X\;.
\label{h437}
\end{eqnarray}
\smallskip

>From \reff{this}, \reff{step1} and \reff{h437} we conclude that
$\Phi_{0X}^{(1)}$ is the well known Heisenberg interaction
\begin{equation}
\Phi_{0X}^{(1)} \;=\; \left\{\begin{array}{ll}
 - h \, S_x^3
- \frac{k}{2},  & \hbox{if } X=\{x\} \\[10pt]
\frac{2\,\bigl[|\tas{+}{b}|^2 + |\tas{-}{b}|^2\bigr]}{U} 
\Bigl( S^3_x S^3_y- \frac{P^0_X}{4}\Bigr)  
+ \frac{2 \,( \tas{+}{\ovb}\,\tas{-}{b} +  \tas{-}{\ovb}\,
\tas{+}{b})}{U} \left[ S^1_x S^1_y +  S^2_x S^2_y \right]  
&\hbox{if } X \hbox{ is a pair of n.n.}
\end{array}\right.
\label{mag}
\end{equation}
[$h$ and $k$ being defined in \reff{defhk}], and
$\Phi_{0X}^{(1)}=0$, for all other $X\subset\zed^\nu$.

If the hopping amplitudes depend neither on the direction of the
hopping nor on the spin orientation ---$t_{\pm b} = t_{\pm \ovb} =
t$--- then the original Hamiltonian $H_\Lambda$ \reff{gen} is the
standard one-band Hubbard Hamiltonian, and the interaction \reff{mag}
is that of an {\em{isotropic}} Heisenberg model.  This model is
difficult to treat rigorously because of its SU(2) symmetry, and it is
not surprising that the symmetric Hubbard model lies outside the scope
of the methods we discuss in Section \ref{lowte}.  For high asymmetry,
$\left|\tas{+}{b}\right| \ll \left|\tas{-}{b}\right|$, the effective
interaction \reff{mag} corresponds to an antiferromagnetic Ising
interaction, perturbed by a small spin-flip term.  This interaction
exhibits a first-order phase transition to a N\'eel phase.  This
applies in particular to the Falicov Kimball model, for which
$\tas{+}{b}=0$ and $\tas{-}{b}=t$.

>From the discussion of Section \ref{firstpert} we know that the other
terms of the transformed interaction $\Psi^{(1)}$ are small.
The terms corresponding to the rest of $\Psi^{(1)}_0$, called
$\Omega^{(1)}$ in \reff{om}, involve only transitions within the
excited band, while the ``off-diagonal'' remainder is of second order
in $t$ [equation \reff{expon}].

\subsection{Second-order perturbation theory} 
\label{hubsec}

Let us now pursue the block diagonalization process to one further
order.  The objective is to find the classical interaction
$\Phi_0^{(2)} $ of \reff{intd}.  General theory tells us that the
remainder is small [bound \reff{expon2}].  In analogy to the
first-order choice, we take $\Phi_0^{(2)} = [\Psi_0^{(2)}]^{00}$,
where $\Psi_0^{(2)}$ is the block-diagonal interaction giving rise to
$K^{(2)}$ in \reff{fla.20}.  Therefore
\begin{equation}
  \label{fla.25}
  \Phi_0^{(2)}  = \Phi_0^{(1)} + \varphi_0^{(2)}\;,
\end{equation}
where $\Phi_0^{(1)}$ was determined in the previous section [formula
\reff{mag}] and $\varphi_0^{(2)}$ is such that 
$V_3^{00} + V_4^{00}= \sum_Y \varphi_{0Y}^{(2)}$.  

Inspection shows that many of the terms comprising $V_3^{00} +
V_4^{00}$ are zero when the original perturbation $Q$ is just a
nearest-neighbor hopping term.  In fact, $\varphi_{0Y}^{(2)}$ is
nonzero only for the following sets $Y$ (we continue to use the letter
$X$ for (unordered) pairs of nearest-neighbor sites): \medskip

\noindent
\emph{(i) $Y=X$, a nearest-neighbor pair:} 
\be 
\varphi_{0X}^{(2)} \;=\;
\frac{1}{8} P^0_X \,\bigl[ {\rm{ad}} S_{1X} ( {\rm{ad}} S_{1X}
({\rm{ad}} S_{1X} ( Q^{01}_X)))\bigr]\, P^0_X. 
\label{a01}
\ee
\smallskip

\noindent
\emph{(ii) $Y=X \cup X'$, where $X$ and $X'$ are different
nearest-neighbor pairs sharing a site:}  There are two
contributions, 
\be 
\varphi_{0\,X\cup X'}^{(2)} \;=\; \widetilde\varphi_{0\,X\cup X'}^{(2)}
+ \overline{\varphi}_{0\,X\cup X'}^{(2)}\;.
\label{a1}
\ee
The first one is
\be
\widetilde\varphi_{0\,X\cup X'}^{(2)} \;=\;
\frac{1}{8}\, \sum_{C_6} P^0_{X\cup X'}\, \Bigl[{\rm
  ad}S_{1{X}_4}({\rm ad}S_{1{X}_3}
({\rm ad}S_{1{X}_2} (Q^{01}_{{X}_1})))\Bigr]\, P^0_{X\cup X'}
\label{a2}
\ee
where $C_6$ is the set of sequences $(X_1,X_2,X_3,X_4)$ satisfying
one of the following conditions
\be
\left\{\begin{array}{c}
X_1=X_2 =X \,;\, X_3=X_4 =X' \\
X_1=X_3 =X \,;\, X_2=X_4 =X'\\
X_1=X_4 =X \,;\, X_2=X_3 =X'
\end{array}\right\}
\quad \bigcup \quad
\left\{\begin{array}{c}
\hbox{same but}\\
X \,\longleftrightarrow\, X'
\end{array}\right\}\;.
\label{pia.40}
\ee
The second contribution is
\be
\overline{\varphi}_{0\,X\cup X'}^{(2)} \;=\;
\frac{1}{2}\, \sum_{C_4}
P^0_{X\cup X'} \,\Bigl[{\rm{ad}} S_{2({X_4 \cup X_3})} 
\left({Q^{01}_{2({X_1 \cup X_2})}}\right)\Bigr]\, P^0_{X\cup X'},
\label{a37}
\ee
with $C_4$ being the set of sequences $(X_1,X_2,X_3,X_4)$ satisfying
one of the following conditions
\be
\left\{\begin{array}{c}
X_1=X_3 =X \,;\, X_2=X_4 =X'\\
X_1=X_4 =X \,;\, X_2=X_3 =X'\\
\end{array}\right\}
\quad \bigcup \quad
\left\{\begin{array}{c}
\hbox{same but}\\
X \,\longleftrightarrow\, X'
\end{array}\right\}\;.
\label{pia.41}
\ee
\smallskip

\noindent
\emph{(iii) $Y=X \cup X' \cup X'' \cup X'''$ forming
a unit square of the lattice:}  For concreteness, let us assume that,
in the sequence $X,X',X'',X'''$, each set shares a site with the next
one, and $X'''$ shares a site with $X$.
Again, there are two contributions 
\be
\varphi_{0\,X\cup X'\cup X''\cup X'''}^{(2)} \;=\; 
\widetilde\varphi_{0\,X\cup X'\cup X''\cup X'''}^{(2)}
+ \overline{\varphi}_{0\,X\cup X'\cup X''\cup X'''}^{(2)}\;.
\ee
The first term is the four-bond version of \reff{a2}:
\be
\widetilde\varphi_{0\,X\cup X'\cup X''\cup X'''}^{(2)} \;=\;
\frac{1}{8}\, \sum_{C_{24}} P^0_{X\cup X'\cup X''\cup X'''}\,
\Bigl[{\rm ad}S_{1{X}_4}({\rm ad}S_{1{X}_3}
({\rm ad}S_{1{X}_2} (Q^{01}_{{X}_1})))\Bigr]\, P^0_{X\cup X'\cup
X''\cup X'''}
\label{a4}
\ee
where 
\be
C_{24} \;=\; \Bigl\{
(X_1,X_2,X_3,X_4) \,:\, X_1, X_2, X_3, X_4 
\in \{ X, X', X'', X'''\}\Bigr\}.
\ee
The second term is the four-bond version of \reff{a37}:
\be
\overline{\varphi}_{0\,X\cup X'\cup X''\cup X'''}^{(2)} \;=\;
\frac{1}{2}\, \sum_{\widetilde C_4}
P^0_{X\cup X'\cup X''\cup X'''} \,\Bigl[ 
{\rm{ad}} S_{2({X_4 \cup X_3})} 
\left({Q^{01}_{2({X_1 \cup X_2})}}\right)\Bigr]\, 
P^0_{X\cup X'\cup X''\cup X'''},
\label{a33}
\ee
where $\widetilde C_4$ is the set of sequences $(X_1,X_2,X_3,X_4)$
satisfying one of the conditions
\be
\Bigl\{ X_1=X, X_2=X', X_3=X'', X_4 =X'''\Bigr\} \; \bigcup\;
\Bigl\{ 3 \hbox{ cyclic permutations} \Bigr\}.
\ee
\medskip

Each of the operators defining the terms \reff{a01}--\reff{a33}
can be expressed in terms of a product of creation and annihilation
operators.  The expressions thus obtained can be simplified by using 
the spin operators defined in \reff{spinop}.  To avoid
unenlighteningly complicated formulas, we write the final
result only in the direction-independent case, 
$t_\pm^b = t_\pm^{\ovb} = t_\pm$ .
\medskip

\noindent
>From \reff{mag} it follows that  if
$t_{\pm b} = t_{\pm \ovb} = t_\pm$, and for  $Y=\nn{xy}$,
\be
\Phi_{0\,\{x,y\}}^{(1)} = \frac{2\,(\tua^2 + \tda^2)}{U} \Bigl(S_x^3 S_y^3 
- \frac{P^0_{\{x,y\}}}{4}\Bigr) + \frac{4\,\tua \tda}{U}\, S_x^{\perp}
S_y^{\perp}, 
\ee
whereas from  \reff{a01}--\reff{a33} we obtain the following expressions:
\medskip

\noindent
\noindent
\emph{(i') For $Y=\nn{xy}$},
\be
\varphi_{0\,\{x,y\}}^{(2)} =
- \frac{2}{U^3} \,(\tua^4 + \tda^4 + 6\,\tua^2 \tda^2 ) 
\,\Bigl(S_x^3 S_y^3 - \frac{P^0_{\{x,y\}}}{4}\Bigr)
\,-\, \frac{8}{U^3} \,(\tua^3 \tda + \tda^3 \tua)
\, S_x^{\perp} S_y^{\perp}.
\label{f1}
\ee
We omit local projectors, because of property \reff{nonec}, and we
denote
\be
S_x^{\perp} S_y^{\perp} \;:= \;S_x^1 S_y^1 +  S_x^2 S_y^2 \;=\;
\frac{1}{2}\,(S_x^+ S_y^- + S_x^- S_y^+)\;.
\label{sperp}
\ee
\smallskip

\noindent
\noindent
\emph{(ii') For $Y=X \cup X'$, with $X= \nn{xy}$ and $X' = \nn{yz}$}:
\bea
\widetilde\varphi_{0\,\{x,y,z\}} &=& 
- \, P^0_{\{x,y,z\}} \,\Bigl[
\Bigl( \frac{2\,(\tua^4 + \tda^4)}{U^3}\, + 
\frac{4\, \tua^2 \tda^2}{U^3}\, \Bigr)\, \Bigl(S_x^3 S_y^3 
+ S_y^3 S_z^3 - S_x^3 S_z^3 - \frac{1}{4}\Bigr)
\nonumber\\ 
&& {}+  \frac{4 \,(\tua^3 \tda + \tda^3 \tua)}{U^3} 
\, \bigl(S_x^{\perp} S_y^{\perp}
+ S_y^{\perp} S_z^{\perp}\bigr) \nonumber\\
&& {}- \frac{8 \tua^2 \tda^2}{U^3}\, S_x^{\perp} S_z^{\perp}
\Bigr]\,P^0_{\{x,y,z\}}
\label{f2}
\eea
and
\bea
\overline\varphi_{0\,\{x,y,z\}} &=& 
P^0_{\{x,y,z\}} \,\Bigr[ \frac{2\,(\tua^4 + \tda^4)}{U^3} 
\Bigl(S_x^3 S_z^3 - \frac{1}{4}\Bigr)
+ \frac{8 \tua^2 \tda^2}{U^3} \,\Bigl(S_x^3 S_y^3  + S_y^3 S_z^3 -
S_x^3 S_z^3 - \frac{1}{4}\Bigr)\nonumber\\
&&{}- \frac{4\, \tua^2 \tda^2}{U^3}\, S_x^{\perp} S_z^{\perp}
+ \frac{4 \,(\tua^3 \tda + \tda^3 \tua)}{U^3}
\,\bigl(S_x^{\perp} S_y^{\perp}
+ S_y^{\perp} S_z^{\perp}\bigr) \Bigr]\,P^0_{\{x,y,z\}}.
\label{f3}
\eea
\smallskip

\noindent
\noindent
\emph{(iii') For $Y=X \cup X' \cup X'' \cup X'''$, with 
$X= \nn{xy}$, $X' = \nn{yz}$, $X''= \nn{zw}$ and $X''' = \nn{zw}$}:
\bea
\lefteqn{
\widetilde\varphi_{0\,\{x,y,z,w\}}^{(2)} \;=\;
P^0_{\{x,y,z,w\}}\,\Bigl\{\frac{8\,(\tua^4 + \tda^4)}{U^3} \,
S_x^3 S_y^3 S_z^3 S_w^3} \nonumber\\
&& {}- \frac{2\,(\tua^4 + \tda^4)}{U^3}\, 
\Bigl(S_x^3 S_y^3  + S_y^3 S_z^3 
+ S_z^3 S_w^3  + S_w^3 S_x^3 -  S_x^3 S_z^3  - S_y^3 S_w^3-
\frac{1}{4}\bigr) 
\nonumber\\
&& {}+ \frac{8}{U^3} \, (\tua^3 \tda + \tda^3 \tua) 
\Bigl[(S_y^{\perp} S_z^{\perp})\,S_x^3 S_w^3 
+ (S_x^{\perp} S_w^{\perp})\,S_y^3 S_z^3 
+ (S_x^{\perp} S_y^{\perp})\,S_z^3 S_w^3 + 
(S_z^{\perp} S_w^{\perp})\,S_x^3 S_y^3 \nonumber\\
&& {}- \frac{1}{4} \Bigl( S_x^{\perp} S_y^{\perp} + S_y^{\perp} S_z^{\perp}
+ S_z^{\perp} S_w^{\perp} + S_w^{\perp} S_x^{\perp}\Bigr)\Bigr]\nonumber\\
&& {}+ \frac{8\, \tua^2 \tda^2}{U^3}\, \Bigl(S_x^+ S_y^- S_z^+ S_w^- 
+   S_x^- S_y^+ S_z^- S_w^+  \Bigr)\nonumber\\
&& {}- \frac{4\,\tua^2 \tda^2}{U^3}\, \Bigl[ 4\,
(S_x^{\perp} S_z^{\perp})\,S_y^3 S_w^3 + 4\,
(S_y^{\perp} S_w^{\perp})\,S_x^3 S_z^3 
- S_x^{\perp} S_z^{\perp} - S_y^{\perp} S_w^{\perp}\Bigr] 
\Bigr\}\,P^0_{\{x,y,z,w\}}\nonumber\\
\label{fz.5}
\eea
and
\bea
\lefteqn{
\overline\varphi_{0\,\{x,y,z,w\}}^{(2)} \;=\;
P^0_{\{x,y,z,w\}}\,\Bigl\{\Bigl[\frac{4\,(\tua^4 + \tda^4)}{U^3}\, 
\Bigl(8\, S_x^3 S_y^3 S_z^3 S_w^3 -   S_x^3 S_z^3  - S_y^3 S_w^3 \Bigr)}
\nonumber\\
&& {}+ \frac{32}{U^3}\, (\tua^3 \tda + \tda^3 \tua) \,
\Bigl[(S_z^{\perp} S_w^{\perp})\,S_x^3 S_y^3 
+ (S_x^{\perp} S_y^{\perp})\,S_z^3 S_w^3 
+ (S_y^{\perp} S_z^{\perp})\, S_x^3 S_w^3 
+ (S_x^{\perp} S_w^{\perp})\,S_y^3 S_z^3 \Bigr]
\nonumber\\
&& {}+ \frac{32 \,\tua^2 \tda^2}{U^3} \,\bigl(S_x^+ S_y^- S_z^+ S_w^- 
+   S_x^- S_y^+ S_z^- S_w^+  \bigr)\nonumber\\
&& {}- \frac{8\,\tua^2 \tda^2}{U^3} \Bigl[ 8\,(S_x^{\perp} S_z^{\perp})\,S_y^3 S_w^3 
+ 8\, (S_y^{\perp} S_w^{\perp})\,S_x^3 S_z^3
+ S_x^{\perp} S_z^{\perp} + S_y^{\perp} S_w^{\perp}
\Bigr]\Bigr\}\,P^0_{\{x,y,z,w\}}\;.
\label{fz.6}
\eea
\medskip

Two particularly interesting limits are the following ones.
\medskip

\noindent
\emph{The symmetric model:}  Setting $t_+ = t_- = t$ in the
above expressions for $\Phi^{(2)}_0$ and grouping operators which
have the same support, we arrive at the expression for the effective
Hamiltonian of the standard one-band Hubbard model in the low energy
sector, to fourth order in the hopping amplitudes:
\bea
\lefteqn{
\bigl[H^{(2)}_{\rm Hub}\bigr]_{\rm eff} \;=\;
\bigl[K^{(2)}_{\rm Hub}\bigr]^{00}} \nonumber\\
&=& - \, \sum_x\Bigl(h S^3_x + \frac{k}{2}
P^0_x\Bigr)\nonumber\\ 
&&{}+  \Bigl(\frac{4t^2}{U} - \frac{16t^4}{U^3}\Bigr)
\sum_{\nn{xy}} \Bigl({\bf{S}}_x \cdot {\bf{S}}_y -
\frac{P^0_{\{x,y\}}}{4}\Bigr) + \frac {4t^4}{U^3} 
\sum_{\nn{xyz}}P^0_{\{x,y,z\}}\,\Bigl({\bf{S}}_x . {\bf{S}}_z - 
\frac{1}{4}\Bigr) \,P^0_{\{x,y,z\}} \nonumber\\
&& {}- \frac {4t^4}{U^3} \sum_{{
\left\{\scriptstyle w \, z \atop  x \, y\right\}}}
P^0_{\{x,y,z,w\}}\,
\Bigl({\bf{S}}_x \cdot {\bf{S}}_y
+ {\bf{S}}_y \cdot {\bf{S}}_z + {\bf{S}}_z \cdot {\bf{S}}_w 
+ {\bf{S}}_w \cdot {\bf{S}}_x\nonumber\\
&& \qquad\qquad\qquad
{}+ {\bf{S}}_x \cdot {\bf{S}}_z + {\bf{S}}_y \cdot {\bf{S}}_w - 
\frac{1}{4}\Bigr) \,P^0_{\{x,y,z,w\}}\nonumber\\
&& {}+   \frac {80t^4}{U^3} \sum_{
\left\{\scriptstyle w \, z \atop  x \, y\right\}}
\Bigl[({\bf{S}}_x . {\bf{S}}_y)
({\bf{S}}_z . {\bf{S}}_w) + ({\bf{S}}_x . {\bf{S}}_w)
({\bf{S}}_y . {\bf{S}}_z) - ({\bf{S}}_x . {\bf{S}}_z)
({\bf{S}}_y . {\bf{S}}_w)\Bigr],\nonumber\\
\ \label{ll.2}
\eea
where the symbol $\nn{xyz}$ denotes that $x$ and $y$ are nearest
neighbor sites and so are $y$ and $z$, while
$\left\{\scriptstyle w \, z \atop  x \, y\right\}$ stands for
four sites around a plaquette.  Expression \reff{ll.2} has been
derived previously.  It has 
been reported, for instance in \cite{bul66}, \cite{tak77} (where the
hopping parameter is allowed to depend on the direction), in
\cite{chaspaole78} (together with higher-order corrections, up to 
eighth order) and in \cite{macgiryos88}.  Our contribution amounts to
showing that if one systematically uses local projections, unlike in the
first three references, the  remainder can be controlled
rigorously. This is the new result, which follows from our results in
\cite{datetal96}. 
\medskip 

\noindent 
\emph{The Falicov-Kimball limit:} Setting $t_+=0, t_-=t$, one obtains
the effective interaction for the Falicov-Kimball model to fourth
order.
\bea \bigl[H^{(2)}_{\rm FK}\bigr]_{\rm eff} &=&
\bigl[K^{(2)}_{\rm FK}\bigr]^{00} \nonumber\\
&=& - \, \sum_x\Bigl(h S^3_x + \frac{k}{2}
P^0_x\Bigr)\nonumber\\
&&{}+ \Bigl(\frac{2t^2}{U} - \frac{2t^4}{U^3}\Bigr) \sum_{\nn{xy}}
\Bigl(S_x^3 S_y^3 -
\frac{P^0_{\{x,y\}}}{4}\Bigr) \nonumber\\
&&{}- \frac{2t^4}{U^3} \, \sum_{\nn{xyz}} P^0_{\{x,y,z\}}\, \Bigl(
S_x^3 S_y^3 + S_y^3 S_z^3 - 2\, S_x^3 S_z^3\Bigr)\,P^0_{\{x,y,z\}}\nonumber\\
&&{}- \frac{2\,t^4}{U^3} \, \sum_{\left\{\scriptstyle w \, z \atop x
    \, y\right\}} P^0_{\{x,y,z,w\}} \Bigl(S_x^3 S_y^3 + S_y^3 S_z^3 +
S_z^3 S_w^3 + S_w^3 S_x^3 + S_x^3 S_z^3 + S_y^3 S_w^3 \Bigr)
P^0_{\{x,y,z,w\}}\nonumber\\
&& {}+ \, \frac{40\,t^4}{U^3} \, \sum_{\left\{\scriptstyle w \, z
    \atop x \, y\right\}} S_x^3 S_y^3 S_z^3 S_w^3 \;.
\label{pia.21}
\eea
For the purposes of determining (classical) ground states we can
ignore the projection operators, in which case the Hamiltonian becomes
\bea
\lefteqn{ - \, \sum_x\Bigl(h S^3_x + \frac{k}{2}
P^0_x\Bigr)
+  \Bigl(\frac{2t^2}{U} - \frac{18t^4}{U^3}\Bigr)
\sum_{\nn{xy}} \Bigl(S_x^3 S_y^3 -
\frac{1}{4}\Bigr)}\nonumber\\
&&{} + \frac{4t^4}{U^3} \, \sum_{x,y \atop{|x-y| =2}}
S_x^3 S_y^3 
+ \frac{6\,t^4}{U^3} \, \sum_{x,y \atop{|x-y| = \sqrt{2}}}S_x^3 S_y^3
\, + \, \frac{40\,t^4}{U^3} \, 
\sum_{\left\{\scriptstyle w \, z \atop  x \, y\right\}}
S_x^3 S_y^3 S_z^3 S_w^3,
\label{pia.11}
\eea 
This effective classical Hamiltonian and its ground states, were
studied in \cite{ken94}.  In \cite{gruetal97} a more general
expression was obtained involving hopping amplitudes that depend on
the direction: $\tas{\ }{[yx]}=\tas{*}{[yx]}=|\tas{\ 
  }{[yx]}|e^{i\theta_{xy}}$, $\theta_{xy}=-\theta_{yx}$.  Such complex
hopping amplitudes describe the influence of a magnetic field.  Our
expansion methods provide systematic control of remainder terms.
\bigskip

\subsection{Higher-order perturbation theory}

While straightforward, in principle, the computation of 
higher-order transformed interactions is a
tedious and error-prone process.  (In fact, already the second order
computations lead to monstrous expressions if  hopping is allowed
to depend on the direction as well as  spin.)  The use of computer codes
 to perform symbolic algebra becomes mandatory.  To encourage
industrious readers, we  sketch  some basic elements
of the algebra needed to treat the resulting expressions.

The different terms arising are products of two types of operators:  (i)
creation and destruction operators, and (ii) projections.  We can
therefore classify the relevant terms in three categories.

\subsubsection{Products of creation and destruction operators}
The algebraic manipulations of these operators are well known:  one
uses the anti-commut\-ation relations \reff{anticom1}--\reff{anticom2}
to obtain expressions in terms of the occupation numbers
$n_{x\sigma}=c^\dagger_{x\sigma}c_{x\sigma}$ and the spin operators
$S_x^{\pm}$ defined in \reff{spm}.  Furthermore, differences of
occupation numbers can be written in terms of $S_x^3$ via
\reff{sx3}.  

\subsubsection{Products of projections}
There is only one equation to consider:
\begin{equation}
P^i_x\,P^j_x \;=\; P^i_x\,\delta_{ij}\;, \quad i,j\;=\;0,1 \ .
\label{pipj}
\end{equation}
A similar equation holds for projections on any set $X$.
Another pair of potentially useful relations are
\begin{equation}
P^0_X\, P^0_{\widetilde X} = P^0_{\widetilde X} \quad , \quad
P^1_X\, P^1_{\widetilde X} = P^1_X \quad , \quad
\label{pxx}
\end{equation}
whenever $X\subset \widetilde X$.  

\subsubsection{Products of projections and creation- and destruction
operators}
This is the non-trivial part of the algebra.  Perhaps, its most important
relations are the intertwining relations 
\begin{equation}
\begin{array}{ccc}
c_{x\sigma} \, P^0_x = P^1_x \, c_{x\sigma}  &\ &
c_{x\sigma} \, P^1_x = P^0_x \, c_{x\sigma}\\[5pt]
c^\dagger_{x\sigma} \, P^0_x = P^1_x \, c^\dagger_{x\sigma}  &\ & 
c^\dagger_{x\sigma} \, P^1_x = P^0_x \, c^\dagger_{x\sigma} \;,
\end{array}
\label{intt}
\end{equation}
for arbitrary $\sigma=\pm$.  They allow a purely algebraic
verification of which terms are zero. We remark that projections
of the form $P^1_X$, with $X$ containing more than one site, are not
directly suited for the use of \reff{intt}, because these projections
are \emph{not} products of single-site projections. We then use that 
\begin{equation}
P^1_X \;=\; 1 - P^0_X \;=\; 1- \prod_{x\in X} P^0_x\;.
\label{ump}
\end{equation}

As an example, let us sketch the algebraic proof that 
\begin{equation}
\bigl[{\rm ad}S_{1\sigma X}(Q^{01}_{\sigma' X'})\bigr]^{00} \;=\; 0
\label{xxn}
\end{equation}
for $X=\nn{xy}$ and $X'=\nn{yz}$, with $x\neq z$; a fact only
parenthetically mentioned after \reff{this}.  Indeed, the left-hand
side of \reff{xxn} is a combination of products of the form
\begin{equation}
P^0_{\{x,y,z\}}\, P^0_{\{x,y\}}\, c^\dagger_{y\sigma}\,c_{x\sigma}\,
P^1_{\{x,y\}} \, P^1_{\{y,z\}} \, c^\dagger_{z\sigma'}\,c_{y\sigma'}\,
 P^0_{\{y,z\}}\,  P^0_{\{x,y,z\}}\;,
\label{manyp}
\end{equation}
plus various $x \,\longleftrightarrow\, y$ and $y
\,\longleftrightarrow\, z$ permutations.
We first notice that the middle projections $P^1_X$
can be removed.  To see this we use \reff{ump} and then move, for
instance, the resulting factors $P^0_y$ to the far left and far right
through the intertwining relations \reff{intt}.  Such factors emerge
as $P^1_y$ and cancel with the corresponding factors $P^0_y$ at both
ends [relation \reff{pipj}].  We are left with
\begin{equation}
P^0_{\{x,y,z\}}\, c^\dagger_{y\sigma}\,c_{x\sigma}\,
c^\dagger_{z\sigma'}\,c_{y\sigma'}\, P^0_{\{x,y,z\}}\;,
\label{lessp}
\end{equation}
where we have also used \reff{pxx}.  To see that such a term is zero,
use again the intertwining relations to show, for instance, that as
there is an odd number of creation and annihilation operators involving
the site $z$, the leftmost $P^0_z$ can be written as $P^1_z$ on the
right.  It then annihilates with the corresponding $P^0_z$.

\noindent
There are many other useful relations involving products of
projections and creation and destruction operators.  Besides the
previous formulas \reff{nonec}, \reff{hoy.1}, \reff{hoy.2} and
\reff{fla.5}/\reff{fla.6}, we mention
\begin{eqnarray}
n_{x\sigma}\, P^0_x &=& [1-n_{x(-\sigma)}] \,P^0_x
\label{hoy.10}\\
n_{x\sigma}\, P^1_x &=& {n_x\over 2} \,P^1_x\,
\label{hoy.11}
\end{eqnarray}
and
\begin{eqnarray}
c^\dagger_{x\sigma'}\,c^\dagger_{x\sigma}\,P^0_x &=& 0
\label{hoy.12}\\[5pt]
c^\dagger_{x\sigma}\,c_{x(-\sigma)}\,P^1_x &=& 0
\label{hoy.13}
\end{eqnarray}
for any $\sigma, \sigma'=\pm$.  
By the intertwining relations \reff{intt} the identities 
\reff{hoy.10}--\reff{hoy.13} can also be written with the
projections on the right or  on the left and the right.

\section{Low-temperature phase diagrams}\label{lowte}

\subsection{Foreword on quantum Pirogov-Sinai theory}
\label{forpir}

The purposes of Sections \ref{pert} and \ref{perthub} was to construct
a unitary operator, $U^{(n)}(\lambda)$, with the property that the
conjugated Hamiltonian
\begin{eqnarray}
  \label{z.70}
   H^{(n)}_\Lambda (\lambda) &=&  U^{(n)}_\Lambda (\lambda) \,
H_\Lambda (\lambda) \, {U^{(n)}_\Lambda (\lambda)}^*\nonumber\\
&=& H^{(n)}_{0\Lambda} (\lambda) + V^{(n)}_\Lambda (\lambda)
\end{eqnarray}
of some quantum lattice system is ``block-diagonal'' to order
$|\lambda|^{n+1}$.  By this we mean that the matrix elements of 
$V^{(n)}_\Lambda (\lambda)$ between ground states, or approximate
ground states (low-energy states) of $H^{(n)}_{0\Lambda} (\lambda)$
are of order $n+1$.  Our method to construct $U^{(n)}_{\Lambda}
(\lambda)$ has been (a convergent form of) analytic perturbation
theory. 

In attempting to gain some insight into the structure of ground states
and low-temperature equilibrium states of the Hamiltonian
$H^{(n)}_{\Lambda} (\lambda)$, it is tempting to argue that one should
neglect the perturbation $V^{(n)}_\Lambda (\lambda)$ and study the
ground states and low-temperature equilibrium states of
$H^{(n)}_{0\Lambda} (\lambda)$ ---hoping that they are close to those
of $H^{(n)}_{\Lambda} (\lambda)$.  Unfortunately, this hope is
ujustified, in general.

It is a well known fact that arbitrarily small perturbations can have
a drastic effect on low-temperature phase diagrams.  This is typically
the case when the truncated part has an infinite degeneracy.  For
example, when there is a continuous symmetry, states are usually
grouped in bands with gapless intraband excitations.  An arbitrarily
small additional term can split the lower bands and hence change the
phase diagram.  Such models are hard to treat rigorously; in
particular, no such treatment is available for the (symmetric) Hubbard
model, \emph{despite our control on the perturbation series of Section
  \ref{perthub}}.  The case in which the truncated interaction has an
infinite degeneracy not connected with a continuous symmetry is
usually more tractable.  Often, entropy effects (quantum and
classical) are the determining aspect, and there exist techniques to
deal with them, as we will illustrate below.

The simplest situation arises when the truncated interaction has a
zero-temperature phase diagram involving only finite degeneracies.  In
favorable cases, we expect that each of these ground states gives rise
to a corresponding ground state, and a low-temperature state, of the
full, untruncated interaction.  The detection and description of these
phases, however, requires some additional machinery.  Indeed, even at
zero temperature, where only energy plays a role, we cannot rely on
plain diagonalization procedures. First, there is no hope of achieving
exact block diagonalization of the whole \emph{family} of Hamiltonians
$\left\{H_\L\right\}$ needed to pass to the thermodynamic limit,
barring miracles (that is, relatively trivial and uninteresting
situations).  And, second, the estimation of the effect of the
``undiagonalized'' remainders is a delicate matter.  The remainder
term, $V^{(n)}_\Lambda (\lambda)$, makes a negligible contribution to
the \emph{energy density}.  But it may have a decisive \emph{quantum
  entropic} effect: It may alter the overall wave function by allowing
``virtual quantum transitions'' between classical configurations.  The
situation is even more involved at nonzero temperatures, due to the
appearance of thermal fluctuations.
\medskip

\noindent
\paragraph{Quantum Pirogov-Sinai theory} \mbox{}
\medskip

Its bare-bones version \cite{datferfro96,borkotuel96} applies to
systems with a leading part that is ``classical'', in the sense of
being diagonal in a tensor product basis (Section \ref{prelim}).  It
is based on \emph{contour expansions}.  The Duhamel expansion is used
to write partition functions and expectations as sums over
$(d+1)$-dimensional piecewise cylindrical surfaces ---the $(d+1)$-st
axis corresponding to the inverse temperature (imaginary time) axis.
Such {\it ``contours''} have cylindrical pieces, corresponding to
Peierls-like contours for the classical part of the interaction, and a
finite number of section changes caused by the quantum perturbation.
The objective of the theory is to prove that, in the thermodynamic
limit, these contours are sparse and far apart.  This yields a precise
mathematical representation of states as ``classical ground states
modified by quantum and thermal fluctuations''.  In fact, both types
of fluctuations are treated on the same footing. So quantum
Pirogov-Sinai theory provides a unified description of ground- and
low-temperature states.

In brief, the theory establishes sufficient conditions for the energy
to dominate over further quantum and thermal entropic effects.  As a
consequence, the latter can only cause small (and smooth) deformations
of the zero-temperature phase diagram of the classical part.  That is,
the phase diagram remains stable under quantum perturbations and small
raises of temperature.  This does not mean that all phases remain
stable throughout the phase diagram.  For instance, small entropic
effects can tilt the balance in favor of particular phase(s) of a
coexistence region and cause the disappearance of the remaining ones
when the thermal or quantum perturbation is switched on.  But the
theory assures that the unfavored phases reappear for slightly
modified parameter values, that is, the perturbed system will also
exhibit a coexistence manifold, albeit a little shifted and deformed.
Moreover, the theory provides a criterion to determine which phases
are stable, and, in fact, it presents a picture of what ``stable
phase'' means.  [As a matter of fact, below we shall resort only to
this stability criterion.  The part on the stability of phase diagrams
will not be useful, as coexistence lines will, in all cases, involve
infinite degeneracies.]

As we discuss in more detail below, the conditions on the classical
part required by the theory are of two types: First, it must lead to
finite ground-state degeneracies within the region of interest, and,
second, it must satisfy the \emph{Peierls condition} which roughly
requires the energy cost of an excitation to be proportional to the
area of its boundary.  The first requirement forces us to stay away
from regions of infinite degeneracy.  So, in principle, we would be
unable to deal with the line $\mu_+=\mu_-$ of the phase diagram of
Figure \ref{fig1}.  We shall see below that the quantum hopping
actually helps us there.  Similarly, models with low-cost energy
excitations, like the balanced model, are out of reach.  In fact, in
\cite{frorey96} it is shown how quantum hopping brings this model
within reach.

When applicable, the theory does a complete job:  It
accounts for entropic effects and provides a full description of 
states.  
Nevertheless, the theory has a limited scope In fact, it
seems to be applicable only when nothing interesting happens as a
result of the perturbation.  The hidden card is
our perturbation technique.  In many instances, it produces a partially
diagonalized interaction with a leading classical part incorporating
crucial quantum effects.  If such a leading part satisfies the right
hypotheses, the subsequent application of Pirogov-Sinai theory will
prove the existence of phases exhibiting  truly quantum-mechanical
features.  In fact, a slight extension of Pirogov-Sinai is needed in
these cases, showing that transitions within the excited band have a
negligible effect, regardless of the order.  Such an extension, based
on a more careful definition of contours that distinguishes
``high-to-high'' transitions from the rest, is presented in
\cite{datetal96}. 
Below, we  prove that hopping produces 
quantum symmetry breaking among the ground states of the on-site 
Hubbard interaction, which brings the degeneracy down to a finite
one.  This is an instance of ``quantum entropic selection''.
Another example, presented in \cite{frorey96}, is the 
``quantum restoration of the Peierls condition'' taking place in the
balanced model.

Incidentally, the thermal entropic selection is a better known
phenomenon, which, in some cases, can trigger a complicated pattern of
phases with cascades or  
staircases of phase transitions.  This effect has been the object of
extensions of Pirogov-Sinai theory based on the notion of
\emph{restricted ensembles}
\cite{brisla85,sla85,grusut88,brisla89,kotetal93}.   
The extension of this theory to quantum systems is  a
promising direction of research.  The work in \cite{kotuel98} can be
considered as a step in this direction, though it does not treat genuinely
degenerate quantum interactions and only resorts to restricted
ensembles as an alternative way to exhibit degeneracy
breaking.

Quantum Pirogov-Sinai theory acts, thus, as the link needed to pass
from diagonalization ---at the level of interactions--- to phases.  It
is a powerful tool that can be used as a black box: Once the system is
seen to be thermodynamic within its range of applicability there is
nothing else to do; the theory provides a flexible and complete
description of the resulting phases.  We know of no other comparably
general and versatile approach to the study of quantum statistical
mechanical phase diagrams.  We see the methodological difference
between quantum mechanics and quantum statistical mechanics.  In the
former, the ultimate goal is to diagonalize, \emph{exactly}, the
Hamiltonian of some 
system.  In the latter, an exact diagonalization is impossible. The
goal is then to do the \emph{minimal} perturbative diagonalization
needed to bring the system within the reach of some theory enabling us
to determine the phase diagram, like our Pirogov-Sinai approach.  Once such a
theory takes over, we are done: All the required information is at our
disposal; any further diagonalization is a waste of time.

\subsection{Summary and hypotheses}
\label{sumhyp}

\subsubsection{Scope of the theory}
Quantum Pirogov-Sinai theory applies to interactions of the form
\reff{inter}, that is,
\begin{equation} 
\Phi = \phicl + Q \;, 
\label{fs.hqv} 
\end{equation} 
where $\phicl = \left\{\phicl_{X}\right\}$ is a finite range classical
interaction (classical in the sense of Section \ref{prelim}), and $Q =
\left\{Q_X\right\}$ is a possibly quantum interaction.  The
operators comprising both parts depend on a finite family of
perturbation parameters 
$\lambda=\{\lambda_1, \ldots, \lambda_k\}$.  For our applications below,
$\phicl$ will be the leading part $\Phi^{(n)}_0$ of a transformed
interaction of an appropriate order $n$.
The theory has two consequences:

\begin{itemize}
\item[(PS1)] \emph{Stability of states}: Under suitable hypotheses, it
  determines the different periodic states of the full interaction,
  for small values of the parameters $\lambda$ and small temperatures.
  Each of these states can be traced, in fact, to some ground state of
  $\phicl$, from which it differs in the addition of a diluted gas of
  thermal and quantum excitations.  In such a case, the corresponding
  ground state of $\phicl$ is said to be \emph{stable} at the given
  temperature and values of $\lambda$.

\item[(PS2)] \emph{Stability of phase diagrams}: It establishes
  (sufficient) conditions under which the zero-temperature phase
  diagram of the classical part remains ``stable'' under the addition
  of the quantum part, $Q$, and/or increasing temperature.  That is,
  the manifolds where different phases coexist are, for small
  temperatures and quantum perturbations, smooth and small
  deformations of the corresponding coexistence manifolds of ground
  states of the classical part.
\end{itemize}

The theory requires two sets of hypotheses: (i) the classical leading
part must have a finite ground-state degeneracy and satisfy the
Peierls condition explained below, and (ii) the matrix elements of the
quantum part must be sufficiently small.

\subsubsection{Hypotheses on the classical part}
\label{hypclas}

Let us discuss the hypotheses on the classical part first.  The finite
degeneracy refers to the presence of only a finite number of
\emph{periodic} ground states for $\phicl$.  In the present setting
($m$-potentials with minimal bond energies normalized to zero) these
are configurations $\omega_1, \ldots, \omega_p$ such that
$\phicl_{X}(\omega_i)=0$ for all bonds $X$.  The Peierls condition
refers to the well known generalized Peierls contours
\cite{pirsin75,pirsin76,sin82}.  They are constructed by means of
\emph{sampling plaquettes}
\begin{equation}
  \label{t.sam}
W_x^{(a)}:= \{ y\in \zed^\nu : |y_i - x_i | \le a, \, {\rm{for}} \,
1 \le i \le \nu\}  \;, \hbox{ for } x\in\zed^\nu\;.
\end{equation}
The set of sampling plaquettes where a configuration $\omega$ does not
coincide with any of the ground states $\omega_i$ is called the
\emph{defect set} of $\omega$.  The \emph{contours} of $\omega$ are
pairs $\gamma=(M,\omega_M)$, where $M$ ---the \emph{support} of
$\gamma$--- is a maximally connected (with respect to intersections)
component of the defect set.  The radius $a$ of the sampling
plaquettes must be larger than the range of $\phicl$ and the period of
each of the $\omega_i$, so that, knowing the set of contours, one can
univocally reconstruct the configuration $\omega$.  In particular,
when $\phicl$ is a transformed interactions $\Phi^{(n)}_0$, $a$ must
be larger than the radius $R$ of the plaquettes used to define
protection zones (Section \ref{prelim}).

For models whose interactions and excitations are determined by
nearest-neighbor conditions, like the Ising model, one can use the
``thin'' Peierls contours \cite{peierls36.f,dob65.f,gri64.f} traced
with segments midway between pairs of adjacent sites.  But in more
general cases one must resort to definitions as above to ensure a
one-to-one correspondence between configurations and families of
contours.  The energy of a configuration is, in the present setting,
the sum of energies, $E(\gamma)$, of contours $\gamma$.  Here,
$E(\gamma)$ is the energy of the configuration having $\gamma$ as its
only contour.  A model satisfies the \emph{Peierls condition} if each
contour has an energy proportional to the cardinality of its support,
$s(\gamma)$.  That is,
\begin{equation}
E(\gamma) \;>\; \kappa s(\gamma)
\label{t.1}
\end{equation}
for some $\kappa>0$ called the \emph{Peierls constant}.

In this paper, we consider a more detailed Peierls condition based on
the distinction of two levels of excitations ---low- and high-lying---
of the classical part.  The distinction is made through a suitably
defined family of local projections $P^0_Y$.  With respect to this
family, a configuration $\omega$ exhibits a \emph{high-lying
  excitation} in $Y$ if $P^0_Y\,\omega \neq \omega$.  For the
applications of this paper, these projections are exactly those
defined in Section \ref{prelim}, because the low-lying excitations of
$\Phi^{\rm cl}=\Phi_0^{(n)}$ are among the local ground states of the
original $\Phi_0$.  The \emph{high-energy defect set} of a
configuration $\omega$ is the union of the plaquettes $W^{(a)}_x$
where it exhibits a high-lying excitation.  A system satisfies a
\emph{two-level Peierls condition} if for each contour $\gamma$ we
have
\begin{equation} 
E(\gamma)  \;>\; \kappa \, \size(\gamma) + 
D\, \size(\gamma^{\rm high})
\label{szp.10}
\end{equation} 
where $\gamma^{\rm high}$ is the part of the high-energy defect set
contained in $\gamma$.  This condition means, in particular, that
high-energy excitations lead to an additive \emph{excess} energy
measured by $D>0$.

Given that $\phicl$ is allowed to depend on the perturbation
parameters $\lambda$, such a dependence is also expected for the
Peierls constants $\kappa$ and $D$.

\subsubsection{Hypotheses on the quantum part}

First, we decompose the operators of the quantum interaction,
$Q=\left\{Q_X\right\}$, into 
``low$\to$low'' ($\ell \ell$), ``low$\to$high'' ($\ell {\rm h}$),
``high$\to$low'' (${\rm h}\ell$) and ``high$\to$high'' (${\rm h}{\rm
h}$) components:
\begin{equation}
\begin{minipage}{7cm}
\begin{eqnarray*}
  \label{szp.1}
  Q^{\ell \ell}_X &\bydef& P^0_X\,Q_X\, P^0_X\\
  Q^{\ell {\rm h}}_X &\bydef& ({\bf 1}_X - P^0_X)\,Q_X\, P^0_X\\
  Q^{{\rm h} \ell}_X &\bydef& P^0_X\, Q_X\,({\bf 1}_X - P^0_X)\\
  Q^{{\rm h} {\rm h}}_X &\bydef& ({\bf 1}_X - P^0_X)\,Q_X\, 
      ({\bf 1}_X - P^0_X)\;. \label{szp.1.1}
\end{eqnarray*}
\end{minipage}
\label{t.4m}
\end{equation}

The conditions on the quantum part are that there exist sufficiently
small numbers $\varepsilon_{\alpha\gamma}$ and $\delta$ such that, for
$\alpha,\gamma = \ell, {\rm h}$
\begin{equation}
  \label{z.1}
  \left\|Q^{\alpha \gamma}_X\right\| \;\le\;
  \varepsilon_{\alpha\gamma}\, \delta^{\size(X)}\;.
\end{equation}
The region of validity of the two-level quantum Pirogov theory
depends on the parameter
\begin{equation} 
\eta \;\bydef\; \max \Biggl(
{\varepsilon_{\ell\ell} \delta\over\kappa} \,,\, 
\delta \sqrt{{\varepsilon_{{\rm h}\ell}\varepsilon_{\ell{\rm h}} \over
    \kappa (\kappa + D)}} \,,\,  
{\varepsilon_{{\rm h}{\rm h}} \delta\over \kappa + D} \,,\,
{\varepsilon_{\ell{\rm h}}\delta\over \kappa + D} \,,\,
{\varepsilon_{{\rm h}\ell}\delta\over \kappa + D}
\Biggr)\;.
\label{shg.1} 
\end{equation} 
Quantum Pirogov-Sinai theory converges provided $\eta$ is sufficiently
small.

The original (one-level) quantum Pirogov-Sinai theory discussed in
\cite{datferfro96,borkotuel96} corresponds to taking
$D=\varepsilon_{{\rm h}\ell}=\varepsilon_{\ell{\rm h}}=
\varepsilon_{{\rm h}{\rm h}}=0$.

Interactions obtained as a result of the partial block-diagonalization
procedure of Section \ref{pert}, contain terms proportional to powers
of $\lambda/D$; ($\lambda$ measures the strength of the quantum
perturbation, and $D$ is proportional to a typical
energy-denominator).  Condition \reff{z.1} thus imposses the bound
\begin{equation}
  \label{z.40}
  \max \Bigl(\lambda\,,\, {\lambda\over D}\Bigr) \;<\; \delta\;,
\end{equation}
where $\delta$ is as in \reff{z.1}; ($\delta$ must be chosen small
enough for the perturbation expansions to converge).  The size of the
${\rm h}\ell$-matrix elements of the quantum perturbation,
$\{Q_X\equiv Q^{(n)}_X (\lambda)\}$, in the Hamiltonian
$H^{(n)}_{\Lambda} (\lambda)$ is estimated by
\begin{equation}
  \label{z.45}
  \|Q^{{\rm h}\ell}_X\| \;\sim\; \|Q^{\ell{\rm h}}_X\| \;\sim\;
{\lambda^{n+1}\over D^n} \,
  \delta^{\size(X)-n-1}\;.
\end{equation}
Comparing \reff{z.45} to \reff{z.1} we find that
\begin{equation}
  \label{z.50}
  \varepsilon_{{\rm h}\ell} \;=\;   \varepsilon_{\ell{\rm h}}
\;=\; \OO\Bigl( {\lambda^{n+1}\over D^n \delta^{n+1}} \Bigr)\;.
\end{equation}
As the ``high$\to$high'' part of the original interaction survives
unaltered, 
\begin{equation}
  \label{z.55}
  \varepsilon_{{\rm h}{\rm h}} \;=\; \OO\Bigl( {\lambda\over\delta}
  \Bigr)\;.  
\end{equation}

\subsection{Stability of phases}

\subsubsection{The basic criterion}

Our criterion for the stability of phases is based on the use of
cluster-expansion technology to construct some objects $f'_1,\ldots,
f'_k$ ---where $f'_i$ is associated to the ground state $\omega_i$ of
$\phicl$--- called \emph{truncated free energy densities}.  They are
defined provided that
\begin{equation}
\varepsilon \;\bydef\;
\max\Bigl( e^{-\beta\OO(\kappa)}\,,\, \eta\Bigr) \;<\; \varepsilon_0
\;, 
\label{t.30}
\end{equation}
for some (small) constant $\varepsilon_0$ that depends on parameters
like the range of $\phicl$, dimensionality of the lattice, size of the
sampling plaquette, and dimension of the on-site Hilbert space.  Within
the convergence region, the truncated free energies are analytic
functions of $\beta$ and of any parameter on which the interaction has
an analytic dependence.  They determine the stability of phases in the
following way.

\paragraph{Stability criterion}
\label{stabph}
\emph{Assume that $\lambda$ is such that $\phicl(\lambda)$ has a
  finitely degenerate ground state and satisfies the Peierls condition
  \reff{szp.10}, and assume that $Q(\lambda)$ satisfies hypotheses
  \reff{z.1}}.  Then if, for some values of $\beta$ and $\lambda$
within the region \reff{t.30},
\begin{equation}
\re f'_p(\underline\mu) \;=\; \min_{1\le q\le k} 
\re f'_q(\underline\mu)\;,
  \label{err.1}
\end{equation}
for some $p$ with $1\le p \le k$, the following holds:
\begin{itemize}
\item[(i)]  $f'_p(\underline\mu)$ coincides with the true free-energy
  density of
  the system;
\item[(ii)]  the infinite-volume limit
\begin{equation}
\lim_{\Lambda\nearrow\zed^d}
{\tr^{\omega_p}_\Lambda \,{\bf A}
\, e^{-\beta {\bf H}^{\omega_p}_\Lambda} \over
\tr^{\omega_p}_\Lambda \, e^{-\beta {\bf H}^{\omega_p}_\Lambda} }
\;\defby\; \langle {\bf A}  \rangle^{\omega_p}_{\beta\,\lambda}
\label{ffx.1}
\end{equation}
exists for any local operator ${\bf A}$; 
\item[(iii)]
\begin{equation}
\left| \langle {\bf A} \rangle^{\omega_p}_{\beta\,\lambda} -
\langle \omega_p | {\bf A} | \omega_p \rangle \right | \;\le\; \|{\bf
A}\| |X| \OO(\varepsilon) \ , 
\label{ffx.2}
\end{equation}
for any operator ${\bf A}\in {\cal A}_{X}$, ($X$ is a finite subset of
the lattice).
\end{itemize}
The notation $H^{\omega_p}_\Lambda$ stands for the Hamiltonian with
``external condition'' $\omega_p$, that is, the sum of all $\Phi_X$
with $X\cap\Lambda\neq\emptyset$, but allowing only matrix elements
between vectors that coincide with $\omega_p$ outside $\Lambda$.
The symbol $\tr^{\omega_p}_\Lambda$ indicates a trace over the space of
such vectors.
\medskip

Note that, in \reff{ffx.1}, $H^{\omega_p}_{\Lambda}$ is the
``effective'' Hamiltonian, rather than the original Hamiltonian
defining the model. The effective Hamiltonian and the original one are
unitarily equivalent, the unitary conjugation being given by an
operator $U^{(n)}_\Lambda (\lambda)$ as in \reff{z.80}, \reff{z.85},
\reff{z.90}.  In order to select a stable phase, we impose boundary
conditions $\omega_p$ outside boxes $\Lambda$ on the \emph{effective
  Hamiltonians} rather than the original Hamiltonians.  One could
reconstruct ``boundary conditions'' for the original Hamiltonians
corresponding to the boundary conditions that we impose on the
effective Hamiltonians.  (Typically, ``boundary conditions'' for the
original Hamiltonian will constrain configurations not only outside
$\Lambda$ but also insided $\Lambda$ but near $\partial\Lambda$.  We
shall not attempt to provide details concerning the map from boundary
conditions for the original Hamiltonians to boundary conditions for
the effective Hamiltonians.  Boundary conditions do \emph{not} have an
operational meaning here (corresponding, e.g.\ to certain experimental
conditions), but are \emph{mathematical devices to select stable
  phases} that should be chosen in as convenient a way as possible.
In fact, this is the conventional role of boundary conditions in
statistical physics.  When we describe a magnetic material in terms of
a quantum Heisenberg model we are working with an \emph{effective
  Hamiltonian}; (there are no explicit exchange interactions present
in the fundamental Schr\"odinger Hamiltonian!).  But we do not
hesitate to impose boundary conditions directly on the Heisenberg
Hamiltonian, rather than on the fundamental Schr\"odinger Hamiltonian
of the material, in order to select stable pure phases (e.g.\ the
direction of spontaneous magnetization).

Enough!  Let's get down to business!
\medskip

We shall say that there is a {\em stable} $\omega_p$-phase whenever
\reff{err.1} is satisfied.

We shall not enter here into the technicalities of the actual proof of
this criterion, which can be found in \cite{datferfro96}. Instead, we
present some general comments on the basic construction. 

The above criterion is proven by resorting to a low-temperature
expansion obtained by iteration of Duhamel's formula for the
exponential of the sum of non-commuting matrices.  This adds an extra
continuous variable ranging from 0 to $\beta$.  There is one such
expansion for each ground state of $\phicl$, involving a ``sum'' of
terms labelled by the sites of the lattice (``spatial variables'') and
the continuous ``inverse temperature'' or ``time'' variable.  Every
term can be labelled by a piecewise cylindrical surface contained in
$\zed^d\times[0,\beta]$, which can be decomposed into connected
components called \emph{quantum contours}.  Outside the surface, the
space is filled with the corresponding ground state of $\phicl$, and
the contours are transition regions whose interiors are occupied by
different ground states of $\phicl$.  The contours have spatial
sections formed by the usual classical contours described above, which
grow cylindrically in the ``time'' direction until there is a sudden
change of section due to the action of a quantum bond.  The hypotheses
on the quantum part ensure that the weight of this contour decays
exponentially with its area: The exponential cost of each ``vertical''
cylindrical piece is given by the Peierls condition \reff{szp.10},
while each ``horizontal'' change of section is penalized by a term
\reff{z.1} according to the type of transition involved.  Each contour
has a finite number of section changes.

We must distinguish between ``short'' and ``long'' contours.  The long
contours traverse the whole ``time'' axis from zero to $\beta$.
Therefore, their cost is at least $\exp(-\const\beta)$, and they
disappear in the limit $\beta\to\infty$.  In particular, a purely
classical system would have \emph{only} contours of this type which,
in fact, would be straight cylinders with no section changes.
``Short'' contours appear and/or disappear at intermediate values of
the continuous integration variable.  Thus they involve some changes
of sections, and their cost is proportional to $\delta$.  Such
contours survive the limit $\beta\to\infty$ and contribute to the
quantum ground state of the full interaction $\Phi$.

We see that, in this approach, thermal and quantum effects are put on
a similar footing.  They are both sources of entropy associated to
different contour geometries.  A low-temperature equilibrium state can
be visualized as a ``sea'' configured of the corresponding ground state
of $\phicl$ plus the fluctuations represented by contours.  In
particular, the short contours can be interpreted as ``ground state
fluctuations''.  At low temperatures and small values of $\delta$,
these fluctuations are dilute, because their large energy cost
overwhelms the ``entropy gain''.  Thus expectations in such a state
differ little from expectations in the associated ground state of
$\phicl$.

\subsubsection{Stability and symmetries}

Two types of symmetries can be distinguished.  First, there are
symmetries that leave each term of the interaction (or of an
equivalent form of it) invariant.  Such a symmetry is associated, for
each finite volume $\Lambda$, to a unitary operator $S_\Lambda$ such
that 
\begin{equation}
S_\Lambda H^{\omega_p}_\Lambda S^{-1}_\Lambda \;=\; 
H^{S\omega_p}_\Lambda\;.
\label{sim.1}
\end{equation}
[The vector $S \omega_p$ is easy to visualize at the level of
classical configurations.  In a quantum statistical mechanical
formalism, it is defined, for instance, via limits of 
$S^{-1}_\Lambda \omega_p$.]  Examples of these symmetries are the spin-flip
symmetry of Ising models in zero field, or the particle-hole symmetry
of the Hubbard model at half filling.

On the other hand, there are symmetries associated to operations on
the lattice, like translations and rotations.  They map each term of
the interaction into a different term, but leave the whole interaction
invariant.  Each such symmetry is defined by a bijection
$T:\zed^d\to\zed^d$ that yields natural bijective maps between
configurations ---$(T\omega)_x=w_{T^{-1}x}$--- and between local
operators ---$(T\Phi)_X=\Phi_{T^{-1}X}$ (for notational simplicity we
denote the different transformations with the same symbol).  The
latter map does not leave Hamiltonians invariant, but connects
Hamiltonians corresponding to translated volumes:
\begin{equation}
T H^{\sigma_p}_\Lambda T^{-1} \;=\; 
H^{T\sigma_p}_{T^{-1}\Lambda}\;.
\label{sim.2}
\end{equation}

In general, a symmetry, $R$, can be a composition, $ST$, of symmetries of
the previous two types.  

The criterion of stability of phases presented above respects
symmetries.  Indeed, we say that a symmetry $R$ \emph{connects} a
family of configurations $\{\omega_1,\ldots,\omega_\ell\}$ if
\begin{equation}
R^i\,\omega_1 \;=\; \omega_{i+1} \quad \hbox{ for } 1\le i\le \ell-1\;.  
\label{s.s1}
\end{equation}
Then, under the hypotheses of the preceding subsection, we have:
\begin{equation}
\begin{minipage}{10cm}
\noindent
\emph{
Ground states of $\phicl$ connected by a symmetry of $\Phi$ are
either all stable or all unstable. 
}
\end{minipage}\label{sim.3}
\end{equation}

We briefly sketch a proof of this fact, as it is not explicitly
given in \cite{datferfro96} or in \cite{datetal96}.  The key
object to look at is the partition function $\Xi_p(V)$ for  
piecewise cylindrical, bounded regions, $V$, of $\zed^d \times
[0,\beta]$.  This object is defined, through the contour expansion, as
the sum of all allowed contour configurations inside $V$ compatible
with the boundary condition $\omega_p$.  For regions $V$ of the form 
$\Lambda\times[0,\beta]$, with $\Lambda$ a finite region of $\zed^d$,
one has
\begin{equation}
\Xi_p(\Lambda) \;=\;
\tr^{\omega_p}_\Lambda \, e^{-\beta {\bf H}^{\omega_p}_\Lambda} \;.
\label{sim.5}
\end{equation}
Looking carefully into the definition of $\Xi(V)$, it is not hard to
conclude that its behavior with respect to symmetries is similar to
that of \reff{sim.5}.  Namely, if $S$ a symmetry of the first type,
\begin{equation}
\Xi_{S\omega_p}(V) \;=\; \Xi_{\omega_p}(V)\;,
\label{sim.6}
\end{equation}
and, if $T$ is a symmetry of the second type
\begin{equation}
\Xi_{T\omega_p}(V) \;=\; \Xi_{\omega_p}(T^{-1} V)\;.
\label{sim.7}
\end{equation}
Here $T^{-1}V$ refers to the space-time region $V$ transformed by the
map $T^{-1}$.

These partition functions are involved in the following
characterization of stability:  A ground state $\omega_p$ of $\phicl$
is stable in the region \reff{t.30} if, and only if, there exists
some constant $C$ such that 
\begin{equation}
\left|{\Xi_{\omega_j}(V)\over\Xi_{\omega_p}(V)}\right| \;\le\;
\exp(C\left|\partial V\right|)\;,
\label{sim.4}
\end{equation}
for every bounded region $V\subset\zed^d\times [0,\beta]$ and $1\le j
\le k$.  Here $|\partial V|$ indicates the area of the external 
boundary of $V$.

The criterion \reff{sim.3} is now easy to verify:  Assume that some
ground states $\omega_1,\ldots,\omega_\ell$ are connected by a symmetry.
If the symmetry is of the first type, then by \reff{sim.6}
\begin{equation}
\Xi_1(V) \;=\; \cdots \;=\; \Xi_\ell(V)\ ,
\label{sim.10}
\end{equation}
for all piecewise cylindrical $V\subset\zed^d\times[0,\beta]$.  Hence
\reff{sim.4} is either verified by all $p=1,\ldots,\ell$ or by none.

Consider then a symmetry of the second type, and assume, for
concreteness, that $\omega_\ell$ is stable.  Then \reff{sim.4} is
verified for $p=\ell$ and all $j$, in particular for $j=q$ where
$\omega_q={T^{-(\ell-q)}\omega_\ell}$ ($1\le q\le \ell$) [see
\reff{s.s1}].  But then, from \reff{sim.7} we have
\begin{equation}
\left|{\Xi_{\omega_\ell}(V)\over\Xi_{\omega_q}(V)}\right| \;=\;
\left|{\Xi_{\omega_{\ell-q}}(T^{-(\ell-q)}V)\over\Xi_{\omega_\ell}
    (T^{-(\ell-q)}V)}\right| 
\;\le\; \exp(C\left|\partial V\right|)\;.
\label{sim.9}
\end{equation}
The last inequality follows from the assumed stability of
$\omega_\ell$ and the fact that $|\partial (T V)|=|\partial V|$.
Therefore, for any $1\le q \le \ell$, and any $1\le j \le k$,
\begin{equation}
\left|{\Xi_{\omega_j}(V)\over\Xi_{\omega_q}(V)}\right| \;=\;
\left|{\Xi_{\omega_j}(V)\over\Xi_{\omega_\ell}(V)}\right| \,
\left|{\Xi_{\omega_\ell}(V)\over\Xi_{\omega_q}(V)}\right| 
\;\le\; \exp(2C\left|\partial V\right|)\;,
\label{sim.15}
\end{equation}
where the inequality follows from the stability of $\omega_\ell$
[eq.\ \reff{sim.4}] and from \reff{sim.9}.  We conclude that all the
ground states $\omega_1,\ldots,\omega_\ell$ are stable (with an
exponential constant $2C$).

\subsection{Stability of phase diagrams}
\label{stabdiag}

More generally, one studies families of interactions $\Phi_{\mu}$
parametrized by a finite set of parameters
$\mu=\{\mu_1,\ldots,\mu_s\}$ ---typically fields or chemical
potentials--- and one is interested in determining the corresponding
\emph{phase diagram}, that is, in obtaining a catalogue of the
different phases present for different values of $\mu$.  If for the
values of $\mu$ under consideration the system satisfies the
hypotheses described above, one can apply quantum Pirogov-Sinai theory
and determine the phase diagram on the basis of condition
\reff{err.1}, now involving $\mu$-dependent truncated free energies
$f'_p$. The cluster expansion tells us that the difference between
these truncated free energies and the corresponding energy densities
is of the order of the parameter $\varepsilon_{\beta\lambda}$
introduced in \reff{t.30}.  Therefore, given the smoothness properties
of $f'_p$, one would expect that the phase diagram, \emph{for $\beta$
  and $\lambda$ in the region} \reff{t.30}, differs little from the
phase diagram of $\phicl$ at zero temperature.

This can, indeed, be proven, under some minor additional hypotheses:
Chiefly, (i) bounds similar to \reff{z.1} but involving the partial
derivatives $\partial Q_{\mu\, X} /\partial\mu_i$, and (ii) a
hypothesis of \emph{regularity} of the phase diagram of $\phicl_\mu$.
The latter roughly means that the parameters $\mu$ completely break
degeneracies among the ground states, so the Gibbs phase rule is
satisfied.  The detailed hypotheses can be found in \cite[Section
5.2]{datetal96}.  The conclusion is: \medskip

\noindent\emph{If the hypotheses sketched above are satisfied for an
open region ${\cal O}$ of the space of parameters $\mu$, then the
phase diagram for $\beta$ and $\lambda$ in a region of the form
\reff{t.30} is regular and is a smooth deformation of the 
zero-temperature phase diagram of  
$\Phi^{{\rm cl}}_{\underline\mu}$ in ${\cal O}$.  The displacement of
the different coexistence manifolds is of the order of
$\varepsilon_{\beta\lambda}$.}

\section{Phase diagram for the Falicov-Kimball regime}
\label{phase}

We now apply the phase-diagram technology described in the previous
section to asymmetric Hubbard models, whose perturbation expansion has
been discussed in Section \ref{perthub}.  Let us denote
\begin{eqnarray}
  \label{z.5}
  t &=& \sup_{[x,y]}\left|\td{[x,y]}\right|\\
\label{z.10}
  \tua &=& \sup_{[x,y]}\left|\tu{[x,y]}\right|
\end{eqnarray}
(with a slight abuse of notation).  Let us see how
further features of the phase diagram appear as we consider higher
orders in the perturbation expansion.

\subsection{Phase diagram to order $t^0$}

To this order, the classical leading interaction correspond to a (formal)
Hamiltonian 
\begin{equation}
H^{{\rm cl}\, (0)} \;=\; U \sum_{x } n_{x +} \, n_{x -} 
-  \mu_{ +} \sum_{x } n_{x +} -
\mu_{ -} \sum_{x } n_{x -}\;.
\label{f.1}
\end{equation}
We see from Figure \ref{fig1} that it trivially satisfies the
hypotheses of Section \ref{hypclas} ---in fact for the one-level
theory--- as long as one avoids the lines where the different regions
with a unique ground state intersect.  In the latter we loose the
required finite degeneracy.  The conclusion is, therefore, that
different ground states remain stable for chemical potentials in the
open regions of uniqueness.  The range \reff{t.30} shrinks to zero as
we approach a coexistence manifold, because so do the different
Peierls constant (which are of the order of the energy of the lowest
excitation).  This phase stability remains true upon the addition of
electron hopping or \emph{any} quantum perturbation which is
translation-invariant and satisfies $\eta\le \varepsilon_0$
(for instance small ion-hopping).

\subsection{Phase diagram  to order $t^2/U$}
\label{phase2}

To this order we can exhibit more features of the phase diagram within
the shaded region in Figure \ref{shaded}.  The transformed interaction
was obtained in Section \ref{hubfirst} [Formula \reff{mag}].  It has a
leading classical part 

\begin{equation}
H^{{\rm cl}\,(1)} \;=\; U \sum_{x } n_{x +} \, n_{x -} +
\frac{2}{U} \sum_{\nn{xy}} |\td{[x,y]}|\,
\Bigl( S^3_x S^3_y - \frac{P^0_{\{x,y\}}}{4} \Bigr)
- {h} \sum_{x} S^3_x\;,
\label{f.10}
\end{equation}
whose periodic ground-state configurations are given in Figure
\ref{fig4a}.  This interaction satisfies a two-level Peierls
condition.  The low-lying excitations have one particle per site and
involve a Peierls constant
\begin{equation}
  \label{z.15}
\kappa\;\sim\;4t^2/U\pm h + \OO(t^4/U^3)\;.  
\end{equation}
The excess energy for configurations in
the non-singly occupied band gives an extra Peierls rate 
\begin{equation}
  \label{z.20}
D\;=\;\max (U-\mu_0 \,,\, \mu_0) \;=\;
\OO(U)
\end{equation}
($\mu_0$ defines the boundaries of the shaded region in Figure
\ref{shaded}; for concreteness we are considering $\mu_0\sim U$).
We see that the leading part \reff{f.10}
satisfies the hypotheses required for the classical part in Section
\ref{hypclas} as long as
\begin{equation}
  \label{z.15.1}
4t^2/U\pm h + \OO(t^4/U^3) \;>\; 0\;,
\end{equation}
that is, except within bands of width $\OO(t^4/U^3)$ centered at the
lines $h=\pm 4 t^2/U$. Moreover, the summability of the transformed
interaction is guaranteed only if $D\gg \kappa$. This condition
determines the limits of the shaded region in Figure \ref{shaded}.

The ``low$\to$low'' transitions must include at least one ionic jump,
hence a factor $\tua$.  Therefore [see eg.\ the quantum correction in
\reff{mag}],
\begin{equation}
  \label{z.25}
\varepsilon_{\ell\ell} = \OO\Bigl({\tua\over t}\,\cdot\, {t^2\over
  U\delta}\Bigr)\;,
\end{equation}
where $\delta$ has been chosen sufficiently small to guarantee the
convergence of the partially diagonalized interaction.  From this,
\reff{z.50} (with $n=1$) and \reff{z.55} we see that the quantum
perturbation satisfies hypotheses \reff{t.30} if
\begin{equation}
  \label{z.30}
{\tua\over t} \,,\,  {t\over U} < \widetilde\varepsilon_0 \;,
\end{equation}
where $\widetilde\varepsilon_0$ is a small number that shrinks to zero
when $\kappa$ tends to zero (and at the boundaries of the shaded
region in Figure \ref{shaded}).  

We conclude that quantum Pirogov-Sinai theory [together with the
symmetry criterion \reff{sim.3}] proves that for \emph{nonzero} but
small values of $t$, the phase diagram of the model is as in Figure
\ref{fig4a}.  The ground states correspond to the configurations shown
in the figure plus quantum fluctuations (=short space-time contours).
Furthermore, they remain stable for small temperatures (long contours
appear sparingly) and small values of $\tua$ (new contours appear,
rarely, involving the new type of quantum transitions).  This picture
is valid outside an excluded $\OO(t^4/U^3)$-vicinity of the lines
$h=\pm2 t^2/U$ where the Peierls constant $\kappa$ cannot be
guaranteed to be positive, and inside the shaded region of Figure
\ref{shaded} where $\kappa \ll D$.  This stability persists under the
addition of arbitrary quantum perturbations satisfying \reff{z.1} that
do not break translation invariance.

\subsection{Phase diagram up to order $t^4/U^3$}

In order to analyze in more detail the excluded regions of widths
$\OO(t^4/U^3)$, we need to consider the next order of perturbation
theory.  For the direction-independent case ($t_-^b\equiv t$), the
leading classical interaction, to this order, is given by the terms in
\reff{pia.11}.  At this point, the hardest part of the analysis is, by
far, the determination of the ground states of this complicated
classical part.  Fortunately, this has been done before, see
\cite{ken94,gruetal97}: the ground states correspond to the
configurations depicted in Figure \ref{fig4b}.  We see
that the classical part, to fourth order, satisfies the hypotheses
required by Pirogov-Sinai theory, except in the vicinity of four lines
of infinite degeneracy.  In regions at a distance $\OO(t^6/U^5)$ from
these lines, the Peierls constant $\kappa$ is of order $t^4/U^3$.  The
gap $D$ is still given by \reff{z.20}.

Moreover, the quantum part has coefficients $\varepsilon_{{\rm
    h}\ell}$ and $\varepsilon_{\ell{\rm h}}$ given by \reff{z.50} with
$n=2$, and $\varepsilon_{{\rm h}{\rm h}}$ given by \reff{z.55}, while
\begin{equation}
  \label{z.60}
\varepsilon_{\ell\ell} = \OO\Bigl({\tua\over t}\,\cdot\, {t^4\over
  U^3\delta^4}\Bigr)
\end{equation}
[see eg.\ the quantum corrections in \reff{f1}--\reff{fz.6}].  

As a consequence, the stability criterion of Section \ref{stabph} is
applicable, for large $\beta$ and $t$ satisfying \reff{z.30} with a
smaller $\widetilde\varepsilon_0$. Together with the symmetry
considerations \reff{sim.3}, this implies the stability of the ground
states of Figure \ref{fig4b} ---except at the excluded bands of
width $\OO(t^6/U^5)$ around the coexistence lines--- under the
addition of temperature, ionic hopping [within the limits impossed by
\reff{z.30}], and any other translation- and rotation-invariant
quantum perturbation satisfying \reff{t.30}.  The phase diagram of
Figure \ref{fig4b} is thus obtained.

A similar analysis can be performed for cases in which the hopping
depends on the direction.  In particular, in the presence of magnetic
flux, ie.\ if $\tas{-}{[yx]}=t \exp(i\theta_{yx})$ with $\theta_{yx}
\in (0,2\pi)$, our block-diagonalization procedure yields, to order
$t^4/U^3$, an effective classical interaction with terms given in
\cite[Formula (3.11)]{gruetal97}.  As shown in this reference, the
phase diagram for the interaction with flux involves the same ground
states as in the flux-less case, but with deformed manifolds of
infinite degeneracy.  The Pirogov-Sinai approach proves the stability
of these ground states, under quantum and thermal perturbations,
except in excluded regions of width $t^6/U^5$ around the lines of
infinite degeneracy.

\section{Perturbation expansion for the 3--band \
 Hubbard \ model}
\label{3band}

In all the models considered in the previous section, 
the Coulomb interactions between the particles were strictly on--site.
In this section we apply our perturbation method to the $3$--band
Hubbard model, as an example of how to proceed when the leading
interaction has nonzero range.

\subsection{The original interaction}

\begin{figure}
\vspace{1cm}
\epsfxsize=\hsize
\epsfbox{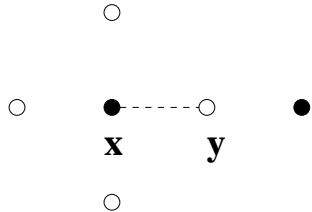}
\vskip 0.2in
\caption{Set $B_X$ formed by the bond $\nn{xy}$ and its protection
  zone for the 3-band Hubbard model}
\label{box3}
\end{figure}

The Hamiltonian $H_{0\Lambda}$ in \reff{3hub} is of unit range, since
it consists of both on--site and nearest neighbor interactions. This
necessitates the use of ``protection zones'' [see Section
\ref{prelim}] for the proper treatment of the local character of the
operators that arise in our perturbation expansion. We define
$R$-plaquettes, $W_x$, through \reff{wx} with $R=1$, and sets $B_X$
through \reff{bx}. In particular, for $X := \nn{xy},\,x\in A, y\in B$,
$|x - y| = 1$, $B_X$ is given by the set of sites shown in Figure
\ref{box3} (or a rotation and/or reflection of it).

It consists of two copper sites and four oxygen sites.
We restrict our attention to parameter values in the following ranges:
\be
U_d > \epsilon_d \,\,;\Delta > 0\,\,;\,\, U_d \gg  
U_p >U_{pd} > 0\,\,;\,\, U_d \gg  \Delta
\,\,;\,\, t_{pd} >0\,\,;\,\, U_d \gg t_{pd},
\label{choice}
\ee
and normalize the ground state energy of $H_{0\Lambda}$ to zero. The 
  unperturbed Hamiltonian is given in terms of an interaction $\Phi_0 =
  \{\Phi_{0Y}\}$, nonzero, if $Y = \nn{xy}$ denotes a pair of nearest
  neighbor sites such that 
$x\in A$ and $y\in B$. The interaction $\Phi_{0X}$ is defined as follows.
\be
 \Phi_{0Y}\;= \; \left\{\begin{array}{ll}
 U_d \, n^d_{x +} \, n^d_{x -}  +    \epsilon_d\,  (n_{x}^d - 1)  
& \hbox{if } Y=\{x\}\,,\, x\in A \\[5pt]
  U_p \,  n^p_{y +} \, n^p_{y -} +  \epsilon_p \, n_{y}^p
& \hbox{if } Y=\{y\}\,,\, y\in B \\[5pt]
U_{pd} \, n^d_{x} \, n^p_{y} 
& \hbox{if } Y=\nn{xy}\,,\, x\in A, y\in B 
\end{array}\right.
\label{3bphi0}
\ee 
For parameter values satisfying \reff{choice}, the energy of a
configuration on $Y$, with respect to the interaction $\Phi_{0Y}$, is
minimum (and equal to zero) when there is a single hole at $x$ and
none at $y$. Hence in a ground state configuration of $H_{0\Lambda}$,
each oxygen site is empty while each copper site is singly occupied by
a hole, the total number of holes in $\Lambda$ being equal to the
total number of copper sites, $|A|$. However, the ground state of
$H_{0\Lambda}$ has a $2^{|A|}$-fold spin degeneracy.

The interaction corresponding to the $3$-band Hubbard Hamiltonian
$H_\Lambda$ \reff{3hub}, with the normalization introduced above, is
given by $\Phi = \{\Phi_{Y}\}$ where \be \Phi_Y = \Phi_{0Y} + Q_Y,
\label{phix3}
\ee
and $Q_Y=0$ unless $Y$ is a pair of nearest-neighbor sites, in which
case 
\be
Q_{\{xy\}} = t_{pd}\, [p^{\dagger}_{y\sigma} 
d_{x\sigma} + d^{\dagger}_{y\sigma} p_{x\sigma}].
\label{qb3}
\ee
We define suitable projection operators on the Hilbert space
$\HH_\Lambda$ as in Section \ref{prelim}. Let $P^0_{B_X}$ denote an operator 
which projects onto (local) ground states of the interaction
$\Phi_0$.  The projection operators
$P^1_{B_X}$ and $P^2_{B_X}$ are defined by \reff{p1} and \reff{p2} of
Section \ref{prelim}. Using the resulting partition of unity 
\be
{\bf{1}} =  P^0_{B_X} +  P^1_{B_X} + P^2_{B_X}\ ,
\label{unity3}
\ee 
we can decompose the perturbation interaction $Q_X$ of \reff{qb3}
as in \reff{qdecom}.  It is clear from the structure of the lattice
[Figure \ref{fig5}] and the definition \reff{qb3} of $Q_X$ that
$Q^{00}_{B_Y} =0$.  Hence 
\be Q_X = Q^{01}_{B_X} + Q^{R}_{B_X},
\label{v3decom}
\ee 
where $Q^{01}_{B_X}$ and $Q^{R}_{B_X}$ ---defined in \reff{q01}
and \reff{qr}--- are linear in the hopping amplitude $t_{pd}$. 
Also, due to our choice of ground-state energy normalization,
\begin{equation}
  \label{pia.30}
  \Phi_{0B_X}^{00} \;=\; 0\;.
\end{equation}

\subsection{First-order perturbation for the 3-band Hubbard model}

As explained in Section \ref{pert}, we first search for a unitary
transformation $U^{(1)}(t_{pd})= \exp(t_{pd} S_1)$ with 
\be
S_1 := \sum_X S_{1B_X},
\ee 
which eliminates the first--order off--diagonal terms $Q^{01}_{B_X}$
of the perturbation interaction $Q$.  From \reff{jf} we have that
\be
S_{1B_X} \;=\;  P^0_{B_X} \frac {Q_{B_X}}{E_0 - E_1}  P^1_{B_X}
+ P^1_{B_X} \frac {Q_{B_X}}{E_1 - E_0}  P^0_{B_X}\,
\label{3jf}
\ee
where
\be 
E_{0} = 0 \quad {\rm{and}} \quad E_{1} = U_{pd} + \Delta,
\ee
with
\be
\Delta := \epsilon_p - \epsilon_d\;.
\label{tdelta}
\ee
Hence
\be
S_{1B_X} \;=\;  {1\over U_{pd} + \Delta} \,
\Bigl\{ P^1_{B_X} \,Q_{B_X}\,  P^0_{B_X} - 
 P^0_{B_X} \,Q_{B_X}\,  P^1_{B_X} \Bigr\}
\label{pia.31}
\ee
for $X \equiv \nn{xy}$, $x\in A$ and $y\in B$.
A straightforward calculation along the lines sketched in Section
\ref{firstpert} shows that the transformed classical interaction has
nonzero terms only for nearest-neighbor pairs $X$.  These terms are
[recall \reff{pia.30}]:
\bea
\Phi^{(1)}_{0B_X} &=& 
\frac{1}{2} P^0_{B_X} \,{\rm ad}S_{1{B_X}}(Q^{01}_{{B_X}})\, P^0_{B_X}
\nonumber\\
&=&\, -\, \frac{t_{pd}^2}{(U_{pd} + \Delta)}  P^0_{B_X}\,\Bigl[
\sum_{\sigma=+, -}  d^{\dagger}_{x\sigma} p_{y\sigma}p^{\dagger}_{y\sigma} 
d_{x\sigma}\Bigr] \, P^0_{B_X}\nonumber\\
&=&\,- \,\frac{t_{pd}^2}{(U_{pd} + \Delta)}P^0_{B_X} \sum_{\sigma=+, -}
\Bigl(n_{x\sigma}^d
(1 - n_{y\sigma}^p)\Bigr)P^0_{B_X} \nonumber\\
&=&\,- \,\frac{t_{pd}^2}{(U_{pd} + \Delta)} P^0_{B_X}.
\label{phi003}
\eea
In the last line we used the identity
\be
P^0_{B_X} \Bigl(\sum_{\sigma=+, -}
n_{x\sigma}^d
(1 - n_{y\sigma}^p)\Bigr)P^0_{B_X} = P^0_{B_X} \Bigl(\sum_{\sigma=+, -}
n_{x\sigma}^d\Bigr)  P^0_{B_X} = 1.
\ee

We conclude that, to this order, the change in the classical
interaction amounts to an irrelevant shift in the (local) ground-state
energy that does not introduce any new effect.  In particular, it
fails to reduce the spin degeneracy.  We need to go to the next order
of our perturbation scheme to find non-trivial contributions.

\subsection{Second-order perturbation for the 3-band Hubbard model}

>From Section \ref{secondpert} we obtain that, in second-order 
perturbation theory, the leading classical part,
$\Phi_0^{(2)} = [\Psi_0^{(2)}]^{00}$, is of the form
\begin{equation}
  \label{pia.25}
  \Phi_0^{(2)}  = \Phi_0^{(1)} + \varphi_0^{(2)}\;,
\end{equation}
where the nonzero terms of the latter are
\begin{equation}
  \label{pia.50}
\varphi_{0B_X}^{(2)}
  \;=\; \frac{1}{8}\, P^0_{B_X} \,
\Bigl[ {\rm{ad}} S_{1B_X} ( {\rm{ad}} S_{1B_X}
({\rm{ad}} S_{1{B_X}} ( Q^{01}_{B_X})))\Bigr] \,P^0_{B_X}\;,
\end{equation}
for $X$ a nearest-neighbor pair, and
\begin{eqnarray}
  \varphi_{0B_Y}^{(2)} &=& 
\frac{1}{8}\, \sum_{C_6}P^0_{B_Y} \,\Bigl[{\rm
  ad}S_{1{{B_X}}_4}({\rm ad}S_{1{{B_X}}_3}
({\rm ad}S_{1{{B_X}}_2}
(Q^{01}_{{{B_X}}_1})))\Bigr]\, P^0_{B_Y} \nonumber\\
&& {}+  \frac{1}{2} \, \sum_{C_4}P^0_{Y} \,\Bigl[{\rm{ad}}
S_{2({{B_X}_4 \cup {B_X}_3})} 
\left({V^{01}_{2({{B_X}_1 \cup {B_X}_2})}}\right)\Bigr]\, P^0_{Y}\;,
  \label{pia.55}
\end{eqnarray}
for $Y=X \cup X'$, where $X=\nn{xy}$, $X'=\nn{yz}$ are pairs of
nearest neighbor sites with $x,z\in A$ and $y\in B$.  The assignments
$C_6$ and $C_4$ in the last two sums are those defined in
\reff{pia.40} and \reff{pia.41}, respectively.

The one-bond contribution \reff{pia.50} is again an uninteresting
energy shift: 
\be 
\varphi_{0B_{\{x,y\}}}^{(2)} \; =\;
 \frac{t_{pd}^4}{(U_{pd} + \Delta)^3}\,  P^0_{B_{\{x,y\}}} \;.
\label{op1}
\ee 
The two-bond contribution \reff{pia.55} can be cast in a more
familiar-looking form by making use of spin operators at copper sites,
[see \reff{spinop}]:
\begin{equation}
{\bf S}_x := \sum_{s, s'=+,-} d^\dagger_{x s} 
{\bf S}_{s s'} d_{x s'}\;.
\label{copspinop}
\end{equation}
One obtains
\be 
\varphi_{0B_{\{x,y,z\}}}^{(2)} \;=\; 
\Bigl[\frac{2\,t_{pd}^4}{(U_pd + \Delta)^3}\Bigr]\,
P^0_{B_{\{x,y,z\}}} + \,{J_{\rm{eff}}} \,
P^0_{B_{\{x,y,z\}}} \,\Bigl({\bf{S}}_x.{\bf{S}}_z - \frac{1}{4} \Bigr)\, 
 P^0_{B_{\{x,y,z\}}}\;,
\label{last}
\ee
with
\be
J_{\rm{eff}} \;:=\; \frac{4 t_{pd}^4}{(U_{pd} + \Delta)^2}\, \left[ 
\frac{1}{U_d} +   \frac{2}{2\Delta + U_p}\right]\;.
\label{jeff}
\ee

>From \reff{pia.25}, \reff{phi003}, \reff{op1} and \reff{last} we
recover the known fact
that the effective Hamiltonian for the $3$-band Hubbard model in the 
low-energy sector, to order $4$ in the hopping amplitude $t_{pd}$, is given
by a $S=1/2$ antiferromagnetic Heisenberg model on the square lattice
of copper sites.  This is the same interaction obtained for the
one-band Hubbard model to order $t^2$ [third line in \reff{ll.2}].
This observation justifies, in part, the use of the simpler one-band
model in studies of magnetic properties of undoped cuprates.  Our
contribution is, once again, to be able to provide convergent
estimates on remainder terms in the transformed interaction.

\section*{Acknowledgements}  We thank Daniel Ueltschi for informing
us about his work with Roman Koteck\'y.  We also
thank Christian Borgs, Jennifer T.\ Chayes and Luc Rey-Bellet for their
interest in our work.  R.~F.\ is grateful to the Institute de Physique
Th\'eorique, EPF-Lausanne, and the Institut f\"ur Theoretische Physik,
ETH-Z\"urich, for hospitality during the completion of this work.  He
also wishes to acknowledge travel support by FAPESP (grant
1998/3366-4).


\end{document}